\documentclass[10pt,journal,twoside]{IEEEtran}
\usepackage{amsthm}
\usepackage{amssymb}
\usepackage[top=0.65in, bottom=0.7in, left=0.55in, right=0.6in]{geometry}
\usepackage{amsmath}
\usepackage{newtxmath}
\usepackage[acronym, nonumberlist]{glossaries}
\makenoidxglossaries

\newglossarystyle{ieeeacronyms}{%
    {\setlength{\parindent}{0pt}%
     \setlength{\parskip}{0pt}}%
    {}%
}
\setglossarystyle{ieeeacronyms}

\usepackage{cite}
\usepackage{tabularx}
\usepackage{epsf}
\usepackage{amsmath}
\usepackage{epsfig}
\usepackage{psfrag}
\usepackage{graphicx}
\usepackage{multirow}
\usepackage{array}
\usepackage{enumerate}
\usepackage{amsthm}
\usepackage{float,lscape}
\usepackage[caption=false,font=small]{subfig}

\usepackage[ruled, vlined, linesnumbered]{algorithm2e}
\usepackage{comment}
\usepackage{color}
\usepackage{scalefnt}

\usepackage{rotating}
\usepackage{breqn}
\usepackage[hyphens]{url}
\usepackage{nomencl}
\usepackage{multicol}
\makenomenclature
\newcommand\Nomenclature[2]{\nomenclature[#1]{#1}{#2}}

\usepackage[normalem]{ulem}
\usepackage{booktabs}

\usepackage{tikz}
\usetikzlibrary{arrows.meta,calc}
\usepackage[T1]{fontenc}
\usepackage{newtxtext,newtxmath}
\usepackage{bm}
\usepackage{hyperref}
\hypersetup{
   colorlinks=true,
   linkcolor=black,
   filecolor=black,      
   urlcolor=black,
   pdftitle={Overleaf Example},
   pdfpagemode=FullScreen,
   }

\makenoidxglossaries

\newacronym{2g}{2G}{second generation}
\newacronym{3g}{3G}{third generation}
\newacronym{3gpp}{3GPP}{3rd Generation Partnership Project}
\newacronym{5g}{5G}{fifth generation}
\newacronym{6g}{6G}{sixth generation}
\newacronym{6lowpan}{6LoWPAN}{IPv6 over low-power wireless personal area networks}

\newacronym{aah}{AAH}{at-home healthcare}
\newacronym{aco}{ACO}{ant colony optimization}
\newacronym{af}{AF}{amplify-and-forward}
\newacronym{ai}{AI}{artificial intelligence}
\newacronym{aircomp}{AirComp}{over-the-air computation}
\newacronym{amr}{AMR}{autonomous mobile robot}
\newacronym{ann}{ANN}{artificial neural network}
\newacronym{aoi}{AoI}{Age of information}
\newacronym{ap}{AP}{access point}
\newacronym{api}{API}{application programming interface}
\newacronym{ar}{AR}{augmented reality}
\newacronym{arq}{ARQ}{automatic repeat request}
\newacronym{auroc}{AUROC}{area under the receiver operating characteristic}
\newacronym{awgn}{AWGN}{additive white Gaussian noise}

\newacronym{bci}{BCI}{brain--computer interface}
\newacronym{bda}{BDA}{big data analytics}
\newacronym{ber}{BER}{bit error rate}
\newacronym{bhtp}{BHTP}{binary hypothesis testing problem}
\newacronym{ble}{BLE}{Bluetooth Low Energy}
\newacronym{bp}{BP}{backpropagation}
\newacronym{bpsk}{BPSK}{binary phase-shift keying}
\newacronym{bs}{BS}{base station}

\newacronym{can}{CAN}{controller area network}
\newacronym{capex}{CAPEX}{capital expenditure}
\newacronym{cc}{CC}{cloud computing}
\newacronym{cdma}{CDMA}{code-division multiple access}
\newacronym{cegc}{CEGC}{cooperative equal-gain combining}
\newacronym{cfar}{CFAR}{constant false-alarm rate}
\newacronym{ch}{CH}{cluster head}
\newacronym{cmrc}{CMRC}{cooperative maximal-ratio combining}
\newacronym{cnn}{CNN}{convolutional neural network}
\newacronym{comp}{CoMP}{coordinated multipoint transmission and reception}
\newacronym{covid19}{COVID-19}{coronavirus disease 2019}
\newacronym{cpms}{CPMS}{cyber-physical manufacturing systems}
\newacronym{cr}{CR}{cognitive radio}
\newacronym{crn}{CRN}{cognitive radio network}
\newacronym{csi}{CSI}{channel state information}
\newacronym{csmaca}{CSMA/CA}{carrier-sense multiple access with collision avoidance}
\newacronym{css}{CSS}{cooperative spectrum sensing}
\newacronym{ctde}{CTDE}{centralized training with decentralized execution}
\newacronym{cv}{CV}{Chair--Varshney}

\newacronym{d2d}{D2D}{device-to-device}
\newacronym{dagsvm}{DAGSVM}{directed acyclic graph support vector machine}
\newacronym{df}{DF}{decision fusion}

\newacronym{daf}{DAF}{decode-and-forward}

\newacronym{dfr}{DFR}{decision-fusion rule}
\newacronym{dl}{DL}{deep learning}
\newacronym{dnn}{DNN}{deep neural network}
\newacronym{drl}{DRL}{deep reinforcement learning}

\newacronym{ecc}{ECC}{error-control coding}
\newacronym{ecg}{ECG}{electrocardiogram}
\newacronym{ed}{ED}{energy detector}
\newacronym{ee}{EE}{energy efficiency}
\newacronym{ef}{EF}{estimate-and-forward}
\newacronym{egc}{EGC}{equal-gain combining}
\newacronym{eh}{EH}{energy harvesting}
\newacronym{embb}{eMBB}{enhanced mobile broadband}
\newacronym{emrc}{EMRC}{enhanced maximal-ratio combining}

\newacronym{ethercAT}{EtherCAT}{Ethernet for Control Automation Technology}

\newacronym{fc}{FC}{fusion center}
\newacronym{fd}{FD}{flexible data rate}
\newacronym{fdma}{FDMA}{frequency-division multiple access}
\newacronym{fedvqcs}{FedVQCS}{federated learning via vector-quantized compressed sensing}
\newacronym{fl}{FL}{federated learning}
\newacronym{fov}{FoV}{field of view}

\newacronym{ga}{GA}{genetic algorithm}
\newacronym{glrt}{GLRT}{generalized likelihood-ratio test}
\newacronym{gmm}{GMM}{Gaussian mixture model}
\newacronym{gnn}{GNN}{graph neural network}
\newacronym{gnss}{GNSS}{Global Navigation Satellite System}
\newacronym{gprs}{GPRS}{general packet radio service}
\newacronym{gps}{GPS}{Global Positioning System}
\newacronym{gsm}{GSM}{Global System for Mobile Communications}
\newacronym{gsn}{GSN}{green social networking}

\newacronym{harq}{HARQ}{hybrid automatic repeat request}
\newacronym{hart}{HART}{Highway Addressable Remote Transducer}
\newacronym{hd}{HD}{high definition}
\newacronym{hdaf}{HDAF}{hybrid decode-amplify-and-forward}

\newacronym{ieee}{IEEE}{Institute of Electrical and Electronics Engineers}
\newacronym{iid}{IID}{independent and identically distributed}
\newacronym{iiot}{IIoT}{industrial Internet of Things}
\newacronym{ils}{ILS}{ideal local sensors}
\newacronym{imdd}{IM/DD}{intensity modulation/direct detection}
\newacronym{imu}{IMU}{inertial measurement unit}
\newacronym{ioe}{IoE}{Internet of Everything}
\newacronym{iomt}{IoMT}{Internet of Medical Things}
\newacronym{iot}{IoT}{Internet of Things}
\newacronym{ip}{IP}{Internet Protocol}
\newacronym{ipv6}{IPv6}{Internet Protocol version 6}
\newacronym{ir}{IR}{ideal relays}
\newacronym{it}{IT}{information technology}
\newacronym{its}{ITS}{intelligent transportation system}
\newacronym{itu}{ITU}{International Telecommunication Union}
\newacronym{isac}{ISAC}{Integrated sensing and communication}

\newacronym{kld}{KLD}{Kullback--Leibler divergence}
\newacronym{knn}{KNN}{$k$-nearest neighbor}

\newacronym{lar}{LAR}{link-adaptive regeneration}
\newacronym{led}{LED}{light-emitting diode}
\newacronym{lidar}{LiDAR}{light detection and ranging}
\newacronym{llr}{LLR}{log-likelihood ratio}
\newacronym{lmpt}{LMPT}{locally most powerful test}
\newacronym{lora}{LoRa}{long range}
\newacronym{los}{LoS}{line of sight}
\newacronym{lpd}{LPD}{low-power device}
\newacronym{lpwa}{LPWA}{low-power wide-area}
\newacronym{lpwan}{LPWAN}{low-power wide-area network}
\newacronym{lrt}{LRT}{likelihood ratio test}
\newacronym{lstm}{LSTM}{long short-term memory}
\newacronym{lte}{LTE}{Long Term Evolution}

\newacronym{m2m}{M2M}{machine-to-machine}
\newacronym{mac}{MAC}{medium access control}
\newacronym{map}{MAP}{maximum a posteriori}
\newacronym{marl}{MARL}{multi-agent reinforcement learning}
\newacronym{mdp}{MDP}{Markov decision process}
\newacronym{mec}{MEC}{multi-access edge computing}
\newacronym[
  shortplural={MEMS},
  longplural={microelectromechanical systems}
]{mem}{MEMS}{microelectromechanical system}

\newacronym{mes}{MES}{manufacturing execution system}
\newacronym{mhtp}{MHTP}{$M$-ary hypothesis testing problem}
\newacronym{mimo}{MIMO}{multiple-input multiple-output}
\newacronym{miot}{mIoT}{massive IoT}
\newacronym{ml}{ML}{machine learning}

\newacronym{mld}{MLD}{maximum-likelihood detector}
\newacronym{mle}{MLE}{maximum-likelihood estimation}
\newacronym{mlla}{MLLA}{max-log with linear approximation}
\newacronym{mlp}{MLP}{multilayer perceptron}
\newacronym{mmtc}{mMTC}{massive machine-type communication}

\newacronym{mmwave}{mmWave}{millimeter wave}
\newacronym{mrc}{MRC}{maximal-ratio combining}
\newacronym{mtc}{MTC}{machine-type communication}

\newacronym{nan}{NAN}{neighborhood area network}
\newacronym{nb-iot}{NB-IoT}{narrowband Internet of Things}
\newacronym{nfc}{NFC}{near-field communication}
\newacronym{nfv}{NFV}{network function virtualization}
\newacronym{nlos}{NLoS}{non-line-of-sight}
\newacronym{noma}{NOMA}{non-orthogonal multiple access}
\newacronym{noniid}{non-IID}{not independent and identically distributed}
\newacronym{ntns}{NTNs}{Non-Terrestrial Networks}

\newacronym{nr}{NR}{New Radio}
\newacronym{np}{NP}{Neyman-Pearson}

\newacronym{ook}{OOK}{on--off keying}
\newacronym{opex}{OPEX}{operating expenditure}
\newacronym{osi}{OSI}{open systems interconnection}
\newacronym{ot}{OT}{operational technology}

\newacronym{p2p}{P2P}{peer-to-peer}
\newacronym{pca}{PCA}{principal component analysis}
\newacronym{pd-noma}{PD-NOMA}{power-domain non-orthogonal multiple access}
\newacronym{pdf}{PDF}{probability density function}
\newacronym{pos}{PoS}{point of sale}
\newacronym{ppg}{PPG}{photoplethysmogram}
\newacronym{profibus}{PROFIBUS}{Process Field Bus}
\newacronym{profinet}{PROFINET}{Process Field Network}
\newacronym{pso}{PSO}{particle swarm optimization}
\newacronym{pmf}{PMF}{probability mas function}

\newacronym{qos}{QoS}{quality of service}

\newacronym{ran}{RAN}{radio access network}
\newacronym{rat}{RAT}{radio access technology}
\newacronym{rf}{RF}{radio frequency}
\newacronym{rfid}{RFID}{radio-frequency identification}
\newacronym{rhs}{RHS}{reconfigurable holographic surface}
\newacronym{ris}{RIS}{reconfigurable intelligent surfaces}
\newacronym{rl}{RL}{reinforcement learning}
\newacronym{rnn}{RNN}{recurrent neural network}
\newacronym{roc}{ROC}{receiver operating characteristic}
\newacronym{roi}{RoI}{region of interest}
\newacronym{rtt}{RTT}{round-trip time}

\newacronym{sar}{SAR}{synthetic aperture radar}
\newacronym{sc}{SC}{selection combining}
\newacronym{scada}{SCADA}{supervisory control and data acquisition}
\newacronym{sdn}{SDN}{software-defined networking}
\newacronym{se}{SE}{spectral efficiency}
\newacronym{sercos}{SERCOS}{serial real-time communication system}
\newacronym{sgd}{SGD}{stochastic gradient descent}
\newacronym{sic}{SIC}{successive interference cancellation}
\newacronym{sinr}{SINR}{signal-to-interference-plus-noise ratio}
\newacronym{sla}{SLA}{service-level agreement}
\newacronym{smrc}{SMRC}{selection maximal-ratio combining}
\newacronym{snr}{SNR}{signal-to-noise ratio}

\newacronym[sort={SpO2}]
{spo2}{\ensuremath{\mathrm{SpO}_2}}{peripheral oxygen saturation}

\newacronym{ss}{SS}{spectrum sensing}
\newacronym{svm}{SVM}{support vector machine}

\newacronym{tcp}{TCP}{transmission control protocol}
\newacronym{tdma}{TDMA}{time-division multiple access}
\newacronym{tti}{TTI}{transmission time interval}

\newacronym{uav}{UAV}{unmanned aerial vehicle}
\newacronym{udn}{UDN}{ultra-dense network}
\newacronym{unb}{UNB}{ultra-narrowband}
\newacronym{uq}{UQ}{uncertainty quantification}
\newacronym{urllc}{URLLC}{ultra-reliable low-latency communication}

\newacronym{v2i}{V2I}{vehicle-to-infrastructure}
\newacronym{v2n}{V2N}{vehicle-to-network}
\newacronym{v2v}{V2V}{vehicle-to-vehicle}
\newacronym{v2x}{V2X}{vehicle-to-everything}
\newacronym{vlc}{VLC}{visible light communication}
\newacronym{vr}{VR}{virtual reality}

\newacronym{wban}{WBAN}{wireless body area network}
\newglossaryentry{wifi}{
  type=\acronymtype,
  name={Wi-Fi},
  text={Wi-Fi},
  plural={Wi-Fi networks},
  sort={Wi-Fi},
  description={wireless networking based on the IEEE 802.11 family of standards}
}
\newacronym{wisun}{Wi-SUN}{wireless smart ubiquitous network}
\newacronym{wlan}{WLAN}{wireless local area network}
\newacronym{wnan}{WNAN}{wireless neighborhood area network}
\newacronym{wpan}{WPAN}{wireless personal area network}
\newacronym{wsn}{WSN}{wireless sensor network}
\newacronym{wwan}{WWAN}{wireless wide area network}

\glsadd{5g}\glsadd{6g}\glsadd{af}\glsadd{ai}%
  \glsadd{aircomp}\glsadd{auroc}\glsadd{awgn}\glsadd{bpsk}\glsadd{cc}%
  \glsadd{cdma}\glsadd{cegc}\glsadd{cfar}\glsadd{ch}\glsadd{cmrc}%
  \glsadd{covid19}\glsadd{cr}\glsadd{crn}\glsadd{csi}\glsadd{cv}%
\glsadd{df}\glsadd{dfr}\glsadd{ecc}\glsadd{ecg}%
  \glsadd{ed}\glsadd{ef}\glsadd{egc}\glsadd{eh}\glsadd{emrc}%
  \glsadd{fc}\glsadd{fdma}\glsadd{fedvqcs}\glsadd{fl}\glsadd{fov}%
  \glsadd{gmm}\glsadd{gnss}\glsadd{hdaf}\glsadd{ieee}\glsadd{iid}%
  \glsadd{ils}\glsadd{imu}\glsadd{iomt}\glsadd{iot}\glsadd{ir}%
  \glsadd{its}\glsadd{lar}\glsadd{led}\glsadd{lidar}\glsadd{llr}%
  \glsadd{los}\glsadd{lrt}\glsadd{map}\glsadd{mem}\glsadd{mhtp}%
  \glsadd{mimo}\glsadd{miot}\glsadd{ml}\glsadd{mlla}\glsadd{mmtc}%
  \glsadd{mrc}\glsadd{mtc}\glsadd{nlos}\glsadd{noma}\glsadd{noniid}%
  \glsadd{ook}\glsadd{pca}\glsadd{pd-noma}\glsadd{pdf}\glsadd{ppg}%
  \glsadd{rf}\glsadd{rhs}\glsadd{ris}\glsadd{rl}\glsadd{roc}%
  \glsadd{roi}\glsadd{sar}\glsadd{sc}\glsadd{sic}\glsadd{sinr}%
  \glsadd{smrc}\glsadd{snr}\glsadd{spo2}\glsadd{ss}\glsadd{tdma}%
  \glsadd{uav}\glsadd{urllc}\glsadd{v2x}\glsadd{vlc}\glsadd{wban}%
  \glsadd{wifi}\glsadd{wsn}%
  \glsadd{isac}\glsadd{ntns}\glsadd{pmf}

\newcolumntype{P}[1]{>{\centering\arraybackslash}p{#1}}
\newcolumntype{M}[1]{>{\centering\arraybackslash}m{#1}}

\graphicspath{{Figures/}} 

\makeatletter
\def\bstctlcite{\@ifnextchar[{\@bstctlcite}{\@bstctlcite[@auxout]}}
\def\@bstctlcite[#1]#2{\@bsphack
 \@for\@citeb:=#2\do{%
   \edef\@citeb{\expandafter\@firstofone\@citeb}%
   \if@filesw\immediate\write\csname #1\endcsname{\string\citation{\@citeb}}\fi}%
 \@esphack}
\makeatother

\makeatletter
\def\bstctlcite{\@ifnextchar[{\@bstctlcite}{\@bstctlcite[@auxout]}}
\def\@bstctlcite[#1]#2{\@bsphack
 \@for\@citeb:=#2\do{%
   \edef\@citeb{\expandafter\@firstofone\@citeb}%
   \if@filesw\immediate\write\csname #1\endcsname{\string\citation{\@citeb}}\fi}%
 \@esphack}
\makeatother
\begin{document}
\scalefont{0.95}
\bstctlcite{IEEEexample:BSTcontrol}
\title{Decision Fusion in Wireless Networks: A Tutorial and Survey of Classical and Learning-Based Methods}

\author{Mohamed~Abdelkader,
        Arafat~Al-Dweik,~\IEEEmembership{Senior~Member,~IEEE,}
        Sobia~Jangsher,~\IEEEmembership{Member,~IEEE,}
        Hamad~Yahya,~\IEEEmembership{Member,~IEEE,}
        Shihab~Jimaa,~\IEEEmembership{Senior~Member,~IEEE}%
\thanks{Mohamed~Abdelkader is with the Department of Computer and Information Engineering, Khalifa University of Science and Technology, Abu Dhabi 127788, UAE, and also with the Department of Electrical Engineering, Alexandria University, Alexandria 21544, EGYPT, (e-mail: \href{mailto:100067003@ku.ac.ae}{100067003@ku.ac.ae}).}%
\thanks{Arafat~Al-Dweik, Hamad~Yahya and Shihab~Jimaa are with the Department of Computer and Information Engineering, Khalifa University of Science and Technology, Abu Dhabi 127788, UAE (e-mail: \href{mailto:dweik@fulbrightmail.org}{dweik@fulbrightmail.org}; \href{mailto:hamad.myahya@ku.ac.ae}{hamad.myahya@ku.ac.ae}; \href{mailto:shihab.jimaa@ku.ac.ae}{shihab.jimaa@ku.ac.ae}).}%
\thanks{Sobia~Jangsheris is with the School of Electronic Engineering, Dublin City University (DCU), Ireland (email:\href{mailto:sobia.jangsher@dcu.ie}{sobia.jangsher@dcu.ie}).}
}
\maketitle

\begin{abstract}
Decision fusion combines local decisions from distributed sensors to support reliable detection of phenomena in wireless networks. The effectiveness of decision fusion depends on sensing accuracy, reporting-channel reliability, and available communications and computation resources. This article surveys decision-fusion techniques for wireless sensor networks and emerging communication systems, bringing together their theoretical foundations and practical implementation requirements. We review network architectures and hard, soft, and quantized reporting strategies, and examine statistical, learning-based, game-theoretic, and heuristic fusion approaches. Advanced fusion techniques, including relay-aware combining, link-adaptive regeneration, branch selection, and likelihood approximations, are discussed with emphasis on their assumptions, information requirements, and limitations. The survey examines tradeoffs among detection performance, computational complexity, communications overhead, energy consumption, and latency. The tradeoffs are considered across representative sensing applications and emerging technologies, including massive machine-type communication, ultra-reliable low-latency communication, uncrewed aerial vehicle-assisted networks, and the massive Internet of Things. The reviewed literature highlights the importance of jointly considering sensing quality, reporting reliability, and fusion processing when selecting a suitable strategy. Finally, we identify research directions in resource-aware optimization, uncertainty calibration, and robust learning-based fusion under heterogeneous data, missing reports, and changing wireless conditions. Particular attention is given to convergence, generalization, and the costs of training and inference.
\end{abstract}

\glsresetall

\section{Introduction}
\Glspl{wsn} is one of the key technologies that is used to collect information from the surroundings using a large number of distributed sensor nodes. The sensor nodes can collect, process, and transmit information over a wireless channel to a data aggregation node, typically denoted as a \gls{fc}, which processes the information sent by the sensors and makes a final global decision. Sensors may cooperate to deliver a consensus decision about an event or phenomenon to the \gls{fc}, a relay, another sensor node, or any combination. \glspl{wsn} can be applied using several architectures, i.e., centralized, decentralized, cluster-based, and cooperative architectures \cite{callaway2003wireless, shorey2005mobile, karl2007protocols, dargie2011fundamentals}. 
 \glspl{wsn} are becoming indispensable for many applications due to their high flexibility, small size, enhanced surveillance coverage, mobility, and cost-effectiveness. Many fields that incorporate \glspl{wsn} include military, \gls{cc}, \gls{iot}, industry, environmental monitoring, \glspl{crn}, medical applications, and home automation. Military applications include target detection, classification, and tracking purposes \cite{hamdi2008bandwidth, tan2010exploiting, amaldi2012design, you2014new, yu2016consensus, ciuonzo2012channel, yu2017distributed}.

Distributed sensing began to emerge early in the literature. The Bayesian detection problem is formulated for multiple sensors, and the optimal decentralized \gls{dfr} is derived in \cite{tenney1981detection}. The \gls{dfr} that combines decisions from multiple sensors is derived in \cite{chair1986optimal}, where the total error probability is minimized. As the need to develop efficient wireless sensors for various applications is continuously increasing, substantial efforts have been made to achieve the optimal performance of distributed wireless sensor systems. In this context, various detection and sensing algorithms have been introduced in the literature, such as energy detection, compressive sensing, matched filtering, cyclostationarity detection, multi-dimensional \gls{ss}, coherent detection, sequential detection, and censoring sensors \cite{zhang2009optimization, lei2010multiple, poh2010quality, lai2010fusion, chen2011maximum}. Efficient data fusion from multiple sensors is a critical task in \glspl{wsn}. Therefore, numerous studies have been proposed in the literature to develop efficient \glspl{dfr} for various detection problems and source signals, including event detection, image fusion, sound fusion, and multimodal sensing. For event detection problem (hypothesis testing problem), which is the main concern of this article, several \glspl{dfr} are classified in this survey depending on system parameters such as, modulation technique, wireless channel conditions, the availability of local sensor detection performance at the \gls{fc}, network architecture, multiple access technique etc. \glspl{dfr} introduced in the literature include \gls{lrt} based, Chair-Varshney, majority voting, $k$-out-of-$n$, $AND$, $OR$, diversity combining, and deflection coefficient \cite{niu2006fusion, chen2006channel, zhang2009optimization, lei2010multiple, poh2010quality, lai2010fusion, chen2011maximum, ahmadi2011distributed, Ciuonzo2014UnknownPd, cheng2012performance, Ciuonzo2013MRC, nevat2014distributed, minotto2014simultaneous, ciuonzo2015massive, rossi2015hmm,goel2016selection, peng2016multimodal, nurellari2016distributed, rossi2016performance, Tarighati2017DHTEH, stefanakis2017perpendicular, zhou2017hyperspectral}.

\subsection{Review of Existing Surveys}
\Gls{df} has attracted significant attention over the last three decades due to its critical role in distributed wireless communications systems. Numerous survey articles have examined specific aspects of \gls{df}, particularly in cognitive radio networks, \glspl{wsn}, \gls{iot}, and \gls{ml}-based wireless communications. However, most existing surveys focus on a single application domain or a particular class of fusion algorithms rather than providing a unified treatment of \gls{df} techniques across next-generation wireless communication systems. This section reviews the most influential survey articles, highlights their contributions, and identifies the research gaps that motivate the present survey.

In \cite{javadi2016}, a tutorial about the \gls{df} over sensor networks is presented. The authors categorized the \glspl{dfr} based on the challenges/limitations of the sensor networks and provided a qualitative comparison of the different schemes. Performance analysis parameters for different schemes are investigated, along with the schemes for various network structures. The article does not cover the \glspl{dfr} from an application perspective. The article was published in $2016$, and since then many advanced techniques have been proposed that need to be categorized accordingly. Also, with the deployment of \gls{5g} networks and industrial \gls{iot}, \gls{df} needs to be analyzed in relation to emerging technologies.

\subsubsection{Surveys on Cooperative \gls{ss}}

Most of the existing survey literature on \gls{df} has focused on \gls{css} in cognitive radio networks. Since cooperative sensing relies on combining local sensing decisions from multiple secondary users, \gls{df} forms the core component of \gls{css}. One of the earliest and most influential surveys was presented in \cite{AKYILDIZ201140}, which provided a comprehensive review of \gls{css} in cognitive radio networks. Cichoń \textit{et al.} \cite{cichon2016} reviewed energy-efficient \gls{css} techniques. The survey provides a taxonomy of energy-aware sensing strategies and highlights the relationship between energy consumption and decision accuracy. The authors of \cite{Arjoune2019} comprehensively examined recent advances in \gls{ss} technologies. The survey reviews both centralized and distributed sensing architectures and discusses hard and soft \glspl{dfr}, as well as emerging \gls{ml} techniques. More recently, surveys have investigated the evolution of dynamic \gls{ss} toward intelligent and adaptive wireless systems. For example, a $2026$ survey reviewed two decades of research in dynamic \gls{ss}. However, the primary emphasis remained on \gls{ss} rather than general \glspl{dfr} \cite{falco2026}.

\subsubsection{Surveys on \glspl{wsn}}
\glspl{wsn} constitute a  prime research area where \gls{df} is extensively studied. Existing surveys generally examine distributed detection, event monitoring, target tracking, fault diagnosis, and environmental sensing. The foundational works on distributed detection and optimal \glspl{dfr} \cite{tenney1981detection,chair1986optimal} were later consolidated in the two-part tutorial \cite{viswanathan1997distributed}, which remains a key reference for \gls{df} in \glspl{wsn}. Generally, most \glspl{wsn} surveys classify information fusion into three hierarchical levels: data, feature, and decision \cite{Wei2025IntegratingSensors,Qian2025MultiSensorFusion}. These surveys discuss statistical \glspl{dfr}, Bayesian inference, voting mechanisms, and \glspl{lrt} while emphasizing communications efficiency and network lifetime \cite{viswanathan1997distributed,song2024energy}. More recent reviews incorporate cluster-based architecture, distributed consensus, and adaptive fusion strategies \cite{song2024energy,zhang2024mary}. Nevertheless, these surveys are generally application-oriented and rarely examine \gls{df} from the broader perspective of modern wireless communication systems that integrate terrestrial, aerial, and satellite networks.

\subsubsection{Surveys on \gls{iot} Information Fusion}
Existing \gls{iot} surveys generally emphasize the challenges posed by heterogeneous devices, massive connectivity, limited computational capabilities, and energy constraints \cite{lilhore2026securefusion,Mittal2026SensorFusionModels}. Recent reviews discuss intelligent data fusion frameworks for smart cities, healthcare, industrial automation, environmental monitoring, and smart agriculture \cite{lilhore2026securefusion,Wei2025IntegratingSensors}. These surveys recognize \gls{df} as an important approach for reducing communications overhead while improving system reliability. However, their discussions often remain at a conceptual level, focusing primarily on \gls{iot} architecture rather than detailed deep analysis of \gls{df} algorithms.

\subsubsection{Surveys on \gls{ml}-Based Fusion}
The integration of \gls{ai} into wireless communications motivated several surveys on \gls{ml} for distributed sensing and cooperative communications \cite{Li2026MultimodalAlignment,Qian2025MultiSensorFusion,Mittal2026SensorFusionModels}. However, existing \gls{ml} surveys generally emphasize learning algorithms rather than \gls{df} architectures.
\subsubsection{Surveys on Secure Decision Fusion}
Existing surveys provide valuable insights into secure \gls{cc} \cite{zhang2025byzantine}, however, they generally treat \gls{df} as a single component of the security framework rather than as the primary research topic. covert-communication work studies fusion from the detector’s side, which security surveys treat only secondarily \cite{covert2023}.

\subsubsection{Surveys on Distributed and Edge Intelligence}
The recent development of mobile edge computing, federated learning, and edge intelligence has shifted research toward distributed \gls{ai}. Recent survey articles examine collaborative inference, distributed learning, semantic communications, \gls{isac}, and \gls{ai}-native wireless networks \cite{Yi2025SemanticAirCompCSS,ciuonzo2025holographic,Li2026MultimodalAlignment}. These surveys discuss adaptive aggregation strategies, confidence estimation, communication-efficient learning, and privacy-preserving collaborative inference \cite{ciuonzo2025score,Mittal2026SensorFusionModels}. However, they mainly focus on \gls{ai} system architectures and rarely provide detailed treatment of classical \gls{df} theory.

\subsubsection{Comparative Analysis of Existing Surveys}
A careful examination of the literature reveals that existing articles can generally be classified into six categories as shown in Table \ref{tab:existing surveys}.
\begin{table}[t]
\centering
\begingroup
\small
\caption{Focus and Limitations of Existing Survey Categories.}
\label{tab:existing surveys}
\setlength{\tabcolsep}{5pt}
\renewcommand{\arraystretch}{1.3}
\begin{tabular}{@{}
  >{\raggedright\arraybackslash}p{(\columnwidth - 4\tabcolsep) * \real{0.2589}}
  >{\raggedright\arraybackslash}p{(\columnwidth - 4\tabcolsep) * \real{0.3459}}
  >{\raggedright\arraybackslash}p{(\columnwidth - 4\tabcolsep) * \real{0.3952}}@{}}
\toprule\noalign{}
\begin{minipage}[b]{\linewidth}\raggedright
\textbf{Category}
\end{minipage} & \begin{minipage}[b]{\linewidth}\raggedright
\textbf{Primary Focus}
\end{minipage} & \begin{minipage}[b]{\linewidth}\raggedright
\textbf{Major Limitation}
\end{minipage} \\
\midrule\noalign{}
1- \gls{css} & Hard/soft fusion, cognitive radio & Limited to \glspl{crn} \\
\bottomrule
2- \glspl{wsn} & Distributed detection and monitoring & Limited application scope \\
\bottomrule
3- \gls{iot} Information Fusion & Smart sensing and heterogeneous devices & Limited algorithmic comparison \\
\bottomrule
4- \gls{ml} & AI-based sensing and learning & Limited discussion of classical fusion \\
\bottomrule
5- Security & Trust, attacks, robust detection & \gls{df} is secondary \\
\bottomrule
6- Edge Intelligence & Distributed \gls{ai} and collaborative inference & Limited theoretical treatment of fusion algorithms \\
\bottomrule\noalign{}
\end{tabular}
\endgroup
\end{table}
Although these surveys have significantly advanced the field, several important limitations remain, which are:\\
1) Most surveys concentrate on a single application domain rather than considering \gls{df} as a general framework applicable across \glspl{wsn}. 2) Classical statistical \glspl{df} and modern \gls{ml}-based fusion techniques are usually reviewed independently, making direct comparison difficult. 3) Relatively, few surveys discuss \gls{df} in emerging wireless technologies such as \gls{isac}, \gls{ris}, cell-free massive \gls{mimo}, \gls{ntns}, \gls{uav}-assisted communications, semantic communications, and \gls{ai}-native \gls{6g} networks. 4) There is limited discussion of unified evaluation metrics that compare fusion techniques with respect to detection accuracy, computational complexity, communication overhead, scalability, robustness, latency, energy efficiency, and security. Finally, most surveys do not provide an integrated perspective that covers the full evolution of \gls{df} from classical Bayesian detection and voting rules to adaptive \gls{ml}, federated intelligence, and trustworthy AI-based fusion frameworks.

Recent surveys have examined \gls{df} within the broader context of multimodal learning and autonomous perception as shown in Table \ref{tab:cross-article-comparison}. The authors in \cite{Li2026MultimodalAlignment} organize alignment and fusion from both structural and methodological perspectives, distinguishing data-level, feature-level, and output-level integration while discussing cross-modal misalignment, data quality, and computational limitations. Focusing on intelligent vehicles, the work in \cite{Wei2025IntegratingSensors} formalizes early, intermediate, and late fusion and reviews techniques that combine sensor-specific predictions to improve the robustness of perception. In \cite{Qian2025MultiSensorFusion}, the authors further examine deep-learning-based architectures, particularly unified bird's-eye-view representations and cross-modal attention. Their discussion highlights the modularity of late fusion, while recognizing that combining final outputs can limit the exploitation of interactions between modality-specific features. From a broader historical perspective, the work in \cite{Mittal2026SensorFusionModels} relates the evolution of sensor-fusion techniques to classical hierarchical models and identifies continuing gaps in higher-level situation and impact assessment, explainability, and trustworthiness. These reviews provide complementary accounts of how heterogeneous information is represented and combined, with decision-level processing forming one part of a wider fusion pipeline. Building on these perspectives, the present survey focuses on \gls{df} in \glspl{wsn}, where local sensing reliability, report quantization, fading, relay errors, and resource constraints must be considered together when evaluating global detection performance.








\begin{table}[t]
\caption{Cross-article comparison. All articles were published in 2025 and 2026.}
\label{tab:cross-article-comparison}
\centering
\footnotesize 
\renewcommand{\arraystretch}{1.1}
\setlength{\tabcolsep}{2pt} 

\def\FS{{\usefont{U}{pzd}{m}{n}\char72}}
\def\ES{{\usefont{U}{pzd}{m}{n}\char73}}

\begin{tabularx}{\columnwidth}{@{} c *{5}{>{\centering\arraybackslash}X} @{}}
\toprule
\textbf{Ref.} &
\textbf{Theory} &
\textbf{AI} &
\textbf{Fusion} &
\textbf{Tractable} &
\textbf{Impact} \\
\midrule

\cite{Wei2025IntegratingSensors} &
\FS\FS\FS\ES\ES &
\FS\FS\FS\FS\ES &
\FS\FS\FS\FS\ES &
\FS\FS\FS\ES\ES &
\FS\FS\FS\FS\FS \\
\addlinespace[3pt]

\cite{Qian2025MultiSensorFusion} &
\FS\FS\ES\ES\ES &
\FS\FS\FS\FS\FS &
\FS\FS\FS\ES\ES &
\FS\FS\ES\ES\ES &
\FS\FS\FS\FS\FS \\
\addlinespace[3pt]

\cite{Li2026MultimodalAlignment} &
\FS\FS\FS\FS\ES &
\FS\FS\FS\FS\FS &
\FS\FS\FS\ES\ES &
\FS\FS\FS\ES\ES &
\FS\FS\FS\FS\FS \\
\addlinespace[3pt]

\cite{Mittal2026SensorFusionModels} &
\FS\FS\FS\FS\FS &
\FS\FS\FS\FS\ES &
\FS\FS\FS\FS\FS &
\FS\FS\FS\FS\ES &
\FS\FS\FS\FS\ES \\
\bottomrule
\end{tabularx}
\end{table}

\subsubsection{Cross-article Comparison}
As shown in Table \ref{tab:research_directions}, the most relevant recent literature is related to \gls{iot}, \glspl{wsn}, \gls{ris}, holographic surfaces, \gls{css}, and \gls{6g}-oriented distributed intelligence. Recent research has extended \gls{df} from conventional \glspl{wsn}/\gls{iot} toward increasingly heterogeneous and intelligent wireless architectures. In \gls{iot}-oriented \glspl{wsn}, recent studies have investigated $M$-ary distributed \gls{df} over multi-hop relay networks and \gls{df} for heterogeneous cyber-physical threat detection \cite{zhang2024mary,song2024energy,lilhore2026securefusion}. In \gls{css}, \gls{df} has evolved to address channel uncertainty, mobility, reporting constraints, and adversarial or Byzantine users \cite{zhang2025byzantine,Yi2025SemanticAirCompCSS}. More recently, \gls{ris} and holographic meta surfaces have been incorporated directly into decision-fusion architectures, enabling channel-aware and energy-efficient distributed detection \cite{ciuonzo2025holographic,zhang2025byzantine}. Emerging approaches further combine cooperative sensing with semantic communication and over-the-air computation, pointing toward communication-efficient distributed intelligence for \gls{6g} networks \cite{Yi2025SemanticAirCompCSS}. Such developments suggest a transition from fixed \glspl{dfr} toward channel-aware, adaptive, secure, and infrastructure-assisted \gls{df} for emerging \gls{6g} systems.

\begin{table}[!t]
\centering
\caption{Recent research directions in \gls{df}.}
\label{tab:research_directions}
\renewcommand{\arraystretch}{1.2}
\setlength{\tabcolsep}{4pt}

\begin{tabularx}{\columnwidth}{|
  >{\raggedright\arraybackslash\bfseries}X|
  c|
  >{\raggedright\arraybackslash}X|}
\hline
Research direction &
\textbf{Ref.} &
\textbf{Why important} \\
\hline\hline

Distributed detection and \gls{df} architectures &
\cite{javadi2016} &
Review of basic detection rules in the context of \glspl{wsn} \\
\hline

\gls{iot}/\glspl{wsn} \gls{df} &
\cite{zhang2024mary} &
$M$-ary distributed \gls{df} over multi-hop \glspl{wsn} \\
\hline

\gls{iot} cyber-physical security &
\cite{lilhore2026securefusion} &
Fusion of heterogeneous \gls{iot}/network decisions \\
\hline

\gls{ris}, \gls{css} &
\cite{zhang2025byzantine} &
\gls{df} under \gls{ris}-assisted cooperative \gls{ss} \\
\hline

\gls{css},  semantic/\gls{6g} communications &
\cite{Yi2025SemanticAirCompCSS} &
Distributed semantic communication + AirComp \\
\hline

\gls{ris}/holographic fusion &
\cite{ciuonzo2025holographic} &
Directly connects \gls{ris}/meta-surfaces with \gls{df} \\
\hline

Channel-aware fusion &
\cite{ciuonzo2025score} &
Joint channel decoding and \gls{df} \\
\hline
\end{tabularx}
\end{table}

\subsection{Motivation and Contribution}
\label{subsec:motivation_contribution}

The surveys discussed above provide valuable insight into different
aspects of \gls{df}. However, the relationship between fusion
techniques, reporting-channel conditions, and emerging wireless
technologies motivates a more integrated review. In particular, the
advantages of a \gls{dfr} should be examined jointly with its requirements for sensing information, channel knowledge, communication
resources, and computation. This survey brings these aspects together
to support the selection of suitable fusion techniques for different
wireless applications. The main contributions of this article are as
follows:

\begin{enumerate}
    \item Reviews the fundamental models, network architectures, and
    reporting strategies of \gls{df}, and classifies statistical,
    learning-based, game-theoretic, and heuristic approaches.

    \item Compares advanced fusion techniques, including relay-aware
    combining, link-adaptive regeneration, branch selection, likelihood
    approximations, and recent multibit and time-aware techniques, with
    emphasis on their information requirements and practical limitations.

    \item Examines the main performance measures used in decision
    fusion, including detection and false-alarm probabilities, decision
    error, receiver operating characteristics, and deflection, together
    with computational complexity and communication overhead.

    \item Reviews representative applications in environmental
    monitoring, healthcare, intelligent transportation, industrial fault
    diagnosis, and precision agriculture, highlighting their different
    sensing and decision requirements.

    \item Discusses \gls{df} in emerging wireless technologies,
    including \gls{mmtc}, \gls{urllc}, \gls{uav}-assisted communication, \gls{vlc},
    \gls{eh}, \gls{noma}, and \gls{miot}.

    \item Identifies research challenges in performance-metric
    optimization and learning-based fusion, particularly convergence,
    implementation complexity, uncertainty calibration, and robustness
    under heterogeneous data and imperfect reporting conditions.
\end{enumerate}

\subsection{Article Organization}
\label{subsec:article_structure}

The rest of this article is organized as follows. The system models, architectures, and reporting strategies are introduced in Section \ref{sec-basicsOfDF}. Section \ref{sec:classification_fusion} presents the classification of fusion approaches, while Section \ref{sec:advanced_fusion} reviews advanced combining techniques and likelihood approximations. The main performance metrics are discussed in Section \ref{sec-PerfMetrics}, followed by representative applications in Section \ref{sec-apps}. Section \ref{sec:decision_fusion_emerging_technologies} examines \gls{df} in emerging wireless technologies. Research challenges and future directions are addressed in Section \ref{sec:future_directions}, with emphasis on performance-metric optimization and the convergence and complexity of learning-based fusion. Finally, Section \ref{sec:conclusion} concludes the article.
\section{Fundamentals of Decision Fusion}\label{sec-basicsOfDF}
\begin{figure}[t]
    \centering
    \includegraphics[width = 1\columnwidth]{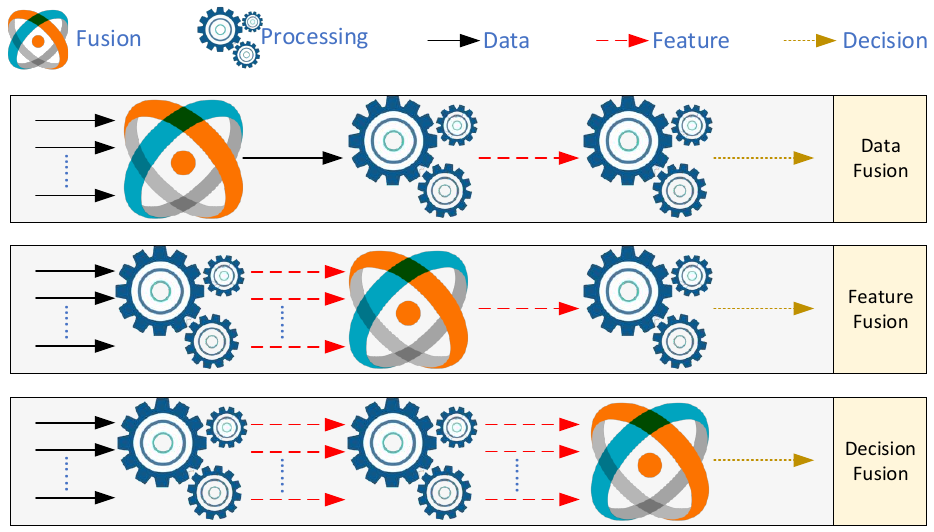}
    \caption{Levels of fusions.}
    \label{fig:fusion}
\end{figure}

The fusion process is defined as the association, correlation, and combination of the information obtained from multiple sensors about a phenomenon to obtain a timely and intelligent decision-making service. 
\Gls{df} can be performed at three levels based on the input-output modes, which are data, feature, and decision, as shown in Fig. \ref{fig:fusion}:
\begin{enumerate}
\item Data fusion: It is low-level fusion where the raw sensor data is combined and then processed to produce a feature vector, which is fed to and processed by the decision layer that computes the global decision \cite{larios2012energy}. 
\item Feature fusion: In this approach, raw data from each sensor is independently processed to extract distinct feature vectors. The extracted vectors are then combined into a unified representation, which serves as the input for global decision-making.
\item Decision Fusion: It is a high-level fusion and a special case of data fusion, in which fusion is performed after distributed decisions are made. The processing is on the raw data, features are extracted, and decisions are taken and forwarded to the fusion block. In this case the initial decisions are derived at lower levels and then fused together.
\end{enumerate}
Next, we discuss a simple model of a \glspl{wsn} to explain the \gls{df} process, and an optimal \gls{dfr} under certain conditions is presented. 

\subsection{Simplified Model}
\begin{figure}[t]
    \centering
    \includegraphics[width=1\columnwidth]{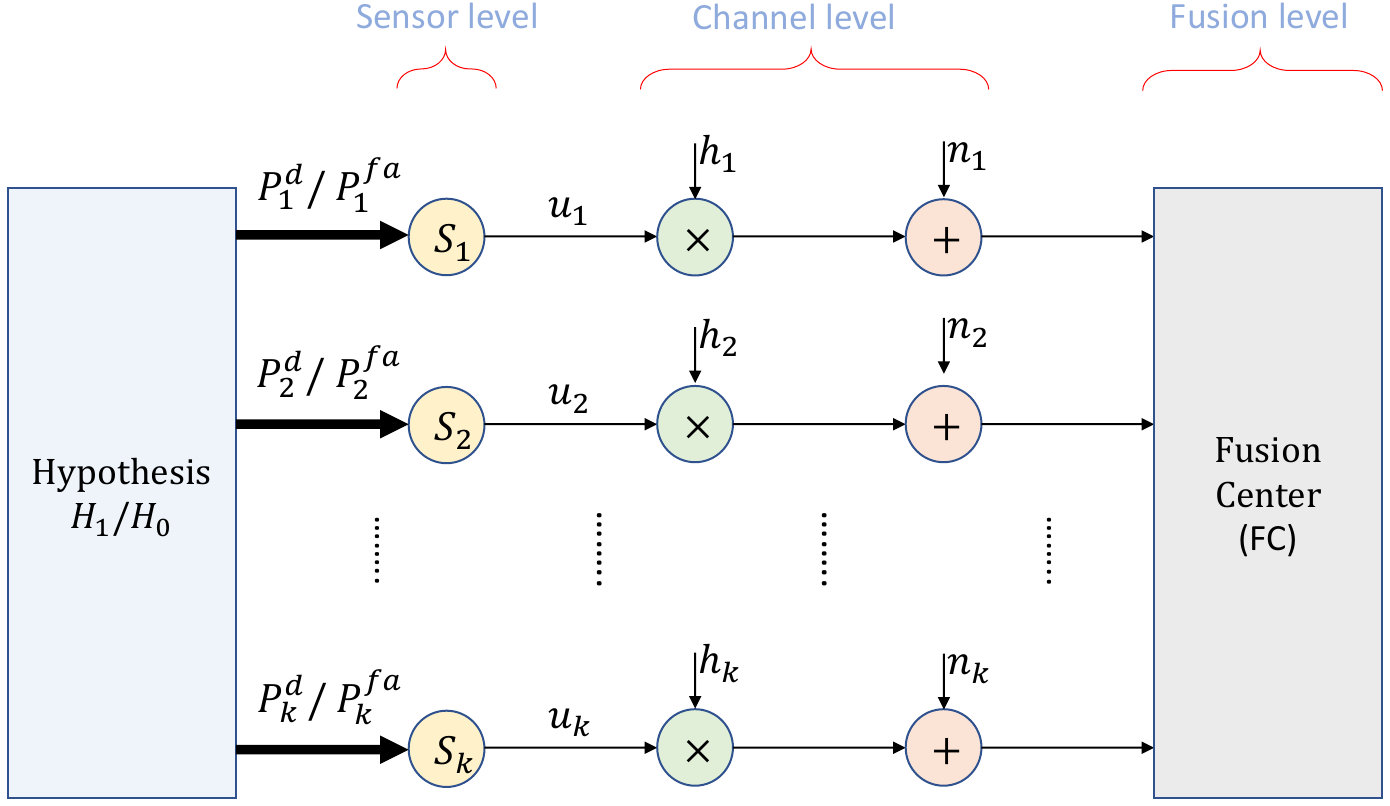}
    \caption{Simplified decision fusion model.}
    \label{fig:basicModel}
\end{figure}

A basic \glspl{wsn} consists of $K$ sensors deployed in a \gls{roi} to collect observations about the same phenomenon, where each sensor makes a decision about the phenomenon separately \cite{chen2004}. In other words, the decision about the occurrence of the event is made locally or independently at the sensor and is transmitted to the \gls{fc}. A simplified \glspl{wsn} with fading and noise  has three layers, as shown in Fig. \ref{fig:basicModel}, for a binary hypothesis case. Next, the steps of binary detection for each layer are:

\subsubsection{Sensor Layer}
Each sensor measures a noisy signal and makes a decision locally. The sensor node may process $T$ observations to get an inference
about the phenomenon before sending information to the \gls{fc}. 
The phenomenon under consideration is framed as a \gls{bhtp}, in which the observations are assumed to be \gls{iid} given each hypothesis. The null hypothesis $\mathcal{H}_0$ represents the absence of the phenomenon, while the alternative hypothesis $\mathcal{H}_1$ represents its presence. For the $k$-th sensor, the \gls{bhtp} is defined as \cite{zhang2009optimization}
\begin{equation}
{\chi}_{k}[t]|{\mathcal{H}_m}=g_{k}[t]x_{m}[t]+n_{k}[t]\label{eq1}%
\end{equation}
where $t$ denotes the time slot (sample), {$\mathcal{H}_m$} denotes the possible
existing hypothesis, $x_{m}[t]$ is the source signal,$\ g_{k}[t]$ is the response of the channel between the source and the sensor, and $n_{k}$\ the \gls{awgn}. Assuming that the processing function employed by the sensor (the type
of sensing) is $G({\cdot})$, the decision of the sensor is
$\mathbf{u}_{k}=G(\boldsymbol{\chi}_k)$ where ${\boldsymbol{\chi}_k}=[{\chi}_{k}[1],{\chi}_{k}[2],\dots,{\chi}_{k}[T]]$. 

 It is practically preferred that sensors send quantized decisions about the
phenomenon instead of the observations themselves 
due to limitations in transmission power and required bandwidth. The low complexity of \glspl{ed} makes them an attractive solution for \glspl{wsn} applications. Moreover, the sensor can process $T$ samples to determine which hypotheses to select, thereby increasing the reliability of local decisions. Therefore, the \gls{bpsk} decision, $\mathbf{u}_{k}\in\{-1$, $1\}$, using \gls{ed} at the sensor can be obtained as \cite{zhang2009optimization}, \cite{chen2017cooperative}, \cite{maleki2013censored},
\begin{equation}
\frac{1}{T}\sum_{t=1}^{T}\left\vert {\chi}_{k}[t]\right\vert ^{2}%
\overset{\mathbf{u}_{k}=1}{\underset{\mathbf{u}_{k}=-1}{{ \gtreqless}}}{\tau}%
_{k}\label{Eq2b}%
\end{equation}
where ${\tau}_{k}$ is a predefined threshold determined based on the false
alarm probability constraint. The detection performance of the $k$th sensor is
characterized by its detection probability $P_{k}^{d}$ and false alarm probability $P_{k}^{fa}$. Assuming that $n_{k}\sim \mathcal{N}(0,\sigma_{n}^{2})$ and $g_{k}\sim\mathcal{CN}(0,\sigma_{g}^{2})$ 
\cite{zhang2009optimization},\cite{chen2017cooperative}, \cite{maleki2013censored}, then
\begin{equation}
P_{k}^{fa}=\Pr\left(  \mathbf{u}_{k}=1|\mathcal{H}_0\right)  =\frac{\Gamma\left(  u,{\tau}%
_{k}/2\right)  }{\Gamma\left(  u\right)  }\label{a}%
\end{equation}%
\begin{equation}
P_{k}^{d}=\Pr\left(  \mathbf{u}_{k}=1|\mathcal{H}_1\right)  =Q_{u}\left(  \sqrt{2\gamma_{k}%
},\sqrt{{\tau}_{k}}\right) \label{b}%
\end{equation}
where $\gamma_{k}$ is the \gls{snr}, {$\mathbf{u}$} is the time
bandwidth factor, $\Gamma\left(  \cdot\right)  $ is the complete Gamma
function, $\Gamma\left(  \cdot\text{,}\cdot\right)  $ is the incomplete Gamma
function, and $Q_{u}\left(  \cdot\text{,}\cdot\right)  $ is the generalized
Marcum Q-function. It can be noted that $P_k^d$ and $P_k^{fa}$ depend mainly on the \gls{snr} and the predefined decision threshold for each sensor. 
The marginal probability of deciding $\mathcal{H}_1$ is, 
\begin{equation}
    \Pr\left({\hat{\mathbf{u}}}_{k}=1\right)=\Pr\left(  {\hat
{\mathbf{u}}}_{k}=1|{\mathcal{H}_0}\right)\Pr(\mathcal{H}_0) 
+\Pr\left(  {\hat{\mathbf{u}}}_{k}%
=1|{\mathcal{H}_1}\right) \Pr(\mathcal{H}_1) 
\end{equation}

\subsubsection{Channel Layer}
All local decisions are transmitted to the \gls{fc} over parallel, independently fading channels, whose statistics depend on the link type (e.g., \gls{vlc} or RF). The noise is \gls{awgn}, uncorrelated across channels. With fading and noise, the signal received at the \gls{fc} is
\begin{equation}
    \mathbf{r}_k=\sqrt{P_k}h_ku_k + \varphi_k
\end{equation}
where $h_k \sim\mathcal{CN}(0,\sigma_{h}^{2})$ is the fading coefficient from sensor $k$ to \gls{fc}, $\varphi_k  \sim\mathcal{CN}(0,\sigma_{\varphi}^{2})$ and $P_k$ is the average transmit power of signal from sensor $k$. \Gls{ecc} may also be applied to ensure the reliability of the transmitted decisions. 

\subsubsection{Fusion Layer}
The received signals from all sensors $\mathbf{r}\in \mathbb{C}^{1\times K}=[\mathbf{r}_1,\mathbf{r}_2, \ldots, \mathbf{r}_K]$ are processed at the \gls{fc} to produce the global decision. Different \glspl{dfr}, depending on the application, are applied at the \gls{fc} to ensure a reliable decision. Apart from $\mathbf{r}$, other parameters such as \gls{csi}, detection probability $P_k^d$, and $P_k^{fa}$ are also required at the \gls{fc} to compute a decision. These probabilities characterize the reliability of each local decision and determine the weighting of each sensor in the optimal \gls{dfr}.

\subsection{Multihop Model}
\begin{figure*}[t]%
\centering
\includegraphics[width=6.5233in] {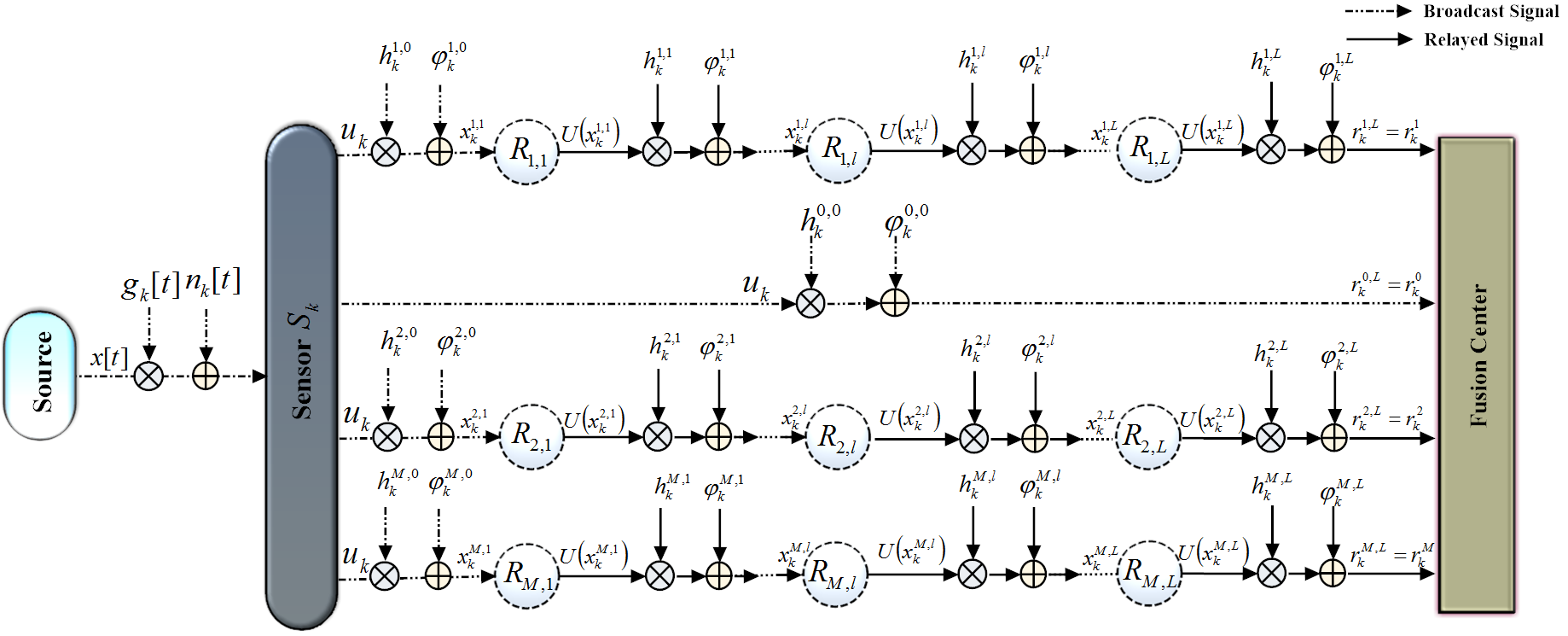}%
\caption{An example of \gls{df} system contains the source signal with one detecting sensor and multi-hop relaying transmission.}%
\label{fig:multihop}%
\end{figure*}
Fig. \ref{fig:multihop} shows an example of a cooperative network with the signal source, one sensor, and $M\times L$ relays. Each sensor communicates with the \gls{fc} in two ways, direct path and $M$ multi-hop relaying paths with $L+1$ hops each, as illustrated in Fig. \ref{fig:multihop}. Moreover, the source, sensor, relay, or the \gls{fc} may use multiple antennas, but single-antenna devices are assumed for analysis tractability. Moreover, the transmission between the source and sensor (secondary user) is considered as a hop, therefore, the total number of hops for the direct signal is $2$, and for
the relayed signals is $L+2 $ hops. Next, the individual steps involved in binary detection across all layers are examined.

\subsubsection{Sensor layer}
The sensor broadcasts its own decision $\mathbf{u}_{k}$ in the broadcasting phase to the \gls{fc} and the first group of the cooperating relays. The cooperating relays process the received symbols and forward them to the next group of relays according to a certain relaying protocol, and so on until the relayed signals reach the \gls{fc}. Although distinct relays are assumed for each sensor, the same relays could be exploited by multiple sensors via orthogonal transmission, e.g., \gls{tdma}, \gls{fdma}, or \gls{cdma}, thereby limiting the total number of relays in the network. Moreover, each sensor is assumed to have knowledge of its dedicated relays via a feedback signal from the \gls{fc} with the assigned route \cite{galluccio2004analytical,saleem2014empirical, wang2014self, park2015multiband}. 
The phenomenon under consideration is the \gls{bhtp}, in which the observations are assumed to be \gls{iid} conditioned on each hypothesis \cite{zhang2009optimization}. For the $k$th sensor, the \gls{bhtp} is defined in \eqref{eq1}.

\subsubsection{Channel Layer}
At the $(m,l)$th relay, the baseband representation of the received signal can be written as%
\begin{multline}
x_{k}^{m,l}=\sqrt{{\alpha}_{k}^{m,l-1}P_{k}^{m,l-1}}{h}_{k}^{m,l-1}%
U(x_{k}^{m,l-1})+{\varphi}_{k}^{m,l-1}, \\ m\in\left\{  1,\ 2,\ \dots
,\ M\right\}  \text{ and }l\in\left\{  0,1,\dots,\ L\right\} \label{Eq4}%
\end{multline}
where $U(x_{k}^{m,l-1})$ denotes the processing at previous relay, i.e., \gls{af}\cite{da2009performance, ngo2011linear, wang2013dynamic, lopez2015average}, \gls{daf} \cite{goel2016selection, Nasri2012EMRC, jain2011exact, sharma2009performance, sharma2011conditionally}, \gls{ef} \cite{ionita2014detectors} or \gls{hdaf} \cite{bai2015performance, feng2017tas}, ${\alpha}_{k}^{m,l-1}$ is a factor that represents the power gain of the sensor/relay, the shadowing and the path loss, ${h}_{k}^{m,l-1}$ is the fading coefficient of the channel between the the $(m,l)$th relay and the previous node, $P_{k}^{m,l-1}$ is the average transmission power, and ${\varphi}_{k}^{m,l-1}$ is the \gls{awgn}. It is worthy to note that $U(x_{k}^{m,0})=\mathbf{u}_{k}\forall m$.

\subsubsection{Fusion Layer}
The \gls{fc} receives $\left(M+1\right)  $ signals from each sensor denoted as ${\overline{\boldsymbol{\mathbf{r}}}}_{k}=\left[  \mathbf{r}_{k}^{0,0},\mathbf{r}_{k}^{1,L}%
,\mathbf{r}_{k}^{2,L},\dots,\mathbf{r}_{k}^{M,L}\right]  $, where $\mathbf{r}_{k}^{0,0}$ represents the direct signal while $\mathbf{r}_{k}^{m,L}$ represents the $M$ relayed signals by the last group of relays (the $L$th group). Thus
\begin{subequations}
\begin{align}
\mathbf{r}_{k}^{m,L} &  =\sqrt{{\alpha}_{k}^{m,L}P_{k}^{m,L}}{h}_{k}^{m,L}U(x_{k}%
^{m,L})+{\varphi}_{k}^{m,L}\text{, } \\
\mathbf{r}_{k}^{0,0} &  =\sqrt{{\alpha}_{k}^{m,L}P_{k}^{0,0}}{h}_{k}^{0,0}\mathbf{u}%
_{k}+{\varphi}_{k}^{0,0}\label{Eq6}%
\end{align}
\end{subequations}
However, the superscript $L$ can be dropped to simplify the notation as follows
\begin{equation}
\mathbf{r}_{k}^{p}=\sqrt{{\alpha}_{k}^{p}P_{k}^{p}}{h}_{k}^{p}U(x_{k}^{p})+{\varphi
}_{k}^{p}\text{, }p\in\left\{  0,1,\ 2,\dots,M\right\} \label{Eq6b}%
\end{equation}
where {the} superscript $p=0$ represents the direct signal, while $p>0$ represents the signals received from the last group relays, and $U(x_{k}^{0})=\mathbf{u}_{k}$.

\subsection{Optimal Likelihood-Based Decision Rule}
\label{sec2-c}
The optimal \gls{df} can be computed given that a) fading channels are known, b) the sensors are conditionally independent given the hypothesis, and c) the detection probability $P^d_k$ and the false alarm probability $P^{fa}_k$ of each sensor are known.  Then, the \gls{lrt} can be expressed as
\begin{equation}
\mathrm{LRT}\triangleq\frac{f\!\left(  \mathbf{r}|\mathcal{H}_1\right)  }{f\!\left(
\mathbf{r}|\mathcal{H}_0\right)  }\overset{\mathcal{H}_1}{\underset{\mathcal{H}_0}{\lesseqgtr}%
}{\tau}_{0}\label{eqn:lrt}%
\end{equation}
where $f\!\left(\mathbf{R}|\mathcal{H}_{1/0}\right)$ is the conditional \gls{pdf} of $\mathbf{R}$ given $\mathcal{H}_1$ or
$\mathcal{H}_0$, and ${\tau}_{0}$ is the system-level threshold that is chosen to
satisfy a certain global $P_{fa}$. The \gls{lrt} is computed as 
\begin{equation}
\label{eqn:optimal}
 \!\!   \frac{f\left(  \mathbf{r}|\mathcal{H}_1\right)  }{f\left(
\mathbf{r}|\mathcal{H}_0\right)  }= {\prod\nolimits_{k=1}^K} \frac{P^d_k\exp{-\frac{(\mathbf{r}_k-h_k)^2}{2\sigma_n^2} }+ (1-P^d_k)\exp{-\frac{(\mathbf{r}_k+h_k)^2}{2\sigma_n^2} }}{P^{fa}_k\exp{-\frac{(\mathbf{r}_k-h_k)^2}{2\sigma_n^2} }+ (1-P^{fa}_k)\exp{-\frac{(\mathbf{r}_k+h_k)^2}{2\sigma_n^2} }}.
\end{equation}
The \gls{llr} is then compared to $\log\left(  \tau_{0}\right)$, to declare the global decision about the hypothesis. Although the \gls{dfr} based on \eqref{eqn:optimal} is optimal \cite{chair1986optimal}, it has some drawbacks in terms of computational complexity and stability. In terms of complexity, the number of multiplications and the number of additions are high. Moreover, a large number of exponentials and logarithms are computed with a broad dynamic \gls{snr} range. The computational instability could be caused by the exponential functions at high \glspl{snr}, where ${\sigma}_{{\varphi}_{k}}^{2}\rightarrow0$, which can lead to system ruination. Finally, it requires the knowledge of $P_{k}^{d}$ and $P_{k}^{fa}$ for all sensors. 

To mitigate the drawbacks of the \gls{lrt} given by \eqref{eqn:optimal}, other suboptimal \glspl{dfr} are introduced in the literature as practical alternatives. These \glspl{dfr} include the two-stage (decode-then-fuse) rule, ideal local sensors based rule, max-log based rule with linear approximation, and other rules based on diversity combining techniques. As shown in \cite{lei2010multiple}, the \glspl{dfr} when ideal sensors are used can be viewed as a weighted sum of the directly received signals from the sensors. Therefore, \glspl{dfr} based on cooperative combining techniques, which can be used at the \gls{fc} to combine the $K\times(M+1)$ signals, are introduced. Many types of diversity combining schemes have been introduced in the literature, i.e., \gls{mrc}, \gls{egc}, or \gls{sc} \cite{goldsmith2005wireless}. However, more recent studies in cooperative diversity have proposed more efficient combiners for relaying networks, i.e., \gls{cmrc}, \gls{smrc}, and \gls{cegc}, in which the conditions of the sensor-to-relay channel are taken into account when evaluating the combining weights.
The performance of the \gls{df} is compromised when a node experiences severe fading or shadowing effects. Most of the studies investigate the performance of \gls{dfr} under Rayleigh, Rician, or Nakagami fading channels.

\subsection{Architecture of Decision Fusion}
The architecture of a \gls{df} system determines how local decisions generated by multiple sensors are exchanged, processed, and ultimately combined to produce a global decision. The choice of architecture has a direct impact on the reliability, scalability, communication overhead, and energy consumption of the network. Depending on how the information flows between sensors and the fusion entity, \gls{df} architectures can generally be classified into centralized, hierarchical (or decentralized), and fully distributed architectures \cite{tenney1981detection,viswanathan1997distributed,chamberland2003decentralized,chamberland2007wireless} as shown in Table ~\ref{tab:decision_fusion_architectures}. 

\textit{1) Centralized Architecture:} In a centralized architecture, all sensors send hard decisions or soft statistics directly to \gls{fc}, which computes the global decision. Because the \gls{fc} incorporates information from all sensors, this architecture generally achieves the best detection performance. However, it requires reliable links between every sensor and the \gls{fc}, which raises communication costs and energy consumption, makes the \gls{fc} a bottleneck as the network grows, and creates a single point of failure. It is therefore best suited to small- to medium-sized networks with sufficient communication resources.

\textit{2) Hierarchical Architecture:} Hierarchical architectures address the scalability limits of centralized fusion by grouping sensors into clusters, where a \gls{ch} combines local decisions before forwarding them to a higher-level \gls{fc}. Such a process reduces the communications overhead, distributes the processing load, and extends network lifetime, making it attractive for large-scale wireless sensor networks, \gls{iot} deployments, and cooperative sensing. However, errors at lower levels can propagate to the global decision, and overall performance depends on the reliability of the \glspl{ch} and their links. Consequently, cluster formation, relay selection, and resource allocation are key design considerations \cite{chamberland2004asymptotic,aldalahmeh2019fusion}.

\textit{3) Distributed Architecture:} In a fully distributed architecture, there is no dedicated \gls{fc}. Sensors exchange information with neighbors and update their local decisions to reach a consensus. Such approach offers high scalability, robustness to node failures, and independence from centralized infrastructure, which suits ad hoc networks, battlefield surveillance, and large-scale \gls{iot}. The drawbacks are the multiple exchange iterations needed for consensus, which increase overhead and latency, and a strong dependence on network connectivity and synchronization \cite{cattivelli2011distributed}. Overall, no architecture is universally optimal: centralized fusion maximizes accuracy, hierarchical fusion balances performance and scalability, and distributed fusion prioritizes robustness and flexibility, thus, the choice depends on the application and system constraints.

\begin{table*}[!t]
\caption{Qualitative Comparison of Decision-Fusion Architectures.}
\label{tab:decision_fusion_architectures}
\centering
\begingroup
\footnotesize
\setlength{\tabcolsep}{4pt}
\renewcommand{\arraystretch}{1.16}
\renewcommand{\tabularxcolumn}[1]{m{#1}}

\begin{tabularx}{\textwidth}{
@{}
>{\raggedright\arraybackslash}m{2.35cm}
*{3}{>{\raggedright\arraybackslash}X}
@{}
}
\toprule

\textbf{Aspect}
& \textbf{Centralized}
& \textbf{Hierarchical}
& \textbf{Fully distributed}
\\
\midrule

\textbf{Information flow}
&
Sensors $\rightarrow$ FC.
&
Sensors $\rightarrow$ CHs $\rightarrow$ upper-level FC.
&
Sensors $\leftrightarrow$ neighboring sensors; no dedicated FC.
\\
\midrule

\textbf{Fusion operation}
&
Each sensor performs local processing; one FC combines the received
hard decisions or soft statistics.
&
CHs combine reports within their clusters; a higher-level FC combines
the cluster outputs
\cite{chamberland2004asymptotic,aldalahmeh2019fusion}.
&
Each node updates its decision using local and neighboring
information; repeated exchanges support consensus
\cite{cattivelli2011distributed}.
\\
\midrule

\textbf{Principal benefit}
&
Provides one processing point with access to the reports collected
from all sensors.
&
Distributes the processing burden and supports scalable,
cluster-based aggregation.
&
Reduces dependence on centralized infrastructure and offers
flexibility and tolerance to node failures.
\\
\midrule

\textbf{Communication and energy}
&
Every sensor needs a reporting link to the FC; collecting many
reports can increase communication and energy costs.
&
Local aggregation reduces the number of long-distance transmissions
and can extend network lifetime.
&
Communication occurs between neighbors, but repeated exchanges can
increase the total message overhead.
\\
\midrule

\textbf{Main limitations}
&
The FC can become a bottleneck and a single point of failure;
reliable sensor-to-FC links are required.
&
Lower-level errors can propagate; performance depends on CH
reliability and the links between fusion levels.
&
Consensus can introduce additional latency; performance depends on
network connectivity and synchronization.
\\
\midrule

\textbf{Design priorities}
&
Reliable reporting links, adequate FC capacity, and protection
against FC failure.
&
Cluster formation, relay selection, and resource allocation.
&
Network connectivity, synchronization, and convergence latency.
\\
\midrule

\textbf{Typical deployments}
&
Small- to medium-sized networks with sufficient reporting resources.
&
Large-scale WSNs, IoT deployments, and cooperative sensing systems.
&
Ad hoc networks, battlefield surveillance, and large-scale IoT.
\\

\bottomrule
\vspace{6pt}
\end{tabularx}
\begin{minipage}{\textwidth}
\scriptsize
\raggedright
\par

\end{minipage}
\endgroup
\vspace{-15pt} 
\end{table*}

\subsection{Decision Decoding Style}
Soft and hard decision decoding are considered widely in the literature. In the hard \gls{df} algorithm, one bit of information is exchanged to indicate whether the received energy exceeds a threshold. Whereas, in the case of soft decision \cite{ma2008soft}, the exact energy levels are transmitted to the \gls{fc} to make an informed decision. 
The information reported by a sensor to the \gls{fc} can take different forms depending on how much local information is preserved during transmission. This reporting strategy is commonly referred to as the decision decoding technique. In general, \gls{df} systems employ either hard-decision decoding, soft-decision decoding, or an intermediate quantized representation. The choice of decoding technique directly affects the tradeoff between detection performance, communication overhead, and energy consumption. Next, the differences between hard, soft, and hybrid decision decoding are elaborated further.

\textit{1) Hard-Decision Decoding:} In hard-\gls{df}, each sensor converts its local observation into a binary decision before transmission. Typically, a single bit is used to indicate whether the observed phenomenon is present or absent. The \gls{fc} then combines the received binary decisions to make a Hard-Decision using an appropriate \gls{dfr},such as the \gls{cv}, majority voting, OR, AND, or the $K$-out-of-$N$. The primary advantage of hard-\gls{df} is its low communication cost. Since only one bit is transmitted from each sensor, bandwidth and energy requirements are significantly reduced. For this reason, hard-\gls{df} is widely adopted in battery-powered \glspl{wsn} and low-complexity \gls{iot} applications. However, converting observations into binary decisions inevitably discards part of the available information. As a result, hard-\gls{df} generally exhibits lower detection performance compared to soft-decision approaches, particularly under low-\gls{snr} conditions \cite{chair1986optimal,viswanathan1997distributed}. 

\textit{2) Soft-Decision Decoding:} In soft-\gls{df}, sensors transmit a reliability measure or decision statistic rather than a single binary bit. The information bit may correspond to the received signal energy, likelihood ratio, \gls{llr}, posterior probability, or any other confidence metric generated during local processing. Because the \gls{fc} receives more detailed information, soft-\gls{df} typically achieves superior performance compared to hard-decision techniques. The performance improvement becomes particularly noticeable in challenging environments characterized by fading, shadowing, or low \gls{snr}. The main drawback of soft-\gls{df} is the increased communication overhead. Transmitting real-valued statistics requires additional bandwidth, higher transmission power, and more sophisticated signal processing at both the sensor nodes and \gls{fc}. Consequently, soft-\gls{df} is often preferred in applications where detection accuracy is paramount \cite{ma2008soft,chamberland2007wireless}. 

\textit{3) Quantized and Hybrid Decision Decoding:} To balance the tradeoff between hard and soft fusion, sensors may transmit a multi-bit quantized version of their local decision statistic, giving the \gls{fc} more information than hard decisions while using less bandwidth than fully soft decisions. Censoring-based reporting is another common approach, in which sensors skip transmission when their observations fall within a predefined uncertainty region, reducing communication overhead and extending network lifetime with little loss in detection performance. These schemes are practical compromises between the simplicity of hard-\gls{df} and the performance of soft-\gls{df}\cite{appadwedula2008censoring,nguyen2015bits}.

In summary, the choice of fusion technique depends on the target application: hard decisions favor low complexity and energy efficiency, soft decisions maximize detection performance, and quantized or hybrid approaches offer a flexible trade-off between the two.
\section{Classification of Decision Fusion Techniques}
\label{sec:classification_fusion}
\begin{figure}[t]
\centering
\includegraphics{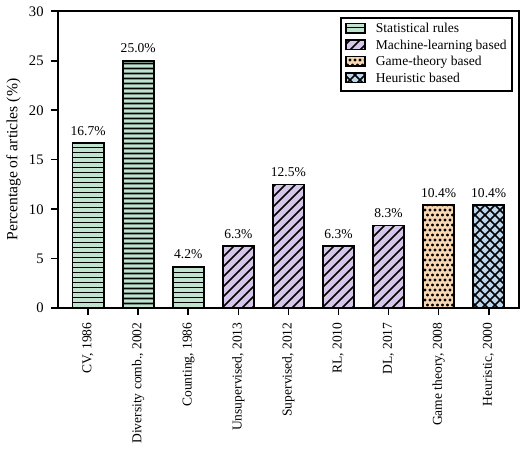}
\caption{Percentage of 48 reviewed articles by decision-fusion approach: statistical rules, machine-learning-based, game-theory-based, and heuristic-based.}
\label{fig:decision_fusion_classification}
\vspace{-15pt} 
\end{figure}

This section classifies the \glspl{dfr} reviewed in this
survey according to the principles used to combine sensing information
at the \gls{fc}. As illustrated in
Fig.~\ref{fig:decision_fusion_classification}, shows the distribution of the reviewed decision-fusion approaches by technique family. Bars show the percentage of 48 surveyed articles that adopt each approach, grouped into statistical rules (\gls{cv}, diversity combining, counting), \gls{ml}-based methods (unsupervised, supervised, \gls{rl}, \gls{dl}), game-theory-based methods, and heuristic-based methods. The year after each label marks the earliest work reviewed for that approach. These are complementary design perspectives rather than mutually exclusive classes. For example,
a statistical \gls{dfr} may use a learned scheduling policy or weights
selected through a game-theoretic formulation. The reporting resolution
(hard, quantized, or soft), network architecture, and communication
technology describe additional dimensions of the same system.

Statistical approaches include the \gls{cv} rule,
diversity-combining schemes, and counting rules. The learning discussion
covers unsupervised, supervised, and reinforcement learning, followed by
deep neural architectures that can be used within these learning
paradigms. Game-theoretic techniques address cooperation, competing
objectives, and adversarial behavior, whereas heuristic techniques introduce
empirical or application-dependent design choices. The following
subsections examine their operating principles, information requirements,
and limitations, emphasizing detection performance, implementation
complexity, communication overhead, and adaptation to changing conditions.

Throughout this section, $\Lambda$ denotes a generic fusion statistic and $\lambda$ its threshold; $H_1$ is declared when $\Lambda \ge \lambda$. $\Lambda$ is called a \gls{llr} only when it equals $\ln [f(\text{observations} \mid H_1) / f(\text{observations} \mid H_0)]$ under the stated observation model, as for the CV rule in \eqref{CV-rule} and the \gls{lrt} in Sec.\ref{sec2-c}. Weighted combinations whose weights are not derived from such a model (\gls{mrc}, \gls{egc}, \gls{sc}, counting rules, and heuristic sums) are called fusion statistics, even if they approximate an \gls{llr} under special conditions.

\subsection{Statistical Fusion Rules}
\gls{df} at the \gls{fc} is essential for aggregating distributed sensor observations to make a reliable global decision. Depending on the available \gls{csi} and local sensing statistics, \glspl{dfr} trade off detection accuracy against communication overhead and computational complexity. This section reviews different categories of statistical \glspl{dfr}.

\subsubsection{Chair--Varshney Fusion}

The \gls{cv} rule is optimal for fixed,
conditionally independent binary decisions when their conditional
probabilities are known. This optimality refers only to the information retained in the decoded decisions. It does not imply equivalence to a detector that observes the continuous-valued received observations
\cite{chair1986optimal,niu2006fusion}. For $K$ received decisions,
the binary-decision \gls{llr} is
\begin{equation}
\Lambda_{\mathrm{CV}}
=
\sum\nolimits_{k:\widehat {\mathbf{u}}_k^{\mathrm{FC}}=+1}
\ln\frac{q_{k,1}}{q_{k,0}}
\;+
\sum\nolimits_{k:\widehat {\mathbf{u}}_k^{\mathrm{FC}}=-1}
\ln\frac{1-q_{k,1}}{1-q_{k,0}}
\underset{\mathcal{H}_0}{\overset{\mathcal{H}_1}{\gtrless}}
\lambda_{\mathrm{CV}}
\label{CV-rule}
\end{equation}
where $q_{k,j}=\Pr(\widehat {\mathbf{u}}_k^{\mathrm{FC}}=+1\mid \mathcal{H}_j)$. Thus, both quantities describe a positive received
decision, but under different hypotheses. Under a symmetric reporting
channel with crossover probability $\epsilon_k^{\mathrm{rep}}$ that is
independent of the hypothesis and the local sensing observation,
\begin{equation}
\begin{aligned}
q_{k,1}
&=\Pr({\widehat u}_k^{\mathrm{FC}}=+1\mid \mathcal{H}_1)
=\epsilon_k^{\mathrm{rep}}
 +(1-2\epsilon_k^{\mathrm{rep}})P_{\mathrm d,k}
\end{aligned}
\label{BER}
\end{equation}
\begin{equation}
\begin{aligned}
q_{k,0}
&=\Pr(\widehat {\mathbf{u}}_k^{\mathrm{FC}}=+1\mid \mathcal{H}_0)
=\epsilon_k^{\mathrm{rep}}
 +(1-2\epsilon_k^{\mathrm{rep}})P_{\mathrm{fa},k}.
\end{aligned}
\label{eq:section3_cv_null_probability}
\end{equation}
The finite logarithms in \eqref{CV-rule} require probabilities strictly
between zero and one, deterministic cases can be handled through their
limits. Reporting errors in cooperative links depend on the relaying
and detection assumptions, as illustrated by the analyses in
\cite{sharma2009performance,sharma2011conditionally,sharma2010performance}.
Asymmetric reporting errors require separate positive-to-negative and
negative-to-positive crossover probabilities instead of the single
$\epsilon_k^{\mathrm{rep}}$ used here.

The \gls{cv} computation does not explicitly require instantaneous \gls{csi} once
the effective probabilities $q_{k,j}$ are supplied. Nevertheless,
demodulation and estimation of these probabilities may still require
channel information. When $q_{k,1}=q_{k,0}$, that decision supplies no
hypothesis-discriminating evidence; equality to $1/2$ is a special case,
not a general sensor assumption. With reliable reporting, \gls{cv} approaches
fusion of the original local decisions. At poorer reporting \gls{snr}, its
performance must be evaluated using the effective probabilities, and
hard demodulation may lose information relative to received-signal
fusion \cite{niu2006fusion}.

Table~\ref{tab:comparison} compares the statistical rules at the level of
the \gls{fc} input and the information needed to implement them. A binary
symbol transmitted by a sensor does not make every receiver a
hard-\gls{dfr}: an \gls{fc} may combine its continuous-valued
channel observations before making a global decision.

\begin{table*}[!t]
\caption{Information Requirements and Interpretation of Statistical Fusion Rules.}
\label{tab:comparison}
\centering
\begingroup
\scriptsize
\setlength{\tabcolsep}{3pt}
\renewcommand{\arraystretch}{1.15}
\begin{tabularx}{\textwidth}{@{}
>{\raggedright\arraybackslash}p{1.15cm}
>{\raggedright\arraybackslash}p{3.35cm}
>{\raggedright\arraybackslash}p{2.65cm}
>{\raggedright\arraybackslash}p{2.45cm}
>{\raggedright\arraybackslash}X@{}}
\toprule
\textbf{Rule} & \textbf{Information required at \gls{fc}} &
\textbf{Input to the statistic} & \textbf{Online processing} &
\textbf{Interpretation and limitation}\\
\midrule
\gls{lrt} & Conditional observation models; channel and sensor parameters when
present in those models & Continuous or discrete observations specified
by the model & Likelihood evaluation; cost depends on model structure &
Optimal for the stated statistical criterion and known model. A soft
received-signal \gls{lrt} need not have the same information as a binary-decision
\gls{lrt}.\\
\specialrule{0.3pt}{1pt}{1pt}
\gls{cv} & Effective probabilities $q_{k,0},q_{k,1}$ & Decoded binary decisions &
Weighted sum; log-weights can be precomputed & Exact binary-decision \gls{llr}
under conditional independence. Reporting errors and sensor mismatch
must be reflected in the probabilities.\\
\specialrule{0.3pt}{1pt}{1pt}
\gls{mrc} & \gls{csi}, powers, noise levels, local operating points
for the sensing-weighted variant & Continuous received samples &
Channel compensation and weighted sum & Conventional \gls{mrc} optimizes
combining \gls{snr} for copies of a common symbol; across nonideal sensors it
is not a general likelihood-optimal detector.\\
\specialrule{0.3pt}{1pt}{1pt}
\gls{egc} & Channel phases in the coherent model & Phase-aligned received
samples & Phase compensation and equal-gain sum & Avoids instantaneous
amplitude weighting, but still needs phase information and calibrated
thresholds.\\
\specialrule{0.3pt}{1pt}{1pt}
\gls{sc} & Branch-quality estimates and the selected branch channel & One
selected received sample per sensor & Branch search followed by combining &
Reduces retained branches, but discards diversity information and remains
sensitive to selection errors.\\
\specialrule{0.3pt}{1pt}{1pt}
Counting & Report alphabet and voting threshold; statistical calibration
for a prescribed false-alarm rate & Decoded binary decisions & Positive-decision
count & Treats decisions equally. It is a \gls{cv} specialization only under
appropriate homogeneous effective probabilities.\\
\bottomrule
\end{tabularx}
\endgroup
\end{table*}

\subsubsection{Diversity-Combining Rules}
Diversity combining offers relatively simple alternatives to full
likelihood evaluation. Their performance depends on the sensing and
reporting model, rather than on a universal \gls{snr} ranking
\cite{Ciuonzo2013MRC,niu2006fusion,varshney2002}. For clarity, the coherent
single-decision model is ${\mathbf{r}}_k=\sqrt{p_k}h_ku_k+n_k^{\mathrm{rep}}$, where
$n_k^{\mathrm{rep}}\sim\mathcal{CN}(0,\sigma_{\mathrm{rep},k}^{2})$ and
$\sigma_{\mathrm{rep},k}^{2}=\mathbb E[|n_k^{\mathrm{rep}}|^2]$ is the
total complex-noise variance.

\paragraph{\gls{mrc}-based rule}
Conventional \gls{mrc} weights observations using their channel gains and
noise levels. When decisions originate from different imperfect sensors,
an additional sensing-reliability factor can be included. The
sensing-weighted \gls{mrc} form considered here is
\begin{equation}
\begin{aligned}
\Lambda_{\mathrm{MRC}}
={}&\frac{1}{K}\sum\nolimits_{k=1}^{K}
(P_{\mathrm d,k}-P_{\mathrm{fa},k})
\frac{\sqrt{p_k}}{\sigma_{\mathrm{rep},k}^{2}}
\operatorname{Re}\{h_k^*{\mathbf{r}}_k\}
&\underset{\mathcal{H}_0}{\overset{\mathcal{H}_1}{\gtrless}}\lambda_{\mathrm{MRC}}.
\end{aligned}
\label{eq:section3_mrc_statistic}
\end{equation}
The conjugation and real part make the statistic consistent with the
complex channel model. Omitting $P_{\mathrm d,k}-P_{\mathrm{fa},k}$ gives
a channel-only combining form. The displayed variant requires the local
operating points as well as reporting \gls{csi}; neither variant is generally
the exact \gls{llr} for heterogeneous nonideal sensors. The factor $1/K$ is a
fixed scale that can be absorbed into the threshold.

\paragraph{\Gls{egc}-based rule}
\Gls{egc} compensates each channel phase and
then assigns equal combining amplitudes:
\begin{equation}
\Lambda_{\mathrm{EGC}}
=\frac{1}{K}\sum\nolimits_{k=1}^{K}
\operatorname{Re}\!\left\{e^{-\mathrm j\angle h_k}{\mathbf{r}}_k\right\}
\underset{\mathcal{H}_0}{\overset{\mathcal{H}_1}{\gtrless}}\lambda_{\mathrm{EGC}}.
\label{eq:section3_egc_statistic}
\end{equation}
Instantaneous channel magnitudes do not appear in the combining
weights, but channel phases are still required. Zero-gain or unavailable
branches are omitted. Equal weights simplify processing, although they
do not represent differences in sensing reliability or received
amplitude. Threshold calibration remains necessary.

\paragraph{\Gls{sc}-based rule}
\Gls{sc} retains one reporting branch for each sensor.
Let $\mathcal M_k$ be its candidate branches and
$\mu_{k,m}=\sqrt{\alpha_{k,m}^{\mathrm{rel}}p_{k,m}}h_{k,m}$ the effective
received amplitude. The relay-power factor is one for a direct or
fixed-power branch. Define
\begin{equation}
\begin{aligned}
\gamma_{k,m}^{\mathrm{rep}}
&=\frac{|\mu_{k,m}|^2}{\sigma_{\mathrm{rep},k,m}^{2}},
\\
m_k^\star
&\in\operatorname*{arg\,max}_{m\in\mathcal M_k}
\gamma_{k,m}^{\mathrm{rep}},
\qquad
\gamma_k^{\mathrm{selected}}=\gamma_{k,m_k^\star}^{\mathrm{rep}}.
\end{aligned}
\label{eq:section3_selection_index}
\end{equation}
The maximizer is a branch index, whereas
$\gamma_k^{\mathrm{selected}}$ is the corresponding \gls{snr}. Ties may be
resolved using the smallest branch index. An observation-domain statistic
is
\begin{equation}
\Lambda_{\mathrm{SC}}
=\sum\nolimits_{k=1}^{K}
\frac{\operatorname{Re}\{\mu_{k,m_k^\star}^{*}{\mathbf{r}}_{k,m_k^\star}\}}
{\sigma_{\mathrm{rep},k,m_k^\star}^{2}}
\underset{\mathcal{H}_0}{\overset{\mathcal{H}_1}{\gtrless}}\lambda_{\mathrm{SC}}.
\label{SC}
\end{equation}
The received samples retain their noise contributions. The candidate set
must be nonempty for each included sensor; otherwise that sensor is
omitted and the threshold is calibrated for the available set. When relays
regenerate a possibly erroneous decision, final-hop \gls{snr} alone does not
describe end-to-end reliability; relay-aware selection is discussed in
Section~\ref{sec:advanced_fusion}. Multiple branches associated with one
sensor are repeated decisions, not additional independent sensors.

\subsubsection{Counting-Based Rules}
Counting rules include OR, AND, majority voting, and a
$\kappa_{\mathrm v}$-out-of-$K$ rule. For bipolar decoded decisions, the
number of positive decisions is
\begin{equation}
N_+
=\sum\nolimits_{k=1}^{K}\frac{1+\widehat {\mathbf{u}}_k^{\mathrm{FC}}}{2}
\underset{\mathcal{H}_0}{\overset{\mathcal{H}_1}{\gtrless}}\kappa_{\mathrm v},
\qquad \kappa_{\mathrm v}\in\{1,\ldots,K\}.
\label{eq:section3_counting_rule}
\end{equation}
The \gls{fc} selects $\mathcal{H}_1$ when $N_+\geq\kappa_{\mathrm v}$. OR uses
$\kappa_{\mathrm v}=1$, AND uses $\kappa_{\mathrm v}=K$, and strict
majority uses $\lfloor K/2\rfloor+1$. The equivalent threshold on the
signed sum is $2\kappa_{\mathrm v}-K$. Counting becomes a monotone
specialization of \gls{cv} for conditionally independent decisions with common
effective probabilities satisfying $q_1>q_0$. Otherwise, equal voting is
a simplicity-driven choice rather than the general \gls{cv} rule
\cite{chair1986optimal,varshney2012distributed}.
The count itself requires little computation, but choosing its threshold
to meet a specified global false-alarm probability still requires
statistical information or validation data. With missing decisions, the
available-decision count and the associated threshold must be adjusted.

\subsection{Machine Learning-Based Fusion Rules}

\gls{ml} estimates a decision function or a supporting
control policy from data and interaction. A feature vector
$\boldsymbol y=[y_1,\ldots,y_{D_{\mathrm f}}]^{\mathsf T}$ may contain
sensor energies, confidence scores, or received-signal features. The \gls{fc}
uses these inputs to infer $\mathcal{H}_0$ (channel idle) or $\mathcal{H}_1$ (channel
occupied) in spectrum-sensing applications \cite{sarikhani2020}.
Adaptability depends on training coverage and model updating; it does
not follow merely from using an \gls{ml} algorithm. The input representation
also determines whether the operation is decision-level, feature-level,
or raw-observation fusion.

\subsubsection{Unsupervised Rules}
Unsupervised learning extracts structure without a complete set of
hypothesis labels. K-means and \gls{gmm} can group
sensing vectors, while \gls{pca} can reduce
feature dimension \cite{thilina2013,khalek2020}. K-means partitions the
training samples into $K_{\mathrm c}$ clusters by minimizing squared
distances to their centroids. A \gls{gmm} instead represents their distribution
using a weighted collection of Gaussian components. More training samples
do not, by themselves, make an arbitrary sensing distribution Gaussian.

Cluster membership does not by itself identify a physical hypothesis: a sample’s assignment to a cluster does not imply that it belongs to a particular hypothesis. A
component-to-hypothesis association must be established using calibration,
limited labels, or justified application knowledge. Similarly, \gls{pca}
produces a representation rather than a complete classifier, and may
remove low-variance features that are informative for detection. The
workflow reviewed in \cite{chen2016} combines offline clustering with
online similarity-based classification and feature extraction.
Table~\ref{tab:unsuper} states the standard clustering objectives and
assignments used to interpret these approaches; the assignments require
a separate mapping to the desired hypotheses.

\begin{table*}[!t]
\caption{Unsupervised Representations and the Additional Step Required for Hypothesis Decisions}
\label{tab:unsuper}
\centering
\begingroup
\scriptsize
\setlength{\tabcolsep}{3pt}
\renewcommand{\arraystretch}{1.25}
\begin{tabularx}{\textwidth}{@{}
>{\raggedright\arraybackslash}p{2.0cm}
>{\raggedright\arraybackslash}p{5.15cm}
>{\raggedright\arraybackslash}p{4.2cm}
>{\raggedright\arraybackslash}X@{}}
\toprule
\textbf{Method / references} & \textbf{Standard fitting objective} &
\textbf{Assignment for a new vector} & \textbf{Decision interpretation}\\
\midrule
K-means\newline\cite{thilina2013}
& $\displaystyle
\min_{\{\mathcal C_j,\boldsymbol c_j\}}
\sum\nolimits_{j=1}^{K_{\mathrm c}}\sum\nolimits_{n\in\mathcal C_j}
\|\boldsymbol y_n-\boldsymbol c_j\|_2^2$
& $\displaystyle
j^\star\in\operatorname*{arg\,min}_{1\leq j\leq K_{\mathrm c}}
\|\boldsymbol y-\boldsymbol c_j\|_2^2$
& The sets $\mathcal C_j$ partition the training indices. Centroid
$\boldsymbol c_j$ identifies a cluster, not a hypothesis without an
additional calibrated association.\\
\specialrule{0.3pt}{1pt}{1pt}
GMM\newline\cite{thilina2013,khalek2020}
& $\displaystyle
\max_{\boldsymbol\theta_{\mathrm{mix}}}
\sum\nolimits_{n=1}^{N_{\mathrm{tr}}}
\ln\!\left[\sum\nolimits_{j=1}^{K_{\mathrm c}}
\omega_j\mathcal N(\boldsymbol y_n;\boldsymbol\mu_j,\boldsymbol\Sigma_j)
\right]$
& $\displaystyle
j^\star\in\operatorname*{arg\,max}_{1\leq j\leq K_{\mathrm c}}
\omega_j\mathcal N(\boldsymbol y;\boldsymbol\mu_j,\boldsymbol\Sigma_j)$
& The weights are nonnegative and sum to one. Covariance eigenvalues require a
positive lower bound or suitable regularization. Components must be mapped
to hypotheses; an unlabeled component index is not a channel-occupancy
label.\\
\bottomrule
\end{tabularx}
\endgroup
\end{table*}

\subsubsection{Supervised Rules}
Supervised learning uses labeled examples to fit a mapping from sensing
or reporting features to the target hypothesis. The techniques reviewed
here include nearest-neighbor and decision-tree classifiers, support
vector machines, and neural networks trained by back-propagation
\cite{thilina2013,supervised2012}. Back-propagation is a training mechanism rather
than a separate hypothesis-testing rule. Likewise, a regression output
requires a specified decision threshold before it becomes a binary
classifier. Performance depends on representative labels, class balance,
and separation between training and evaluation data. A low training loss
does not establish the global false-alarm rate under a new reporting
channel or sensor population.

\subsubsection{Reinforcement Learning-Based Rules}
\Gls{rl} is suitable for systems that repeatedly
interact with a changing wireless environment. An agent learns a policy
from the rewards associated with its actions. The state may include
available decisions, channel conditions, stored energy, and information
freshness, while the action can control activation, reporting power,
fusion weights, or the decision threshold
\cite{Luong2019DRLSurvey,sarikhani2020,rl2010}. \gls{rl} can therefore adapt a
conventional detector rather than replace its complete statistical
structure. A representative hard-decision template is
\begin{equation}
\begin{aligned}
\Lambda_{\mathrm{RL}}(t)
&=\sum\nolimits_{k\in\mathcal D(t)}
w_k^{\mathrm{RL}}(\boldsymbol s_t^{\mathrm{RL}})
\widehat {\mathbf{u}}_k^{\mathrm{FC}}(t)
&\underset{\mathcal{H}_0}{\overset{\mathcal{H}_1}{\gtrless}}
\lambda_{\mathrm{RL}}(\boldsymbol s_t^{\mathrm{RL}}),
\end{aligned}
\label{eq:RL_fusion_statistic}
\end{equation}
where $\mathcal D(t)$ is the set of accepted decoded decisions and
$\boldsymbol s_t^{\mathrm{RL}}$ is the available state representation.
Activation is selected by the policy, but actual decision delivery also
depends on the channel. The expression alone does not specify a reward,
training algorithm, or optimal policy. Partial observability may require
a history-based state or a belief representation.

In decentralized networks, multi-agent \gls{rl} allows sensors to select
local actions while a common objective couples their behavior.
Centralized training with decentralized execution can exploit broader
information during training without requiring it at every sensor during
deployment. Cooperative spectrum-sensing studies illustrate this design
direction \cite{Gao2021MARLCSS}. Its limitations include changing
co-agent policies, exploration cost, delayed knowledge of the true
hypothesis, and the difficulty of attributing a global error to one
local action. Reliability-sensitive deployment consequently requires
careful reward design, representative offline training, and constrained
online adaptation rather than unrestricted exploration.

\subsubsection{Deep-Learning Architectures}
Deep learning uses multilayer neural architectures to learn nonlinear
representations and decision functions. It is not a fourth learning
paradigm disjoint from supervised, unsupervised, and reinforcement
learning. The architectures discussed here can be trained within those
paradigms. When a network estimates hypothesis posterior probabilities,
an equal-misclassification-cost decision is
\begin{equation}
\widehat{\mathcal{H}}=\mathcal{H}_{j^\star},
\qquad
j^\star=\operatorname*{arg\,max}_{0\leq j<J}
p_{\boldsymbol\vartheta}(\mathcal{H}_j\mid\boldsymbol y).
\label{eq:section3_deep_decision}
\end{equation}
Here $\boldsymbol\vartheta$ contains the network parameters and $J$ is
the number of hypotheses, not the number of relays or quantization
levels. For binary \gls{np} operation, a separately calibrated
score threshold may be more appropriate than a posterior argmax.

Multilayer perceptrons can process fixed-length feature vectors,
convolutional networks can exploit structured signal representations,
and recurrent architectures can model temporal dependencies. Recent
applications reviewed in the manuscript include cooperative spectrum
sensing and distributed wireless-signal classification
\cite{Lee2019DeepCooperative,Gao2019DeepLearning,Rajendran2018,deep2017}.
Not all of these applications fuse hard local decisions; some combine
raw or intermediate features. Learned representations can reduce the
need for an explicit analytical likelihood, but they do not remove
assumptions about training distributions, calibration, or deployment
conditions. Data requirements, computational cost, and performance under
channel or sensor-distribution shift therefore remain central concerns.

\subsection{Game-Theoretic Approaches to Decision Fusion}
Game theory models the interaction of agents whose objectives need not
coincide. Sensors may limit cooperation to conserve energy, while an \gls{fc}
may seek better detection and malicious participants may attempt to
corrupt the decision. These interactions concern the design of sensing,
reporting, incentives, and defenses; a utility function is not itself a
fusion statistic. For illustration, the utility of sensor $k$ can be
written as
\begin{equation}
U_k
=c_{\mathrm d}^{U}P_{\mathrm d}
-c_{\mathrm{fa}}^{U}P_{\mathrm{fa}}
-c_{\mathrm E}^{U}E_k
-c_{\mathrm C}^{U}C_k,
\label{eq:section3_game_utility}
\end{equation}
where $E_k$ and $C_k$ are its energy and communication costs. The
nonnegative coefficients express preferences and include the scaling
needed to combine quantities with different units. The probabilities
and costs depend on the agents' actions and the chosen detector.

Noncooperative games model individual strategic choices, whereas
cooperative and coalition-formation approaches describe ways of sharing
resources or benefits. Attacker--defender formulations model adversarial
behavior and may themselves be noncooperative games. These categories
therefore overlap rather than forming a disjoint taxonomy. Studies of
Byzantine \gls{df} and cooperative spectrum-sensing games provide
representative examples \cite{Abrardo2016,Wang2010,Hao2012,game2008}.
The approach is useful for analyzing incentives and stability, but
requires defensible utility functions and a specified equilibrium
concept. A stable equilibrium need not minimize the centralized
probability of error, and computing or tracking it can become costly in
large, changing networks.

Covert communication gives the game-theoretic view different role for fusion, since here the \gls{fc} is on the adversary's side. In \cite{covert2023}, a transmitter and a cooperative jammer act as a single player against an \gls{fc} that combines the reports of colluding energy-detecting wardens, under a finite blocklength. The \gls{fc} fuses either hard decisions, using \gls{cv} weights, or the local energies themselves with \gls{mrc}-type weights. The transmit and jamming powers are randomized by one player and the detection thresholds by the other, and the utility trades the outage probability at the legitimate receiver against $P_{fa} + (1-P_d)$ at the \gls{fc}. The equilibrium is then computed by linear programming. Without a jammer, the authors show that adding wardens reduces covertness. With a cooperative jammer, the gain from extra wardens is negligible once the maximum transmit and jamming powers are large enough, and this holds for both fusion schemes.

\subsection{Heuristic and Application-Dependent Fusion Designs}
Heuristic designs use empirical or simplified choices when an exact
probability model is unavailable or expensive to implement. One example
is a weighted combination of received decisions,
\begin{equation}
\Lambda_{\mathrm{heur}}
=\sum\nolimits_{k=1}^{K}w_k\widehat {\mathbf{u}}_k^{\mathrm{FC}}
\underset{\mathcal{H}_0}{\overset{\mathcal{H}_1}{\gtrless}}\lambda_{\mathrm{heur}}.
\label{eq:section3_heuristic_statistic}
\end{equation}
The form of a weighted sum does not make it heuristic by itself.
Likelihood-derived weights give a statistical rule; weights estimated
from labeled data give a learned rule; and empirical weighting without
such a derivation is a heuristic design. Channel quality, past sensing
performance, trust, or information freshness may guide the weights.
Battery state and communication cost usually affect participation and
decision quality indirectly, rather than making a sensor's observation
intrinsically more accurate.

The resource, reliability, and trust considerations reviewed in
\cite{Liu2020EH,AlJarrah2020,heuristic2000} motivate such design
choices, but do not imply that every technique in those articles belongs to
one heuristic family. Fuzzy logic and evidence-theoretic fusion provide
additional representation and combination frameworks. Bayesian inference
is a probabilistic principle, not inherently a heuristic. Similarly,
genetic algorithms, particle swarm optimization, and ant colony
optimization are search mechanisms for thresholds, participation, or
weights rather than standalone detectors. Their optimization cost may
be substantial even when the final online \gls{dfr} is simple.

The main attraction of an empirical design is flexibility, but the
chosen weights and threshold require validation under the anticipated
sensor and channel conditions. Low computation and reliable detection
cannot be assumed simultaneously without specifying the algorithm and
its operating point. Table~\ref{tab:section3_fusion_summary} summarizes
the resulting distinctions between a fusion statistic, a learning
mechanism, a model architecture, and a supporting design technique.
\begin{table*}[!t]
\caption{Summary of Decision-Fusion Approaches}
\label{tab:section3_fusion_summary}
\centering
\begingroup
\scriptsize
\setlength{\tabcolsep}{3pt}
\renewcommand{\arraystretch}{1.05}
\begin{tabularx}{\textwidth}{@{}
c
>{\raggedright\arraybackslash\hsize=0.65\hsize}X
>{\raggedright\arraybackslash\hsize=1.15\hsize}X
>{\raggedright\arraybackslash\hsize=1.20\hsize}X@{}}
\toprule
\textbf{Ref.} & \textbf{Reviewed approach} &
\textbf{Core operation} & \textbf{Scope and limitations}\\
\midrule
\cite{chair1986optimal,varshney2012distributed,Ciuonzo2013MRC,niu2006fusion}
& Statistical fusion (\gls{cv}, diversity combining, and counting rules)
& \gls{cv} forms the \gls{llr} of decoded binary decisions from their
effective conditional probabilities, including reporting crossover errors
when modeled. \gls{mrc}, \gls{egc}, and \gls{sc} combine received samples
using channel-weighted, phase-only, or selected-branch weights; a
sensing-weighted \gls{mrc} variant also uses local operating characteristics.
Counting rules (OR, AND, majority) compare the number of positive decoded
decisions with a threshold.
& \gls{cv} is exact for binary decisions under conditional independence and
known probabilities, but is not generally equivalent to processing all
received continuous observations. Diversity combining is not automatically
an exact \gls{llr}; its performance depends on \gls{csi}, noise information,
branch correlation, and the source of the local decisions. Counting needs
little online processing and equals \gls{cv} only under homogeneous effective
probabilities, yet its threshold still requires calibration for a prescribed
false-alarm rate.\\
\specialrule{0.35pt}{0.8pt}{0.8pt}
\cite{thilina2013}
& Supervised and unsupervised learning
& Learns class mappings from labels or extracts clusters and lower-dimensional
representations from unlabeled data.
& Clustering and \gls{pca} require an additional hypothesis-decision mechanism.
Supervised performance depends on representative labels and deployment
coverage; neither paradigm defines the reporting resolution.\\
\specialrule{0.35pt}{0.8pt}{0.8pt}
\cite{sarikhani2020,Gao2021MARLCSS}
& \gls{rl}-assisted design
& Learns actions that control sensing, participation, reporting resources,
or fusion parameters through interaction and rewards.
& \gls{rl} is a policy-learning mechanism and may retain a statistical
detector. Exploration, partial observability, reward design, and training
cost require explicit treatment.\\
\specialrule{0.35pt}{0.8pt}{0.8pt}
\cite{Lee2019DeepCooperative,Gao2019DeepLearning,Rajendran2018}
& Deep neural architectures
& Learns nonlinear signal representations and classification scores
using multilayer networks. The cited applications include sensing and
wireless-signal classification.
& An architecture dimension that overlaps supervised, unsupervised,
and \gls{rl} techniques. Feature- or signal-level processing should not be
mislabeled as hard-\gls{df}.\\
\specialrule{0.35pt}{0.8pt}{0.8pt}
\cite{Abrardo2016,Wang2010,Hao2012,covert2023}
& Game-theoretic design
& Models incentives, cooperation, or adversarial interaction using
utilities and equilibrium concepts.
& A strategic design framework, which can use statistical or learned
fusion. Equilibrium stability is not synonymous with globally optimal
detection; realistic utilities and solution costs matter.\\
\specialrule{0.35pt}{0.8pt}{0.8pt}
\cite{Liu2020EH,AlJarrah2020}
& Resource- and reliability-aware design context
& Uses energy availability, reporting quality, or trust to motivate
participation and weighting choices. Empirical instances may be
heuristic.
& These references provide related design contexts, not a single common
heuristic algorithm. Bayesian modeling and evidence frameworks are not
intrinsically heuristic, and metaheuristics optimize rather than define
the final detector.\\
\bottomrule
\end{tabularx}
\endgroup
\end{table*}
\section{Advanced Fusion Techniques}
\label{sec:advanced_fusion}
Advanced fusion techniques address the practical costs of likelihood-based
detection. Although the \gls{lrt} is an important benchmark
under a specified model and decision criterion, its implementation in a
cooperative or multihop wireless sensor network may require reporting \gls{csi}, local operating probabilities, relay-transition
statistics, and repeated nonlinear computation. The objective is therefore
not only reliable detection, but also an acceptable information, signaling,
energy, bandwidth, and computational burden. The approaches reviewed here
include channel-aware combining, relay-power adaptation, branch selection,
ideal-node assumptions, and piecewise-linear likelihood approximations
\cite{AlJarrah2019DFCoopWSN,AlJarrah2020}.

These approaches act at different stages of the same system. \Gls{ils} and \gls{ir}
simplify the statistical model; \Gls{cmrc} and \gls{cegc} change the receiver's combining
weights; \gls{lar} changes relay transmission; and \Gls{smrc} changes the retained
reporting paths. \Gls{mlla} approximates a likelihood calculation. They are not
mutually exclusive alternatives to the statistical, learned, game-theoretic,
and heuristic perspectives of Section~\ref{sec:classification_fusion}.

Recent extensions also change report resolution, receiver hardware, and
parameter acquisition. Recent work studies $M$-ary multihop decision problems as well as
multibit local decision \cite{zhang2024mary,Zhang2025MultiBitDDF}.
Holographic architectures jointly design an analog front end and digital
inference \cite{ciuonzo2025holographic,Ciuonzo2026HolographicTarget}.
Time-aware reporting, blind estimation, semantic representations, and
over-the-air aggregation introduce additional timing, learning, and
communication choices
\cite{Tabella2023TimeAware,Perez2026BlindFusion,Yi2025SemanticAirCompCSS}.
Where a cited work fuses features rather than local hypothesis decisions,
that distinction is retained. A compact formula used to illustrate a common
design principle is not necessarily the exact algorithm of every cited study.

\subsection{Ideal Local Sensors Based Rule}
\label{subsec:ils_rule}

\gls{ils} rule is a useful benchmark for cooperative
\gls{df}. It assumes perfect local sensing,
$P_{\mathrm d,k}=1$ and $P_{\mathrm{fa},k}=0$, so that uncertainty is
introduced only by reporting and relay regeneration. This assumption
separates sensing errors from communication errors, but should not be
interpreted as a general model of a deployed sensor network
\cite{AlJarrah2019DFCoopWSN}. Work on unknown detection probabilities
addresses a related information limitation without making all sensors
perfect; it does not imply that every such study uses the same cooperative
\gls{ils} receiver \cite{Ciuonzo2014UnknownPd}.

Let $m=0$ denote the direct branch and $m=1,\ldots,M$ the relayed
branches associated with sensor $k$. A common coherent reporting model
for the binary schemes in this section is
\begin{equation}
\begin{aligned}
{\mathbf{r}}_{k,m}
&=\mu_{k,m}\widehat {\mathbf{u}}_{k,m}^{\mathrm{rel}}
+n_{k,m}^{\mathrm{rep}},
\\
\mu_{k,m}
&=\sqrt{\alpha_{k,m}^{\mathrm{rel}}p_{k,m}}\,h_{k,m}.
\end{aligned}
\label{eq:ils_received_signal}
\end{equation}
Here $\widehat {\mathbf{u}}_{k,m}^{\mathrm{rel}}$ is the last regenerated binary
symbol, $\widehat {\mathbf{u}}_{k,0}^{\mathrm{rel}}={\mathbf{u}}_k$ is the original sensor
decision, and $h_{k,m}$ includes the final reporting-channel attenuation.
The nominal power is $p_{k,m}$ and the actual radiated power is
$\alpha_{k,m}^{\mathrm{rel}}p_{k,m}$, with
$\alpha_{k,0}^{\mathrm{rel}}=1$. For the fixed-power ILS benchmark,
$\alpha_{k,m}^{\mathrm{rel}}=1$. The receiver noise is
$n_{k,m}^{\mathrm{rep}}\sim\mathcal{CN}(0,\sigma_{\mathrm{rep},k,m}^{2})$,
where $\sigma_{\mathrm{rep},k,m}^{2}$ is its total complex variance.
Define
\begin{equation}
z_{k,m}={\mathbf{r}}_{k,m}\mu_{k,m}^{*},
\qquad
y_{k,m}
=\frac{4\operatorname{Re}\{z_{k,m}\}}
{\sigma_{\mathrm{rep},k,m}^{2}}.
\label{eq:ils_branch_statistic}
\end{equation}
The factor four follows from the coherent \gls{bpsk} log-density ratio under
this complex-noise convention. We condition on the known channel and
power-control state. The branch likelihoods factor only when reporting
noises and regeneration outcomes are independent conditional on the
common source decision. Independence across sensors must likewise hold
conditional on the hypothesis for the subsequent sensor-wise \gls{llr} sum.

Under the \gls{ils} model, the statistic is
\begin{equation}
\Lambda_{\mathrm{ILS}}
=\sum\nolimits_{k=1}^{K}\left(y_{k,0}+\sum\nolimits_{m=1}^{M}\Omega_{k,m}\right)
\underset{\mathcal{H}_0}{\overset{\mathcal{H}_1}{\gtrless}}\lambda_{\mathrm{ILS}},
\label{eq:ils_llr}
\end{equation}
where the threshold is selected for the required global operating
point. With symmetric relay crossover probabilities, the branch
contribution is
\begin{equation}
\begin{aligned}
\Omega_{k,m}
&=\ln\!\left(
\frac{d_{k,m}^{\mathrm{odds}}+\exp(y_{k,m})}
{1+d_{k,m}^{\mathrm{odds}}\exp(y_{k,m})}
\right), d_{k,m}^{\mathrm{odds}}
=\frac{\epsilon_{k,m}^{\mathrm{rel}}}
{1-\epsilon_{k,m}^{\mathrm{rel}}}.
\end{aligned}
\label{eq:ils_omega}
\end{equation}
The probability $\epsilon_{k,m}^{\mathrm{rel}}$ describes a regenerated
symbol being different from ${\mathbf{u}}_k$, with the same crossover in both
directions. For a multihop path it must describe the entire regeneration
chain. It is not the error probability of the final noisy sample at the
\gls{fc}. Asymmetric relay errors require two transition probabilities and a
correspondingly different likelihood.

The nonlinear contribution can be approximated by
\begin{equation}
\begin{aligned}
\Omega_{k,m}&\approx\widetilde\Omega_{k,m}
=\operatorname{clip}(y_{k,m};c_{k,m}^{\mathrm{rel}}),
\\
c_{k,m}^{\mathrm{rel}}
&=\ln\!\left(
\frac{1-\epsilon_{k,m}^{\mathrm{rel}}}
{\epsilon_{k,m}^{\mathrm{rel}}}\right),
\qquad 0<\epsilon_{k,m}^{\mathrm{rel}}<\tfrac12,
\end{aligned}
\label{eq:ils_clip}
\end{equation}
where
\begin{equation}
\operatorname{clip}(x;c)
=\max\{-c,\min\{x,c\}\},
\qquad c\geq0.
\label{eq:clip_definition}
\end{equation}
For fixed relay statistics, the clipping levels can be computed offline.
The unit-slope central segment is an approximation: the exact slope of
\eqref{eq:ils_omega} at zero is $1-2\epsilon_{k,m}^{\mathrm{rel}}$.
Thus, unit-slope clipping is most natural for reliable relays. An
error-free relay gives $\Omega_{k,m}=y_{k,m}$, while a crossover of
one half gives zero evidence. These limits should be used directly
rather than evaluating a singular clipping level.

The clipped rule avoids repeated nonlinear operations, but a mismatched
ideal-sensor assumption may leave substantial sensing error even when
reporting links are strong. Score-based tests and blind parameter learning
provide different alternatives to perfect prior sensor knowledge
\cite{ciuonzo2025score,Perez2026BlindFusion}. They should not be
classified as equivalent \gls{ils} approximations merely because they also
reduce information or implementation requirements.
\subsection{Cooperative Maximal-Ratio Combining Based Rule}
\label{subsec:cmrc_rule}

A sensor decision and its regenerated copies need not have equal
reliability. Channel-only \gls{mrc} can give excessive influence to a strong
final link carrying an incorrectly regenerated symbol. \Gls{cmrc} incorporates the preceding-hop quality into the combining weight
\cite{Wang2007CooperativeDemodulation}. Channel-aware \gls{mimo} decision-fusion
work provides a related \gls{mrc} design context, but its receive architecture
should be distinguished from a relay network \cite{Ciuonzo2013MRC}.

Using the model in \eqref{eq:ils_received_signal}, the branch-weighted
combiner used throughout this section is
\begin{equation}
\Lambda_{\mathrm{BW}}
=\sum\nolimits_{k=1}^{K}\sum\nolimits_{m=0}^{M}
w_{k,m}^{\mathrm{br}}
\frac{\operatorname{Re}\{\mu_{k,m}^{*}{\mathbf{r}}_{k,m}\}}
{\sigma_{\mathrm{rep},k,m}^{2}}
\underset{\mathcal{H}_0}{\overset{\mathcal{H}_1}{\gtrless}}\lambda_{\mathrm{BW}}.
\label{eq:branch_weighted_combiner}
\end{equation}
The rules below differ in how $w_{k,m}^{\mathrm{br}}$ is chosen and
where reliability control is applied; the threshold is calibrated
separately for each rule. The branch weight $w_{k,m}^{\mathrm{br}}$ is
distinct from a sensor-level fusion weight. \Gls{cmrc} with fixed relay powers
uses $\alpha_{k,m}^{\mathrm{rel}}=1$ and $w_{k,0}^{\mathrm{br}}=1$. For a
two-hop relay, a normalized weighting example for \gls{cmrc} is
\begin{equation}
w_{k,m}^{\mathrm{br}}
={\min\{\gamma_{k,m}^{\mathrm{sr}},
                 \gamma_{k,m}^{\mathrm{rf}}\}}
*\frac{1}{\gamma_{k,m}^{\mathrm{rf}}},
\qquad m\geq1.
\label{eq:cmrc_weight}
\end{equation}
Here $\gamma_{k,m}^{\mathrm{sr}}$ is the sensor-to-relay \gls{snr} and
$\gamma_{k,m}^{\mathrm{rf}}=p_{k,m}|h_{k,m}|^2/
\sigma_{\mathrm{rep},k,m}^{2}$ is the nominal relay-to-\gls{fc} \gls{snr} before
any power scaling. Consequently, the effective second-hop \gls{snr} is
$\alpha_{k,m}^{\mathrm{rel}}\gamma_{k,m}^{\mathrm{rf}}$.
A branch with zero usable received power is omitted and assigned zero
combining weight instead of evaluating a zero denominator.

This weighting prevents a strong final channel from concealing a weak
first hop, but remains a simplified reliability representation rather
than a general exact \gls{llr}. Obtaining preceding-hop \gls{csi} creates signaling
and tracking costs that grow with the number of paths. Recent multihop
and multibit fusion studies explicitly model the complete symbol
transitions and develop reduced-complexity alternatives
\cite{zhang2024mary,Zhang2025MultiBitDDF}. The two-hop expression
\eqref{eq:cmrc_weight} should not be extended to an arbitrary relay chain
without specifying an end-to-end reliability model.

\subsection{Link-Adaptive Regeneration Based Rule}
\label{subsec:lar_rule}

\gls{lar} moves part of the reliability control
from the \gls{fc} to the relay. The relay adjusts its transmit power according
to the decoding quality of its first hop and the quality of its reporting
link, so that an uncertain regenerated decision is not necessarily sent
at full nominal power \cite{Wang2008LAR,Vien2009SmartRelaying}.
For relay $m$, the transmitted symbol is
\begin{equation}
x_{k,m}^{\mathrm{tx}}
=\sqrt{\alpha_{k,m}^{\mathrm{rel}}p_{k,m}}
\widehat {\mathbf{u}}_{k,m}^{\mathrm{rel}},
\qquad 0\leq\alpha_{k,m}^{\mathrm{rel}}\leq1.
\label{eq:lar_signal}
\end{equation}
The factor $\alpha_{k,m}^{\mathrm{rel}}$ is a power coefficient;
$p_{k,m}$ must not already include this scaling.

\subsubsection{LAR With Instantaneous CSI}
With instantaneous \gls{csi} for both hops, the coefficient can be written as
\begin{equation}
\alpha_{k,m}^{\mathrm{rel,inst}}
=\frac{\min\{\gamma_{k,m}^{\mathrm{sr}},
                 \gamma_{k,m}^{\mathrm{rf}}\}}
{\gamma_{k,m}^{\mathrm{rf}}}.
\label{eq:lar_inst}
\end{equation}
Using the nominal second-hop \gls{snr} defined above, this limits its scaled
value to the weaker hop. The adaptation is fast, but requires sufficiently
current second-hop information at the relay, typically through feedback.

\subsubsection{LAR With Average Second-Hop CSI}
A lower-overhead form replaces the instantaneous second-hop \gls{snr} by its
average $\overline\gamma_{k,m}^{\mathrm{rf}}$:
\begin{equation}
\alpha_{k,m}^{\mathrm{rel,avg}}
=\frac{\min\{\gamma_{k,m}^{\mathrm{sr}},
                 \overline\gamma_{k,m}^{\mathrm{rf}}\}}
{\overline\gamma_{k,m}^{\mathrm{rf}}}.
\label{eq:lar_avg}
\end{equation}
The first-hop decoding condition still controls the coefficient, while
frequent second-hop feedback is avoided. At the \gls{fc}, the combiner in
\eqref{eq:branch_weighted_combiner} is applied with
$w_{k,m}^{\mathrm{br}}=1$, because reliability control is performed at
the relay through the coefficient $\alpha_{k,m}^{\mathrm{rel}}$ contained
in $\mu_{k,m}$. The direct branch uses $\alpha_{k,0}^{\mathrm{rel}}=1$.
The \gls{fc} must know or estimate this effective gain; shifting adaptation
to the relay does not remove all receiver-side channel requirements.

\gls{lar} is attractive when simple relay-side adaptation can reduce the
burden on the \gls{fc}. Its benefit remains sensitive to the age and accuracy
of the \gls{csi}. In a long relay chain, a single local scaling factor is not a
complete description of regeneration errors. Multihop and multirelay
fusion therefore motivates extending this principle through an explicit
end-to-end transition model \cite{zhang2024mary,Zhang2025MultiBitDDF}.

\subsection{Enhanced Maximal-Ratio Combining Based Rule}
\label{subsec:emrc_rule}

Enhanced \gls{mrc} (EMRC) combines relay-side regeneration with \gls{fc}-side
weighting. Adjustable coefficients represent both reporting hops using
instantaneous and average channel information \cite{Nasri2012EMRC}.
Within the common observation model, \gls{emrc} applies the combiner in
\eqref{eq:branch_weighted_combiner} with the weights
$w_{k,m}^{\mathrm{br}}$ from \eqref{eq:emrc_weight}, while the relay
uses $\alpha_{k,m}^{\mathrm{rel}}$ from \eqref{eq:emrc_alpha}.
The following two-coefficient parameterization is retained as an
illustrative survey model of this design principle; it is not asserted
to reproduce every optimized EMRC law or its performance guarantee:
\begin{subequations}
\label{eq:emrc_coefficients}
\begin{align}
\alpha_{k,m}^{\mathrm{rel}}
&=\frac{\min\{\gamma_{k,m}^{\mathrm{sr}}\delta_{k,m}^{\mathrm{sr}},
\overline\gamma_{k,m}^{\mathrm{rf}}\delta_{k,m}^{\mathrm{rf}}\}}
{\overline\gamma_{k,m}^{\mathrm{rf}}},
\label{eq:emrc_alpha}\\
w_{k,m}^{\mathrm{br}}
&=\frac{\min\{\overline\gamma_{k,m}^{\mathrm{sr}}/
\delta_{k,m}^{\mathrm{sr}},\gamma_{k,m}^{\mathrm{rf}}\}}
{\gamma_{k,m}^{\mathrm{rf}}}.
\label{eq:emrc_weight}
\end{align}
\end{subequations}
Bars denote \gls{snr} averages before relay scaling. The tuning parameters
are positive, but positivity alone does not enforce the power bound
$\alpha_{k,m}^{\mathrm{rel}}\leq1$. Their feasible set must additionally
satisfy
\[
\min\{\gamma_{k,m}^{\mathrm{sr}}\delta_{k,m}^{\mathrm{sr}},
\overline\gamma_{k,m}^{\mathrm{rf}}\delta_{k,m}^{\mathrm{rf}}\}
\leq\overline\gamma_{k,m}^{\mathrm{rf}}.
\]
For example, $0<\delta_{k,m}^{\mathrm{rf}}\leq1$ is a sufficient
restriction for this particular parameterization. The direct branch
again uses $w_{k,0}^{\mathrm{br}}=\alpha_{k,0}^{\mathrm{rel}}=1$.

Joint relay and \gls{fc} adaptation offers finer control than changing either
stage alone, but requires choosing the parameters for the operating
channels and resource limits. Positive coefficients by themselves do
not establish an optimum. Multibit and multihop operation makes the
transition model and tuning problem larger, motivating reduced-complexity
end-to-end formulations \cite{Zhang2025MultiBitDDF}. Comparisons should
therefore include the cost of coefficient optimization and \gls{csi} acquisition,
not just the final weighted sum.

\subsection{Cooperative Equal-Gain Combining Based Rule}
\label{subsec:cegc_rule}

\Gls{cegc} reduces the \gls{fc}'s instantaneous
channel-amplitude requirements by phase-aligning the reporting branches
and using equal combining gains. Relay adaptation can still use first-hop
\gls{csi} and average second-hop information \cite{Vien2012SmartEGC}.
For the phase-only combining interpretation, the real statistic is
\begin{equation}
\Lambda_{\mathrm{CEGC}}
=\sum\nolimits_{k=1}^{K}\sum\nolimits_{m=0}^{M}
\overline w_{k,m}^{\mathrm{br}}
\operatorname{Re}\!\left\{
e^{-\mathrm j\angle h_{k,m}}{\mathbf{r}}_{k,m}\right\}
\underset{\mathcal{H}_0}{\overset{\mathcal{H}_1}{\gtrless}}\lambda_{\mathrm{CEGC}}.
\label{eq:cegc_statistic}
\end{equation}
Strict equal-gain combining corresponds to
$\overline w_{k,m}^{\mathrm{br}}=1$. A slowly varying average-\gls{snr}
weight can instead account for persistent differences between direct
and relayed paths, giving a weighted \gls{egc} variant that is no
longer strictly equal gain. Weight design and threshold calibration
still depend on the adopted channel model, and neither form universally
outperforms channel-amplitude weighting.

Phase compensation gives the signal term
$|\mu_{k,m}|\widehat {\mathbf{u}}_{k,m}^{\mathrm{rel}}$ and rotated receiver noise.
Here $|\mu_{k,m}|=|h_{k,m}|\sqrt{\alpha_{k,m}^{\mathrm{rel}}p_{k,m}}$.
This amplitude factor is already present in ${\mathbf{r}}_{k,m}$ and is not
multiplied into the sample a second time. Zero-gain or inactive branches
are excluded. A compatible relay-side coefficient is
\begin{equation}
\alpha_{k,m}^{\mathrm{rel}}
=\frac{\min\{\gamma_{k,m}^{\mathrm{sr}},
\overline\gamma_{k,m}^{\mathrm{rf}}\}}
{\overline\gamma_{k,m}^{\mathrm{rf}}},
\qquad m\geq1.
\label{eq:cegc_alpha}
\end{equation}
Relay-power selection and fusion center \gls{egc} are separate operations.

Semantic over-the-air aggregation uses a related superposition mechanism
to reduce orthogonal reporting requirements \cite{Yi2025SemanticAirCompCSS}.
It is not ordinary \gls{egc}: the transmitted representations are learned for
the sensing task. The connection is architectural and communication-related,
not an equivalence between the statistical \glspl{dfr}.
\subsection{Selection Maximal-Ratio Combining Based Rule}
\label{subsec:smrc_rule}

\Gls{smrc} retains a selected relayed branch for each sensor
and combines it with the direct report. It can reduce the number of
active relay transmissions and the associated synchronization and
processing requirements \cite{Ikki2008BestRelaySelection,Yi2008RelaySelectionDiversity}.
Selection must use a reliability measure consistent with the relay
protocol. For the \gls{daf} setting here, we use the following
explicit bottleneck-\gls{snr} surrogate:
\begin{equation}
\gamma_{k,m}^{\mathrm{sel}}
=\min\!\left\{
\gamma_{k,m}^{\mathrm{sr}},
\alpha_{k,m}^{\mathrm{rel}}\gamma_{k,m}^{\mathrm{rf}}
\right\}.
\label{eq:smrc_e2e_snr}
\end{equation}

In contrast, the rational expression
$\gamma_1\gamma_2/(\gamma_1+\gamma_2+1)$ follows from a particular
variable-gain \gls{af} model and must not be treated as a
universal \gls{df} identity. A transition-error-based selection metric is an
alternative when the regeneration probabilities are available.

For nonempty available relay set $\mathcal M_k^{\mathrm{rel}}$, select
\begin{equation}
m_k^\star
\in\operatorname*{arg\,max}_{m\in\mathcal M_k^{\mathrm{rel}}}
\gamma_{k,m}^{\mathrm{sel}}.
\label{eq:smrc_selection}
\end{equation}
Ties can be resolved using the smallest relay index. An implementable
FC statistic is
\begin{equation}
\begin{aligned}
\Lambda_{\mathrm{SMRC}}=\sum_{k=1}^{K}\bigg[&
\frac{\operatorname{Re}\{\mu_{k,0}^{*}{\mathbf{r}}_{k,0}\}}
{\sigma_{\mathrm{rep},k,0}^{2}}
+w_{k,m_k^\star}^{\mathrm{br}}
\frac{\operatorname{Re}\{\mu_{k,m_k^\star}^{*}{\mathbf{r}}_{k,m_k^\star}\}}
{\sigma_{\mathrm{rep},k,m_k^\star}^{2}}\bigg]
\underset{\mathcal{H}_0}{\overset{\mathcal{H}_1}{\gtrless}}\lambda_{\mathrm{SMRC}}.
\end{aligned}
\label{eq:smrc_statistic}
\end{equation}
When no relay is available, the corresponding relay term is omitted.
The samples contain their receiver noise; replacing them by only the
source symbols would not describe the noisy statistic available at the
\gls{fc}. The optional $w_{k,m_k^\star}^{\mathrm{br}}$ represents relay
reliability and equals one for a channel-only selected \gls{mrc} variant.

Selection discards some diversity evidence and is sensitive to errors
in the branch-quality estimates. It also does not turn copies of one
sensor decision into independent sensing observations. Report age,
regeneration errors, and information complementarity may be incorporated
in more task-oriented selection rules, but such extensions require a
specified objective rather than an assumed universal \gls{snr} ordering.

\subsection{Max-Log With Linear Approximation}
\label{subsec:mlla_rule}

The cooperative LRT contains logarithms of sums of likelihood terms.
\gls{mlla} replaces each log-sum by its
dominant term and uses the clipped relay contribution introduced above.
It retains local sensing reliability while reducing online nonlinear
computation \cite{Lei2010MaxLog,AlJarrah2019DFCoopWSN}.
Define the per-sensor evidence variable
\begin{equation}
v_k=y_{k,0}+\sum\nolimits_{m=1}^{M}\widetilde\Omega_{k,m}.
\label{eq:theta_mlla}
\end{equation}
The approximate statistic is
\begin{equation}
\Lambda_{\mathrm{MLLA}}
=\sum\nolimits_{k=1}^{K}\psi_k^{\mathrm{MLLA}}(v_k)
\underset{\mathcal{H}_0}{\overset{\mathcal{H}_1}{\gtrless}}\lambda_{\mathrm{MLLA}},
\label{eq:mlla_llr}
\end{equation}
where $0<P_{\mathrm{fa},k}<P_{\mathrm d,k}<1$,
\begin{equation}
\psi_k^{\mathrm{MLLA}}(v)=
\begin{cases}
c_k^-,&v<v_k^-,\\
v+c_k^0,&v_k^-\leq v\leq v_k^+,\\
c_k^+,&v>v_k^+,
\end{cases}
\label{eq:mlla_piecewise}
\end{equation}
\begin{equation}
c_k^+=\ln\frac{P_{\mathrm d,k}}{P_{\mathrm{fa},k}},
\qquad
c_k^0=\ln\frac{P_{\mathrm d,k}}{1-P_{\mathrm{fa},k}},
\qquad
c_k^-=\ln\frac{1-P_{\mathrm d,k}}{1-P_{\mathrm{fa},k}},
\label{eq:mlla_constants_1}
\end{equation}
\begin{equation}
v_k^+=\ln\frac{1-P_{\mathrm{fa},k}}{P_{\mathrm{fa},k}},
\qquad
v_k^-=\ln\frac{1-P_{\mathrm d,k}}{P_{\mathrm d,k}}.
\label{eq:mlla_constants_2}
\end{equation}

These constants make the three segments continuous and can be computed
offline for a fixed operating point. The notation separates the scalar
evidence $v_k$ from learned model parameters $\boldsymbol\vartheta$,
and saturation levels $c_k^{\pm}$ from reporting-bit counts. A sensor
with $P_{\mathrm d,k}=P_{\mathrm{fa},k}$ supplies no discriminating
evidence. Reversed sensor polarity or deterministic operating points
require a corresponding reformulation or limiting treatment.

Score-based fading-aware fusion provides a different low-complexity
route based on derivatives of a likelihood and Fisher-information
quantities \cite{ciuonzo2025score}. Max-log dominance, clipping,
and a score test are not the same approximation. They share the goal of
reducing computation, but their validity regions, required nuisance
parameters, and threshold calibrations differ.

\subsection{Recent Channel-Aware, Multihop, and Time-Aware Extensions}
\label{subsec:recent_advanced_extensions}

Recent fusion techniques extend both the reporting representation and the receiver
architecture. A common binary plug-in likelihood structure is
\begin{equation}
\begin{aligned}
\Lambda_{\mathrm{CA}}
&=\ln\frac{
f(\boldsymbol r\mid \mathcal{H}_1;
\widehat{\boldsymbol\theta}_{\mathrm{stat}},\boldsymbol\Phi)}{
f(\boldsymbol r\mid \mathcal{H}_0;
\widehat{\boldsymbol\theta}_{\mathrm{stat}},\boldsymbol\Phi)}\underset{\mathcal{H}_0}{\overset{\mathcal{H}_1}{\gtrless}}\lambda_{\mathrm{CA}},
\end{aligned}
\label{eq:recent_channel_aware_llr}
\end{equation}
where $\widehat{\boldsymbol\theta}_{\mathrm{stat}}$ contains estimated
sensor, channel, and transition parameters, and $\boldsymbol\Phi$
represents the adjustable settings of the receiver front end (for example,
a reconfigurable holographic surface), which shape the received observation
$\boldsymbol r$. The function $f$ denotes a density
or probability mass according to the observation. Substituting estimated
parameters does not automatically preserve known-model optimality. A
generalized likelihood-ratio test instead specifies the relevant
hypothesis-dependent maximizations.

Multihop $M$-ary \gls{df} accounts for the adopted source
decision and reporting-path model, while simplified rules relax some
information requirements \cite{zhang2024mary}. Multibit multirelay
fusion likewise relates codeword transitions, instantaneous \gls{csi}, and
receiver complexity \cite{Zhang2025MultiBitDDF}. Here, ``$M$-ary''
follows the cited publication terminology and does not, by itself,
establish the local decision resolution. The hypothesis count, local
report-alphabet size, and relay count $M$ in the binary models above
are separate quantities. Likewise, a forwarding protocol must not be
replaced by a binary regeneration model without checking its assumptions.

Holographic fusion jointly designs a reconfigurable receive surface and
a digital detector. This changes the receiver architecture, not the
meaning of the hypotheses. A large effective aperture can be coupled
to relatively few \gls{rf} chains, with the surface and detector optimized
under channel-knowledge or complexity constraints
\cite{ciuonzo2025holographic}. Unknown-target-location detection
motivates related generalized-likelihood and deflection-based designs
\cite{Ciuonzo2026HolographicTarget}. A survey-level joint-design template
is
\begin{equation}
\begin{aligned}
\underset{\boldsymbol\Phi,\boldsymbol w,\lambda}
{\operatorname{minimize}}\quad&
P_{\mathrm e}^{(g)}(\boldsymbol\Phi,\boldsymbol w,\lambda)
\\
\operatorname{subject\ to}\quad&
\boldsymbol\Phi\in\mathcal S_{\mathrm{RHS}},
&\mathcal C_{\mathrm{FC}}(\boldsymbol\Phi,\boldsymbol w)
\leq C_{\max}.
\end{aligned}
\label{eq:holographic_joint_design}
\end{equation}
The feasible surface set $\mathcal S_{\mathrm{RHS}}$ and the hardware or
computational cost $\mathcal C_{\mathrm{FC}}$ must be specified for the
chosen architecture. This generic problem is not a claim that every
cited receiver minimizes the same error criterion.

Blind learning addresses unknown sensor operating points and reporting
crossover probabilities by estimating them from received decision
sequences, possibly using partial labels or prior information
\cite{Perez2026BlindFusion}. It changes how statistical parameters are
obtained, rather than replacing the need for an identifiable probability
model. Estimates require sufficient data and may need tracking in a
nonstationary environment.

Decision timing provides another source of evidence in distributed
sequential sensing \cite{Tabella2023TimeAware}. To illustrate the
statistical requirement, consider one accepted report per sensor and a
fixed hypothesis during the observation interval. Let $T_k^{\mathrm{rep}}$
be its reception time, $V_k$ its decoded decision, and $f_{k,j}(v,\tau)$
their joint mass or density under $\mathcal{H}_j$. Define
$\mathcal D(t)=\{k:T_k^{\mathrm{rep}}\leq t\}$ and the nonarrival
probability $S_{k,j}(t)=\Pr(T_k^{\mathrm{rep}}>t\mid \mathcal{H}_j)$.
Under conditional independence across sensors, the observation-history
LLR for this illustrative model is
\begin{equation}
\begin{aligned}
\Lambda_{\mathrm{seq}}(t)
={}&\sum\nolimits_{k\in\mathcal D(t)}
\ln\frac{f_{k,1}(V_k,T_k^{\mathrm{rep}})}
{f_{k,0}(V_k,T_k^{\mathrm{rep}})}
+\sum\nolimits_{k\notin\mathcal D(t)}
\ln\frac{S_{k,1}(t)}{S_{k,0}(t)}.
\end{aligned}
\label{eq:time_aware_statistic}
\end{equation}
Silence is informative when its probability differs between hypotheses.
An arrival-only sum omits the second term and is exact only when it
vanishes under the adopted model; otherwise it is an approximation.
Delays, erasures, and censoring must enter the reception-time model.
Repeated-transmission schemes require their own complete history
likelihood rather than an unqualified reuse of the one-report example.

These extensions act on different dimensions: report resolution,
forwarding topology, analog receiver design, parameter estimation, and
decision timing. They can be combined in one system, but their gains
should be assessed together with training, \gls{csi}, calibration, and
optimization costs rather than interpreted as a ranking of exclusive
fusion families.

\subsection{Ideal Relays Based Rule}
\label{subsec:ir_rule}

The \gls{ir} approximation retains imperfect local sensing but
assumes error-free regeneration,
$\epsilon_{k,m}^{\mathrm{rel}}=0$ for all relayed branches
\cite{AlJarrah2019DFCoopWSN}. The copies of one sensor decision can then
be coherently combined before evaluating that sensor's contribution to
the hypothesis likelihood. Assuming branch independence conditional on
${\mathbf{u}}_k$ and the channel state, define similar to \eqref{eq:ils_branch_statistic}. The corresponding statistic is
\begin{equation}
\begin{aligned}
\Lambda_{\mathrm{IR}}
&=\sum\nolimits_{k=1}^{K}
\ln\!\left(
\frac{P_{\mathrm d,k}+(1-P_{\mathrm d,k})e^{-Y_k}}
{P_{\mathrm{fa},k}+(1-P_{\mathrm{fa},k})e^{-Y_k}}
\right)
\underset{\mathcal{H}_0}{\overset{\mathcal{H}_1}{\gtrless}}\lambda_{\mathrm{IR}}.
\end{aligned}
\label{eq:ir_llr}
\end{equation}
Branch independence conditional on a shared decision is not the same
as treating every copy as independent evidence about the hypothesis.
The local operating probabilities appear once per sensor in
\eqref{eq:ir_llr}.

For informative sensors with
$0<P_{\mathrm{fa},k}<P_{\mathrm d,k}<1$, define
\begin{equation}
\begin{aligned}
a_k^{\mathrm{IR}}&=P_{\mathrm d,k}-P_{\mathrm{fa},k},
\\
c_k^+&=\ln\frac{P_{\mathrm d,k}}{P_{\mathrm{fa},k}},
\qquad
c_k^-=\ln\frac{1-P_{\mathrm d,k}}{1-P_{\mathrm{fa},k}}.
\end{aligned}
\label{eq:ir_parameters}
\end{equation}
A clipped tangent approximation to the sensor contribution is
\begin{equation}
\psi_k^{\mathrm{IR}}(Y)=
\begin{cases}
c_k^-,&Y<c_k^-/a_k^{\mathrm{IR}},\\
a_k^{\mathrm{IR}}Y,
&c_k^-/a_k^{\mathrm{IR}}\leq Y\leq c_k^+/a_k^{\mathrm{IR}},\\
c_k^+,&Y>c_k^+/a_k^{\mathrm{IR}},
\end{cases}
\label{eq:ir_linear}
\end{equation}
which gives
\begin{equation}
\Lambda_{\mathrm{IR}}^{\mathrm{lin}}
=\sum\nolimits_{k=1}^{K}\psi_k^{\mathrm{IR}}(Y_k)
\underset{\mathcal{H}_0}{\overset{\mathcal{H}_1}{\gtrless}}\lambda_{\mathrm{IR,lin}}.
\label{eq:ir_linear_llr}
\end{equation}
The slope at the origin is $a_k^{\mathrm{IR}}$ and the saturation values
match the limiting exact contribution. This is an approximation, not an
algebraic identity. Uninformative sensors contribute zero and should be
handled without division by $a_k^{\mathrm{IR}}=0$.

\gls{ir} simplifies relay-parameter acquisition, but can overestimate the
reliability of weak first-hop paths. Explicit multihop transition models
and blind estimation of unknown reporting parameters provide alternatives
\cite{zhang2024mary,Zhang2025MultiBitDDF,Perez2026BlindFusion}.
\gls{ils} and \gls{ir} relax different parts of the sensing-to-decision model;
neither assumption uniformly contains the other.
\subsection{Fast Soft-Decision Fusion}
\label{subsec:fast_soft_decision}

Soft reporting conveys more than one binary decision per sensor, at the
cost of reporting bandwidth, energy, dynamic range, and processing.
The gain depends on how that information is transported through the
reporting channel \cite{ma2008soft,Chaudhari2012HardSoft}.
Fast soft-decision fusion techniques seek lightweight online decision boundaries
using statistical information obtained beforehand
\cite{golvaei2021}. A representative real soft-feature vector is
\begin{equation}
\boldsymbol s_n=[s_{1,n},\ldots,s_{K,n}]^{\mathsf T}
\in\mathbb R^K,
\label{eq:fast_soft_vector}
\end{equation}
where the entries denote the calibrated soft features available at the
FC for observation index $n$. A linear statistic and its decision are
\begin{equation}
\Lambda_{\mathrm{soft}}(n)=\boldsymbol w^{\mathsf T}\boldsymbol s_n,
\label{eq:fast_soft_stat}
\end{equation}
\begin{equation}
\widehat{\mathcal{H}_n}=
\begin{cases}
\mathcal{H}_1,&\Lambda_{\mathrm{soft}}(n)\geq\lambda_{\mathrm{soft}},\\
\mathcal{H}_0,&\Lambda_{\mathrm{soft}}(n)<\lambda_{\mathrm{soft}}.
\end{cases}
\label{eq:fast_soft_decision}
\end{equation}
For a regularized Fisher-type example, let
$\widehat{\boldsymbol\Sigma}_{\mathrm W}$ be the pooled within-class
covariance and $\widehat{\boldsymbol\mu}_j$ the estimated class means.
Then a possible weight vector is
\begin{equation}
\boldsymbol w=
(\widehat{\boldsymbol\Sigma}_{\mathrm W}+\delta_{\mathrm W}\boldsymbol I_K)^{-1}
(\widehat{\boldsymbol\mu}_1-\widehat{\boldsymbol\mu}_0),
\qquad \delta_{\mathrm W}>0.
\label{eq:fast_soft_weight}
\end{equation}
This explicit covariance choice is a survey-level linear-discrimination
example, not a claim that all cited fast-fusion algorithms use the same
estimator. A null-covariance deflection design is a distinct alternative.
Regularization or a diagonal model is useful when training data are
insufficient for a reliable full covariance inverse.

For a nonzero weight vector and an approximately Gaussian null
statistic with positive variance, a moment-based threshold targeting
$\alpha_{\mathrm{FA}}\in(0,1)$ is
\begin{equation}
\lambda_{\mathrm{CFAR}}
=\boldsymbol w^{\mathsf T}\widehat{\boldsymbol\mu}_0
+Q^{-1}(\alpha_{\mathrm{FA}})
\sqrt{\boldsymbol w^{\mathsf T}
\widehat{\boldsymbol\Sigma}_0\boldsymbol w},
\label{eq:fast_soft_cfar}
\end{equation}
where $Q$ is the standard Gaussian upper-tail function and
$\widehat{\boldsymbol\Sigma}_0$ is estimated under $\mathcal{H}_0$. This gives
exact Gaussian-null calibration only with the correct known moments;
estimated moments provide an approximate threshold that must be tested
on independent null observations. A possible tracking recursion is
\begin{equation}
\begin{aligned}
\lambda_{\mathrm{soft}}(t+1)
={}&\lambda_{\mathrm{soft}}(t)
+\eta_{\mathrm{FA}}
[\widehat P_{\mathrm{fa}}^{(g)}(t)-\alpha_{\mathrm{FA}}].
\end{aligned}
\label{eq:fast_soft_adaptive}
\end{equation}
The step size $\eta_{\mathrm{FA}}>0$ controls adaptation. Its error
estimate requires verified null data or a justified null model; the
frequency of positive decisions on arbitrary observations is not a
false-alarm estimate. Stability and finite-sample false-alarm control do
not follow from this recursion alone. These linear designs are
lightweight, but their calibration depends on reliable null statistics.

\subsection{Multibit, Hybrid, and Semantic Reporting}
\label{subsec:multibit_hybrid_reporting}

Soft fusion assumes that real-valued features are available at the FC;
in practice, they must be conveyed over a bandwidth-limited reporting
channel. Multibit reporting offers intermediate resolution between hard
and continuous-valued soft reporting. Recent work considers quantized raw
observations or local likelihoods together with error-prone reporting
channels \cite{Mao2024MultiBitSparse}. A sensor quantizer is represented by
\begin{equation}
q_k^{\mathrm q}=\mathcal Q_k(s_k;b_k^{\mathrm q})
\in\{0,1,\ldots,2^{b_k^{\mathrm q}}-1\},
\qquad b_k^{\mathrm q}\in\mathbb Z_{\geq1}.
\label{eq:multibit_quantizer}
\end{equation}
The bit count $b_k^{\mathrm q}$ specifies the local decision alphabet,
not the number of hypotheses. Increasing it can reduce quantization
loss, but the resulting channel use and transmission errors must also
be considered. Multiple-bit fusion thus extends the representation
without automatically dominating one-bit reporting under every common
resource budget \cite{Aziz2025MultiBitFusion}.

Hybrid-resolution detection allows some sensors to send quantized
values and others to use a full-precision observation model, with
sensor-specific thresholds and bandwidth allocation
\cite{Mao2026HybridBandwidth}. A generic bandwidth constraint is
\begin{equation}
\sum\nolimits_{k=1}^{K}W_k(b_k^{\mathrm q},v_k^{\mathrm{mode}})
\leq W_{\max},
\qquad v_k^{\mathrm{mode}}\in\{0,1\},
\label{eq:hybrid_bandwidth_constraint}
\end{equation}
where mode zero denotes quantized reporting and mode one the specified
soft-reporting format. The function $W_k$ is bandwidth in hertz, not
stored battery energy. A full-precision statistical model still needs an
explicit physical reporting model and does not imply cost-free transport.

Semantic communication and over-the-air computation instead train
compact task-relevant representations and aggregate them over the
multiple-access channel \cite{Yi2025SemanticAirCompCSS}. This is
feature- or representation-level fusion when the transmitted variables
are not local hypothesis decisions. Its classification is therefore
separate from both a hard-decision \gls{cv} test and conventional \gls{egc}.
Detection thresholds should be calibrated under the actual sensing and
reporting conditions; low reconstruction or aggregation error alone
does not establish the desired global false-alarm rate.

Fast linear, multibit, hybrid, and semantic approaches make different
choices about representation and receiver processing. Their comparison
should include detection performance, reporting resources, latency,
training and calibration cost, and channel mismatch. These dimensions
complement the technique families in Section~\ref{sec:classification_fusion}
rather than creating an exclusive hierarchy.

Table~\ref{tab:advanced_fusion_checklist} compares the features of
representative studies reviewed in this section. The checklist highlights
how classical combining and likelihood-based techniques relate to recent
multibit, learned, time-aware, and hardware-aware fusion approaches.
The entries describe the scope discussed in this section rather than
an overall ranking of detection performance. Table \ref{tab:adv_summary} summarizes the core operation, conditions, and limitations of each of the advanced fusion techniques.

\begin{table*}[!t]
\caption{Summary of Advanced Fusion Methods and Complementary Design Mechanisms.}
\label{tab:adv_summary}
\centering
\begingroup
\scriptsize
\setlength{\tabcolsep}{3pt}
\renewcommand{\arraystretch}{1.05}
\begin{tabularx}{\textwidth}{@{}
c
>{\raggedright\arraybackslash\hsize=0.65\hsize}X
>{\raggedright\arraybackslash\hsize=1.15\hsize}X
>{\raggedright\arraybackslash\hsize=1.20\hsize}X@{}}
\toprule
\textbf{Ref.} & \textbf{Method/design role} &
\textbf{Core operation} & \textbf{Conditions and limitations}\\
\midrule
\cite{AlJarrah2019DFCoopWSN}
& \gls{ils}
& Assumes perfect local sensing and models uncertainty in the reporting
and regeneration stages. Clipping provides a further numerical approximation.
& A statistical-model simplification, not a guarantee that actual sensors
are reliable. Symmetric relay transitions and conditional independence
are required for the scalar branch expressions used here.\\
\specialrule{0.35pt}{0.8pt}{0.8pt}
\cite{Wang2007CooperativeDemodulation,Ciuonzo2013MRC}
& \gls{cmrc}
& Weights received relay samples using preceding-hop reliability as well
as final-hop channel information.
& A combining mechanism. Relay-aware weighting and MIMO decision-fusion
architectures are related but not identical; information acquisition and
noise normalization must be included in the comparison.\\
\specialrule{0.35pt}{0.8pt}{0.8pt}
\cite{Wang2008LAR,Vien2009SmartRelaying}
& \gls{lar}
& Adapts relay power using instantaneous first-hop quality and instantaneous
or average second-hop \gls{snr}.
& A reporting-control mechanism used with an \gls{fc} combiner. The nominal and
scaled powers must be distinguished, and the effective gain must be
available to the receiver.\\
\specialrule{0.35pt}{0.8pt}{0.8pt}
\cite{Nasri2012EMRC}
& Enhanced \gls{mrc} (EMRC)
& Coordinates adjustable relay-scaling and receiver-combining coefficients
using information from both hops.
& Parameter tuning and power feasibility are part of the design.
The survey's illustrative coefficient parameterization is not a universal
optimized EMRC law.\\
\specialrule{0.35pt}{0.8pt}{0.8pt}
\cite{Vien2012SmartEGC,Ikki2008BestRelaySelection,Yi2008RelaySelectionDiversity}
& \gls{cegc} and \gls{smrc}
& \gls{cegc} phase-aligns samples with equal combining gains; \gls{smrc} combines a
selected relay path with the direct observation.
& \Gls{egc} aligns signal phases without re-multiplying received amplitudes, which are embedded in the channel samples. Selection requires a consistent reliability metric, the \gls{df} bottleneck expression
in this survey is a declared surrogate.\\
\specialrule{0.35pt}{0.8pt}{0.8pt}
\cite{Lei2010MaxLog,ciuonzo2025score}
& Max-log / \gls{mlla} and score-based tests
& Uses dominant-term likelihood approximations or score-based statistics
to reduce nonlinear processing.
& These are distinct statistical approximations, not interchangeable
first-order rules. Their domains, nuisance information, and threshold
calibration differ.\\
\specialrule{0.35pt}{0.8pt}{0.8pt}
\cite{AlJarrah2019DFCoopWSN,Perez2026BlindFusion}
& Ideal relays and blind parameter estimation
& \gls{ir} neglects regeneration errors; blind learning estimates unknown
sensor/reporting probabilities for a statistical detector.
& An ideality assumption and a parameter-estimation strategy are different
design choices. \gls{ir} can be optimistic; blind estimation requires sufficient
and identifiable data.\\
\specialrule{0.35pt}{0.8pt}{0.8pt}
\cite{Tabella2023TimeAware,zhang2024mary,Zhang2025MultiBitDDF,ciuonzo2025holographic,Ciuonzo2026HolographicTarget}
& Timing, multihop, and holographic extensions
& Incorporates report timing, nonbinary forwarding transitions, or a
configurable analog receiver.
& These concern separate timing, representation, and hardware dimensions.
A history likelihood may need silence information, and repeated copies
must not be treated as independent sensing evidence.\\
\specialrule{0.35pt}{0.8pt}{0.8pt}
\cite{golvaei2021,Mao2024MultiBitSparse,Mao2026HybridBandwidth,Aziz2025MultiBitFusion,Yi2025SemanticAirCompCSS}
& Fast soft, multibit, hybrid, and semantic fusion
& Combines richer sensor representations using linear, quantized,
locally most powerful, or task-oriented receiver processing.
& Reporting resolution is separate from the test family. Calibration,
channel distortion, bandwidth, learning cost, 
semantic features may not be local hypothesis decisions.\\
\bottomrule
\end{tabularx}
\endgroup
\end{table*}

\begin{table*}[!t]
\caption{Feature-Coverage Checklist of Representative Studies on Advanced
Decision Fusion.}
\label{tab:advanced_fusion_checklist}
\centering

\begingroup
\footnotesize
\setlength{\tabcolsep}{2.5pt}
\renewcommand{\arraystretch}{1.18}
\renewcommand{\tabularxcolumn}[1]{m{#1}}

\providecommand{\dfyes}{\ensuremath{\checkmark}}
\providecommand{\dfnr}{\textemdash}
\begin{tabularx}{\textwidth}{
@{}
c
c
>{\raggedright\arraybackslash}X
*{8}{>{\centering\arraybackslash}m{1.08cm}}
@{}
}
\toprule

\textbf{Ref.}
& \textbf{Year}
& \textbf{Reviewed technique}
& \textbf{CA}
& \textbf{RM}
& \textbf{RA}
& \textbf{LC}
& \textbf{MQ}
& \textbf{LE}
& \textbf{TA}
& \textbf{HD}
\\

\midrule

\cite{Wang2007CooperativeDemodulation}
& 2007
& \gls{cmrc}
& \dfyes & \dfyes & \dfnr & \dfyes
& \dfnr & \dfnr & \dfnr & \dfnr
\\
\specialrule{0.25pt}{1.4pt}{1.4pt}

\cite{Wang2008LAR}
& 2008
& Link-adaptive regeneration (LAR)
& \dfyes & \dfyes & \dfyes & \dfyes
& \dfnr & \dfnr & \dfnr & \dfnr
\\
\specialrule{0.25pt}{1.4pt}{1.4pt}

\cite{Ikki2008BestRelaySelection}
& 2008
& Best-relay selection / \gls{smrc}
& \dfyes & \dfyes & \dfyes & \dfyes
& \dfnr & \dfnr & \dfnr & \dfnr
\\
\specialrule{0.25pt}{1.4pt}{1.4pt}

\cite{Lei2010MaxLog}
& 2010
& Max-log \gls{df}
& \dfyes & \dfnr & \dfnr & \dfyes
& \dfnr & \dfnr & \dfnr & \dfnr
\\
\specialrule{0.25pt}{1.4pt}{1.4pt}

\cite{Nasri2012EMRC}
& 2012
& Enhanced \gls{mrc} (EMRC)
& \dfyes & \dfyes & \dfyes & \dfnr
& \dfnr & \dfnr & \dfnr & \dfnr
\\
\specialrule{0.25pt}{1.4pt}{1.4pt}

\cite{Vien2012SmartEGC}
& 2012
& Smart relaying with cooperative \gls{egc}
& \dfyes & \dfyes & \dfyes & \dfyes
& \dfnr & \dfnr & \dfnr & \dfnr
\\
\specialrule{0.25pt}{1.4pt}{1.4pt}

\cite{AlJarrah2019DFCoopWSN}
& 2019
& \gls{ils}, \gls{mlla}, and ideal-relay approximations
& \dfyes & \dfyes & \dfnr & \dfyes
& \dfnr & \dfnr & \dfnr & \dfnr
\\
\specialrule{0.25pt}{1.4pt}{1.4pt}

\cite{golvaei2021}
& 2021
& Fast soft-\gls{df}
& \dfnr & \dfnr & \dfnr & \dfyes
& \dfnr & \dfyes & \dfnr & \dfnr
\\
\specialrule{0.25pt}{1.4pt}{1.4pt}

\cite{Tabella2023TimeAware}
& 2023
& Time-aware sequential \gls{df}
& \dfnr & \dfnr & \dfnr & \dfnr
& \dfnr & \dfnr & \dfyes & \dfnr
\\
\specialrule{0.25pt}{1.4pt}{1.4pt}

\cite{Mao2024MultiBitSparse}
& 2024
& Multibit detection over error-prone reporting links
& \dfyes & \dfnr & \dfnr & \dfnr
& \dfyes & \dfnr & \dfnr & \dfnr
\\
\specialrule{0.25pt}{1.4pt}{1.4pt}

\cite{zhang2024mary}
& 2024
& $M$-ary multihop \gls{df}
& \dfyes & \dfyes & \dfnr & \dfyes
& \dfnr & \dfnr & \dfnr & \dfnr
\\
\specialrule{0.25pt}{1.4pt}{1.4pt}

\cite{Zhang2025MultiBitDDF}
& 2025
& Multibit multihop / multirelay fusion
& \dfyes & \dfyes & \dfnr & \dfyes
& \dfyes & \dfnr & \dfnr & \dfnr
\\
\specialrule{0.25pt}{1.4pt}{1.4pt}

\cite{ciuonzo2025holographic}
& 2025
& Channel-aware holographic fusion
& \dfyes & \dfnr & \dfnr & \dfyes
& \dfnr & \dfnr & \dfnr & \dfyes
\\
\specialrule{0.25pt}{1.4pt}{1.4pt}

\cite{ciuonzo2025score}
& 2025
& Score-based fading-aware fusion
& \dfyes & \dfnr & \dfnr & \dfyes
& \dfnr & \dfnr & \dfnr & \dfnr
\\
\specialrule{0.25pt}{1.4pt}{1.4pt}

\cite{Aziz2025MultiBitFusion}
& 2025
& Multiple-bit hard/soft fusion
& \dfnr & \dfnr & \dfnr & \dfyes
& \dfyes & \dfnr & \dfnr & \dfnr
\\
\specialrule{0.25pt}{1.4pt}{1.4pt}

\cite{Yi2025SemanticAirCompCSS}
& 2025
& Semantic cooperative sensing, over-the-air aggregation
& \dfyes & \dfnr & \dfnr & \dfnr
& \dfnr & \dfyes & \dfnr & \dfnr
\\
\specialrule{0.25pt}{1.4pt}{1.4pt}

\cite{Perez2026BlindFusion}
& 2026
& Blind learning of sensor and reporting parameters
& \dfyes & \dfnr & \dfnr & \dfnr
& \dfnr & \dfyes & \dfnr & \dfnr
\\
\specialrule{0.25pt}{1.4pt}{1.4pt}

\cite{Mao2026HybridBandwidth}
& 2026
& Hybrid quantized/full-precision, bandwidth allocation
& \dfyes & \dfnr & \dfnr & \dfnr
& \dfyes & \dfnr & \dfnr & \dfnr
\\
\specialrule{0.25pt}{1.4pt}{1.4pt}

\cite{Ciuonzo2026HolographicTarget}
& 2026
& Holographic detection of an unknown-location target
& \dfyes & \dfnr & \dfnr & \dfnr
& \dfnr & \dfnr & \dfnr & \dfyes
\\

\bottomrule
\end{tabularx}

\vspace{3pt}

\begin{minipage}{\textwidth}
\scriptsize
\raggedright

\textit{Legend:}
\dfyes\ indicates a feature explicitly discussed for the listed
technique in this section.
\dfnr\ means that the feature is not established in that discussion;
it does not prove that the full article excludes the feature.
\par
\smallskip

\textit{Criteria:}
\textbf{CA}: reporting-channel properties or impairments are considered;
\textbf{RM}: relay or multihop reporting is considered;
\textbf{RA}: relay-power adaptation or relay/branch selection is used;
\textbf{LC}: a simplified \gls{dfr} or reduced-complexity alternative
is explicitly discussed;
\textbf{MQ}: multibit or mixed-resolution local decision is established;
an $M$-ary hypothesis task alone does not establish this feature;
\textbf{LE}: parameters, a decision boundary, or transmitted
representations are learned or estimated from data (not necessarily
using a neural model);
\textbf{TA}: sequential decision timing is used as part of fusion;
\textbf{HD}: a reconfigurable/holographic receiver is jointly designed
with the detector.
\end{minipage}
\endgroup
\end{table*}
\section{Performance Metrics for Decision Fusion}
\label{sec-PerfMetrics}
 The detection performance can be measured using several performance metrics such as detection probability, false alarm probability, \gls{roc} curve, \gls{auroc} curve, probability of detection error, energy consumption, etc. Most of these parameters are interlinked but still provide an important insight about the performance of the \gls{df} algorithm. In this section, we analyze the basic performance metrics for the \gls{df} algorithms along with their merits and demerits. 
\subsection{Probability of Detection and Probability of False Alarm}
The probability of detection, denoted as $P_d$, and probability of false alarm, denoted as $P_{fa}$, are widely used metrics to analyze the performance of \gls{df} algorithms in most applications. Both are system level metrics that are difficult to derive in closed-form for the optimal \gls{dfr}. Therefore, most of the literature considers performing simulation experiments, which might lead to rapid research and development at the expense of reduced accuracy. Moreover, $P_d$ represents the correct prediction of the decision and for example, it represents the correct spectrum occupancy prediction in spectrum sensing applications. Mathematically, $P_d$ for a binary hypothesis can be expressed as
\begin{equation}
P_d= \Pr \left(  \Lambda > \tau_0|\mathcal{H}_1\right)
\end{equation}
where $\Lambda$ is the \gls{llr} of the \gls{dfr}, ${\tau}_{0}$ is the system-level threshold that is chosen to satisfy a certain $P_{fa}$.  

Furthermore, $P_{fa}$ is defined as the incorrect prediction of the decision. Mathematically, $P_{fa}$ for a binary hypothesis can be expressed as
\begin{equation}
P_{fa}= \Pr \left(  \Lambda < \tau_0|\mathcal{H}_1\right)
\end{equation}
The computation of an analytical expression for $P_{fa}$ is difficult for all types of classifiers and is generally computed numerically. In the literature, it is computed with an assumption of identical sensors as it provides an asymptotically optimum solution. For numerical computation, this assumption leads to less or no performance loss.

Using the counting rule and considering $N$ sensors, $P_{fa}$ can be written as follows \cite{varshney2012distributed},
\begin{equation}
P_{fa}=\sum\nolimits_{k=k*}^N {N \choose k} (P_k^{fa})^k (1-P_k^{fa})^{N-k}
\end{equation}
where  $P_k^{fa}$ is the $k$th sensor probability of false alarm. If all sensors are identical, then the system level and sensor level probability of detection and probability of false alarm are identical, i.e., $P_{d}=P_k^{d}$ and $P_{fa}=P_k^{fa}$ \cite{chen2004}. The statistics of the \glspl{dfr} are then the sum of \gls{iid} random variable. In \cite{dey2020}, closed form expression for probability of false alarm and probability of detection for classifiers for a wide band collaborative spectrum sensing problem is discussed. The \gls{df} center is equipped with a large antenna array.
\begin{table}
\caption{Commonly used Performance Metrics.}
\label{tab:common_metrics}
\resizebox{0.48\textwidth}{!}{%
\begin{tabular}{|M{1.3cm}|M{0.6cm}|M{4.5cm}|M{1.2cm}|}
     \hline 
 \textbf{Metric}& \textbf{Ref.} & \textbf{Expression} & \textbf{Analysis}\\
 \hline\hline
    $P_{fa}$ &\cite{Acharya2014NeymanPearson} & $P_{fa}= \Pr \left(  \Lambda < \tau_0|\mathcal{H}_1\right)$ & Intractable\\
     \hline
     $P_d$&\cite{Acharya2014NeymanPearson} & $P_d= \Pr \left(  \Lambda > \tau_0|\mathcal{H}_1\right)$ & Intractable\\
     \hline
    \Gls{roc} curve&\cite{Acharya2014NeymanPearson} & $P_{f\!a}\!\!=\!\!\sum\limits_{k=k*}^N {N \choose k} (P_k^{f\!a})^k (1-P_k^{f\!a})^{N-k}$ & Intractable\\
     \hline
    Deflection &\cite{Acharya2014NeymanPearson} & $ D(\mathbf{w})=\frac{\left(\mathbb{E}({y_c}|\mathcal{H}_1)-\mathbb{E}(y_c|\mathcal{H}_0)\right)^2}{\mathrm{Var}(y_c|\mathcal{H}_0)}$ & Intractable\\
     \hline
\end{tabular}}\label{tab:false}
\end{table}

\subsection{Deflection Coefficient Measurement}
 Deflection coefficient measurement is widely used in the literature to characterize energy detectors in \gls{ss}, where higher deflection coefficient mean better error performance. It is also called the output \gls{snr} \cite{picinbono1995}. To define the deflection coefficient measurement mathematically, consider the following. Assume that the decisions from the sensors are transmitted to the \gls{fc} through a control channel in an orthogonal manner. The control channel is corrupted with normally distributed \gls{awgn} only. Therefore, the linear combination of the decisions at the \gls{fc} can be expressed as follows \cite[Eq. (13)]{quan2008},
\begin{equation}
y_c=\sum\nolimits_{i=1}^Mw_iy_i=\mathbf{w}^T \mathbf{y}
\end{equation}
where $\mathbf{y}=[y_1,y_2,\ldots,y_M]^T\in\mathbb{R}^{M\times 1}$ is the vector of received decisions from the sensors, and $\mathbf{w}=[w_1,w_2,\ldots,w_M]^T\in\mathbb{R}^{M\times 1}$ is the weight vector that determines the contribution of each sensor to the global decision, and $w_i\geq0$. Consequently, the deflection measurement can be expressed as 
\begin{equation}
D(\mathbf{w})=\frac{\left(\mathbb{E}[y_c|\mathcal{H}_1]-\mathbb{E}[y_c|\mathcal{H}_0]\right)^2}{\mathrm{Var}(y_c|\mathcal{H}_0)}\label{eq-defcoef}
\end{equation}
where $\mathbb{E}[\cdot]$ is the expectation and $\mathrm{Var}(\cdot)$ is the variance. Therefore, the computation of the deflection depends on the first and second order statistics, unlike other closed-form performance metrics which need the distribution of the fusion statistic \cite{yan2015decision}.  

In addition, deflection coefficient maximization has been studied in literature \cite{jayakrishan2008}. When Gaussian assumptions are used, maximizing the deflection point exhibits the same behavior as of a likelihood ratio receiver. The deflection measurement for an optimal detector and the counting rule is computed in \cite{jayakrishan2008}. 

A modified deflection coefficient was proposed in \cite{quan2008} to measure the effect of \gls{pdf} on the detection performance. It is defined as follows \cite[Eq. (57)]{quan2008},
\begin{equation}
\widetilde{D}(\mathbf{w})=\frac{\left(\mathbb{E}[\mathbf{w}|\mathcal{H}_1]-\mathbb{E}[\mathbf{w}|\mathcal{H}_0]\right)^2}{\mathrm{Var}(\mathbf{w}|\mathcal{H}_1)}.
\end{equation}
Furthermore, the modified deflection coefficient measurement maximization problem under a unit-norm $\mathbf{w}$ constraint is formulated in \cite{quan2008} as 
\begin{equation}
 \max_{w_i} {\widetilde{D}(\mathbf{w})}, \quad
\textrm{s.t.: }  ||\mathbf{w}||_2^2=1.
\end{equation}

\subsection{ROC Measurement}
The \gls{roc} curve shows all feasible $(P_{fa},P_d)$ tuples, which represent a tradeoff between the probability of detection and the probability of false alarm for every allowable threshold of the decision classifier. In some studies, complementary \gls{roc} is also studied to show the performance evaluation of the \gls{df} technique. In detection theory, the \gls{roc} curve captures the effect of signal detection in a single plot and can provide a measure to compare the different classifiers. For the case of \gls{ss}, studies have plotted \gls{roc} curve for varying \gls{snr}, time-bandwidth product, multipath fading product, shadowing, etc.
\begin{figure}[t]
    \centering
\includegraphics[width = 0.75\columnwidth]{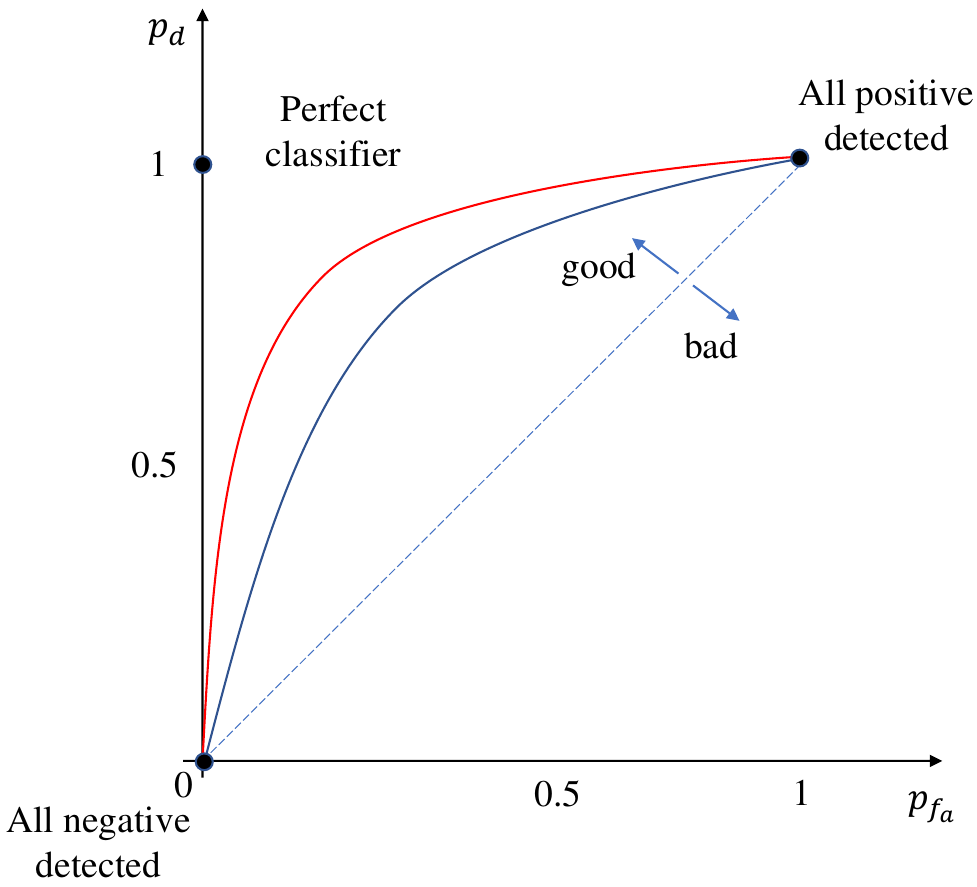}    \caption{\Gls{roc} Curve.}
    \label{fig:roc}
    \vspace{0pt}
\end{figure}
Fig. \ref{fig:roc} shows an example of an \gls{roc} plot. A perfect classifier lies on the coordinate $(0,1)$, whereas the coordinate $(1,1)$ represents a classifier deciding on $\mathcal{H}_0$ for all events, the coordinate $(0,0)$ represents a classifier deciding on $\mathcal{H}_1$ for all events, and finally $(1,0)$ represents the worst classifier that is opposite to the perfect classifier. The dashed blue line represents a reference for a random classifier. As the performance of the classifier moves towards the coordinate $(0,1)$, it becomes better, and as it moves towards $(1,0)$ it gets worse. 

Mostly, \gls{roc} is calculated using a brute-force technique, which is a complex task. It becomes more difficult when comparing multiple classifiers. In the following, the steps to compute the \gls{roc} curve are summarized:
\begin{enumerate}
    \item Simulate or experimentally generate two ensembles of system realizations.
    \item For each realization, calculate the corresponding decision function and compare its value to the threshold $\tau_0$.
    \item Compute the probability of detection and probability of false alarm using the same threshold as in step 2.
    \item Repeat steps 2 and 3 for different threshold values.
\end{enumerate}

In \cite{vladeanu2022}, \gls{roc} for energy detection based spectrum sensing with multiple time slots is studied. For \gls{crn}, the energy detector's performance using \gls{roc} is studied in \cite{gahane2018, chin2016}. Table \ref{tab:common_metrics} compares between the commonly used performance metrics used in \gls{df}.

\subsection{AUROC curve}
The \gls{auroc} is used when it is difficult to visually compare multiple \glspl{dfr}. It is defined as the area covered by the \gls{roc} curve \cite{atapattu2010,attapattu2011}. It is mathematically expressed as \cite[Eq. (4)]{atapattu2010},
\begin{equation}
A(\gamma)=\int_0^1 P_d\left(\gamma, \tau_0 \right)dP_{fa}\left(\tau_0\right) 
\end{equation}
where the value of $A(\gamma)$ varies from $\frac{1}{2}$ to $1$, with $\frac{1}{2}$ being the worst classifier, and $1$ being the perfect classifier. In addition, closed-form expressions of $A(\gamma)$ are derived for the energy detector considering various wireless communication applications. For instance, the authors of \cite{atapattu2010} derived the \gls{auroc} of \gls{mrc} considering perfect channel estimation, channel estimation errors, and channel fading correlations. 

\subsection{Minimum Probability of Error}
In some decision fusion applications, satisfying the global probability of false alarm while maximizing the global probability of detection is not sufficient. Therefore, the target is to minimize the global probability of fusion error, which is defined as \cite[Eq. (8)]{Arafat2020-SJ},
\begin{equation}
    P_e = P_{fa}\Pr(\mathcal{H}_0) + (1 - P_d)\Pr(\mathcal{H}_1) \label{eq-PrFusion-error}
\end{equation}
Consequently, the objective is to jointly find the local and global \gls{df} thresholds, $\tau_k$ and $\tau_0$, that minimize \eqref{eq-PrFusion-error}. To solve this problem, the \gls{csi} for each link, local probabilities of detection, and false alarm are needed. Such an optimization problem is multi-dimensional with an intractable solution. Therefore, sub-optimal \gls{df} approaches are considered in the literature. The trivial solution is to have a zero threshold at the local and global \gls{df}. Such an approach is known as distributed maximum likelihood, which minimizes the local probability of errors only in the case of equally likely events. However, in dynamic scenarios such as spectrum occupancy, it is not optimal. Alternatively, a two-stage approach is adopted in \cite{Arafat2020-SJ}. In the first stage, the received local sensing decision is decoded using a \gls{map} rule to minimize the local error probability. In the second stage, the decoded local decisions are fused using the \gls{cv} rule. This approach is known as the decode-then-fuse approach. Other approaches are discussed in Sec. \ref{sec-basicsOfDF}-C.

\subsection{Kullback-Leibler Divergence}
Kullback-Leibler divergence measures the distance between two probability distributions, and it is defined as \cite[Eq. (6)]{liu2019}
\begin{equation}
    \mathrm{KL}(P||Q) = \sum\nolimits_{x\in{X}}P(x)\log\frac{P(x)}{Q(x)}
\end{equation}
where $P(x)$ and $Q(x)$ are \glspl{pmf} of the null hypothesis observations and the alternative hypothesis observations. In addition, $\mathrm{KL}(P||Q)=0$ indicates identical distributions, while higher $\mathrm{KL}(P||Q)$ indicates higher distinction between the \glspl{pmf} of the hypotheses observations. The optimal detector at the local sensors has been found in \cite{Berber2026-LSP} to maximize a variant of the KL divergence
between the distributions of quantized messages sent to the \gls{fc}.

\subsection{\texorpdfstring{$F_1$}{F1} Score}

In statistical analysis, the $F_1$ score is used to evaluate binary hypothesis testing performance. It is defined as the harmonic mean of precision and recall such that
\begin{equation}
F_1 =
2\frac{\mathrm{precision}\times\mathrm{recall}}
{\mathrm{precision}+\mathrm{recall}}\label{eq-f1}
\end{equation}
where precision and recall are defined as
\begin{align}
\mathrm{precision}
&=\frac{P_d\Pr(\mathcal{H}_1)}
{P_d\Pr(\mathcal{H}_1)+P_{\mathrm{fa}}\Pr(\mathcal{H}_0)}\label{eq-prec}
\\
\mathrm{recall}
&=
P_d.
\label{eq-recall}
\end{align}
By substituting \eqref{eq-prec} and \eqref{eq-recall} in \eqref{eq-f1}, we can write
\begin{equation}
   F_1 =\frac{2P_d\Pr(\mathcal{H}_1)}
{(P_d+1)\Pr(\mathcal{H}_1)+P_{\mathrm{fa}}\Pr(\mathcal{H}_0)}.
\end{equation}
It is important to note that the harmonic mean is used because precision and recall are equally important, and the $F_1$ score is high only when both metrics are high. In addition, the $F_1$ score satisfies $0\leq F_1\leq 1$, where 1 corresponds to perfect precision and recall, and $0$ indicates poor precision and recall performance. Therefore, larger values of $F_1$ are desired.

An important advantage of the $F_1$ score is that it combines the $P_d$ and $P_{fa}$ performances into a single scalar metric. In addition, it strongly penalizes a detector that performs poorly in either precision or recall. Although the conventional $F_1$ score assigns equal weights to precision and recall, a weighted metric such as the $F_\beta$ score can assign different weights to precision and recall.

\subsection{Confusion Matrix}
So far, what has been discussed is binary hypothesis testing. However, when the problem has M-ary hypotheses, $P_d$ and $P_{fa}$ are insufficient to characterize the detector's performance. Therefore, a more expressive performance metric needs to be defined, and that is the confusion matrix. The confusion matrix has a size of $M\times M$, where the rows correspond to ``actual" classes and the columns correspond to ``predicted" classes. Accordingly, the matrix ($m$,$n$)th entry represents $\Pr(\widehat{\mathcal{H}_n}|\mathcal{H}_m)$. Each element of the diagonal represents $P_d$ for the corresponding class.

Furthermore, the off-diagonal elements can be viewed from two complimentary views. Fixing the $m$th row shows how the samples corresponding to the true hypothesis $\mathcal{H}_m$ are distributed with respect to misdetections, where the sum of the off-diagonal elements in the $m$th row gives the probability of misdetection for $\mathcal{H}_m$. Moreover, fixing the $n$th column shows how different true hypotheses contribute to declaring $\mathcal{H}_n$, representing the probability of false alarms for $\mathcal{H}_m$.
\subsection{Complexity and Overhead}
The complexity of the \gls{dfr} can be quantified based on the number of mathematical operations involved at the local sensors and \gls{fc}, such as complex multiplications, additions and logarithmic operations. Furthermore, time-complexity should be quantified for iterative approaches to ensure convergence in real-time. The complexity at the local sensors is crucial because such devices usually have limited power and compute resources. Furthermore, the communication overhead needed to achieve \gls{df} at the \gls{fc} should be highlighted. The aim for researcher should be to design highly spectrum-efficient protocols. This includes channel estimation, propagation of local decisions, local detection probabilities, and local false alarm probabilities.

\section{Applications of Decision Fusion}
\label{sec-apps}
\gls{df} has been applied across a wide range of domains, depending on the application, the fused output may represent a class label, risk level, fault condition, event detection, control action or system-state decision. \gls{df} is therefore particularly valuable in distributed and safety-critical environments, where reliance on a single sensing or decision source may be insufficient. Prominent application areas include autonomous and intelligent systems, health monitoring, industrial monitoring and fault diagnosis, environmental monitoring, security and cybersecurity, and agriculture and precision farming. Table \ref{tab:apps} gives a summary of the applications presented in this article along with the commonly used fusion technique and main objective. 
\begin{table*}[htbp]
\centering
\caption{Applications of \gls{df} Across Different Domains.}
\label{tab:apps}
\renewcommand{\arraystretch}{1.3}
\resizebox{\textwidth}{!}{
\begin{tabular}{p{2cm} p{4cm} p{5cm} p{4cm} p{4cm}}
\hline
\textbf{Application} &
\textbf{Application Example} &
\textbf{Decision Sources} &
\textbf{Typical Fusion Methods} &
\textbf{Main Objective} \\
\hline
Autonomous and Intelligent Systems &
Autonomous vehicles, robotics, \gls{uav}, multi-agent systems &
Camera, LiDAR, radar, navigation modules, intelligent agents &
Weighted voting, probabilistic fusion, fuzzy techniques, learning-based fusion &
Improve perception, navigation, and autonomous decision-making \\
\midrule
Health Monitoring &
Disease diagnosis, physiological monitoring, patient deterioration detection, rehabilitation monitoring &
\Gls{ecg}, \gls{ppg}, \gls{spo2}, blood pressure, temperature, wearable sensors, inertial sensors &
Majority voting, weighted voting, Bayesian fusion, ensemble learning, Kalman filtering, deep learning-based fusion &
Improve diagnostic reliability, reduce false alarms, and enable robust continuous patient monitoring \\
\midrule
Environmental Monitoring &
Wildfire detection, flood monitoring, air-quality assessment, pollution tracking, disaster monitoring &
Satellite imagery, \gls{sar}, optical sensors, \gls{iot} sensors, meteorological stations, \gls{uav}, environmental classifiers &
Weighted voting, Bayesian fusion, Dempster--Shafer theory, ensemble learning, probabilistic fusion, deep learning-based fusion &
Improve event detection, reduce false alarms and missed detections, and provide reliable environmental-state assessment \\
\midrule
Industrial Monitoring and Fault Diagnosis &
Machinery monitoring, predictive maintenance, process fault detection, structural health monitoring, power-system fault diagnosis &
Vibration, acoustic, temperature, pressure, current, voltage, strain sensors, condition-monitoring systems, diagnostic classifiers &
Majority voting, weighted voting, Bayesian fusion, Dempster--Shafer theory, ensemble learning, fuzzy logic, deep learning-based fusion &
Improve fault-detection reliability, identify fault type and severity, reduce false alarms, and support predictive maintenance \\
\midrule
Agriculture and Precision Farming &
Crop disease detection, irrigation management, soil-condition assessment, yield prediction, weed detection, livestock monitoring &
Satellite and \gls{uav} imagery, multispectral and hyperspectral sensors, soil-moisture sensors, weather stations, \gls{iot} devices, agricultural classifiers &
Weighted voting, Bayesian fusion, ensemble learning, fuzzy logic, Dempster-Shafer theory, \gls{ml} and deep learning-based fusion &
Improve crop and soil assessment, support timely agricultural decisions, optimize resource use, and increase monitoring reliability \\
\hline
\end{tabular}
}
\vspace{-15pt}
\end{table*}


\subsection{Environmental Monitoring}
Environmental monitoring is another important application area for \gls{df}, particularly in systems designed for wildfire detection, flood monitoring, air-quality assessment, pollution tracking, and other forms of natural-hazard surveillance. Such systems commonly rely on heterogeneous information acquired from satellite imagery, ground-based sensors, \gls{iot} devices, meteorological stations, unmanned aerial vehicles, and remote-sensing platforms. The resulting decisions may include event occurrence, such as fire/no-fire or flood/no-flood, classification of affected regions, estimation of hazard severity, and generation of early-warning alerts. Since individual sensing modalities can be affected by cloud cover, atmospheric conditions, limited spatial or temporal resolution, sensor noise, and missing observations, combining information or decisions from multiple sources can provide a more reliable representation of environmental conditions \cite{zWANG2024,XU2024}.

\subsection{Health Monitoring}
Health monitoring represents an important application of \gls{df}, particularly with the increasing use of wearable, mobile, and \gls{iomt} devices for continuous observation of physiological and behavioral conditions. The decisions produced in health-monitoring systems range from estimation of physiological parameters to higher-level classification of health conditions. At the sensing level, fusion can be used to improve estimates of heart rate, blood pressure, respiration, and human motion. \gls{df} becomes particularly significant when several sensing or classification modules independently produce assessments of the patient's condition. In this case, the individual outputs, such as normal/abnormal classifications, risk probabilities, activity states or disease predictions, can be combined at the decision level using majority voting, weighted voting, probability averaging, Bayesian inference, ensemble learning, or rule-based approaches.  In\cite{yaldiz2023}, the authors used heart-rate and triaxial-accelerometer measurements together with machine-learning-based anomaly detection to predict impending exertional heat stroke. It demonstrates the value of multimodal monitoring for detecting health deterioration before a critical event occurs. Physiological strain derived from heart rate and physical stress derived from motion information were jointly analyzed, enabling abnormal physiological responses to be distinguished from those expected solely from increased physical activity. A decision-level approach for COVID-19 health prediction combined random forest, gradient boosting, and extreme gradient boosting classifiers using soft voting to obtain the final patient-health prediction is done in \cite{gumaei2022}.

Despite these advantages, reliable \gls{df} for health monitoring remains challenging. Physiological data exhibit considerable inter-person and intra-person variability, while sensor reliability changes according to motion, placement, environmental conditions, and device characteristics. Furthermore, decisions generated from correlated sensors should not necessarily be treated as independent evidence. Effective health-monitoring systems therefore require mechanisms for estimating sensor and classifier reliability and dynamically weighting local decisions. Missing modalities, synchronization errors, limited power and computational resources in wearable devices, privacy requirements, and the need for clinically interpretable decisions must also be considered. 

\subsection{Autonomous and Intelligent Transportation Systems}
In autonomous and intelligent systems, \gls{df} is mainly required to support decisions related to environmental perception, object and obstacle recognition, localization, navigation, trajectory planning, and autonomous action selection \cite{baccari2024, futuretransp6030103}. These decisions are derived from heterogeneous information sources, including cameras, LiDAR, radar, ultrasonic sensors, \gls{gnss} measurements, \gls{imu}, and, in multi-agent scenarios, decisions communicated by other autonomous agents. For example, autonomous vehicles combine visual, range, and motion information to determine the presence and class of surrounding objects, estimate their own position, and select appropriate driving actions, whereas mobile robots and UAVs use similar heterogeneous measurements for localization, mapping, obstacle avoidance, and path planning. The output generated through individual sensors or classifiers are combined through approaches such as voting, weighted decision rule, Bayesian inference and evidence based techniques. Perception and decision support in autonomous vehicles is studied in \cite{Liu2022}. They fuse radar and camera data for target recognition and tracking in severe weather.

\subsection{Industrial Monitoring and Fault Diagnosis}
Modern industrial systems comprise distributed sensors, control units, and communication networks that continuously generate heterogeneous measurements, making them a natural application area for \gls{df}. The resulting decisions include fault classification and localization, anomaly and intrusion detection, and maintenance triggering. Since individual sensors or diagnostic units are prone to noise, communication failures, and limited coverage, \gls{df} combines multiple local assessments into a more reliable global decision.

In industrial plants, local nodes can independently evaluate process variables such as pressure, temperature, or vibration and issue preliminary fault decisions, which are then fused spatially and temporally. The authors in \cite{Tabella2021} proposed such a spatio-temporal framework for quickest fault detection in an oil and gas scenario, where local nodes transmit decisions rather than raw measurements. In \cite{nair2024}, an Extended Kalman Filter-based approach was used for distributed fusion in industrial wireless sensor networks, improving state-estimation accuracy and robustness against redundant or noisy observations. For \gls{iiot} environments, \cite{Liu2021} addressed signal classification through feature fusion and federated learning, enabling collaborative training without centralizing raw data and supporting the identification of abnormal transmission conditions. For industrial control systems, \cite{Xue2024} proposed a real-time intrusion-detection framework explicitly based on \gls{df}, in which the outputs of multiple detection models are combined into a global decision. This improves robustness because individual detectors differ in detection capability and false-alarm behavior across attack types.

Overall, \gls{df} improves the reliability, robustness, and scalability of industrial monitoring, from fault detection and state estimation to \gls{iiot} signal classification and cybersecurity. Open challenges include heterogeneous sensors, communication delays, missing observations, sensor failures, changing operating conditions, conflicting local decisions, and real-time requirements.

\subsection{Agriculture and Precision Farming}

In agriculture and precision farming heterogeneous data from soil sensors, weather stations, remote sensing platforms, \gls{iot} devices, and agricultural robots are combined to support more accurate and timely decisions. Typical applications include irrigation management, fertilization, crop monitoring, disease detection, yield estimation, and autonomous field operations. Lambrinos \textit{et al.} \cite{Lambrinos2019} demonstrated an \gls{iot}-based decision support framework for precision farming in which distributed agricultural measurements are collected and processed to support farm-management decisions. Authors in \cite{Qiu2025} applied multi-sensor fusion to smart fertilization robots, combining environmental and soil-related measurements with Kalman-filter and \gls{lstm}-based processing to improve environmental perception and fertilization decision making. Remote sensing also plays a major role in precision agriculture, where satellite, \gls{uav}, multispectral, hyperspectral, and LiDAR data can be fused to improve crop and field assessment. The authors in \cite{saki2025} review the progression of remote-sensing fusion techniques from conventional approaches to deep-learning and transformer-based techniques for precision agriculture. The authors in \cite{agri2026} combinined multi-modal data streams—including spectral imagery, thermal readings, LiDAR point clouds, and proximal soil sensors— with algorithms ranging from conventional models to deep learning and Transformers. The resulting sensor fusion resolves single-sensor ambiguities to achieve up to 30\% higher predictive accuracy and superior fault tolerance across yield forecasting, soil health mapping, and weed and disease detection.

Overall, fusion in precision farming enables local measurements and model outputs to be integrated into higher-level decisions such as when and where to irrigate, fertilize, inspect, or intervene. Depending on the application, commonly used techniques include Kalman filtering, ensemble learning, deep learning, transformer-based fusion, and decision-support rules. These approaches help reduce the uncertainty associated with individual sensors and improve the efficiency, automation and resource utilization of modern agricultural systems.

\section{Decision Fusion with Emerging Technologies}
\label{sec:decision_fusion_emerging_technologies}

\begin{figure*}[t]
\centering
\includegraphics{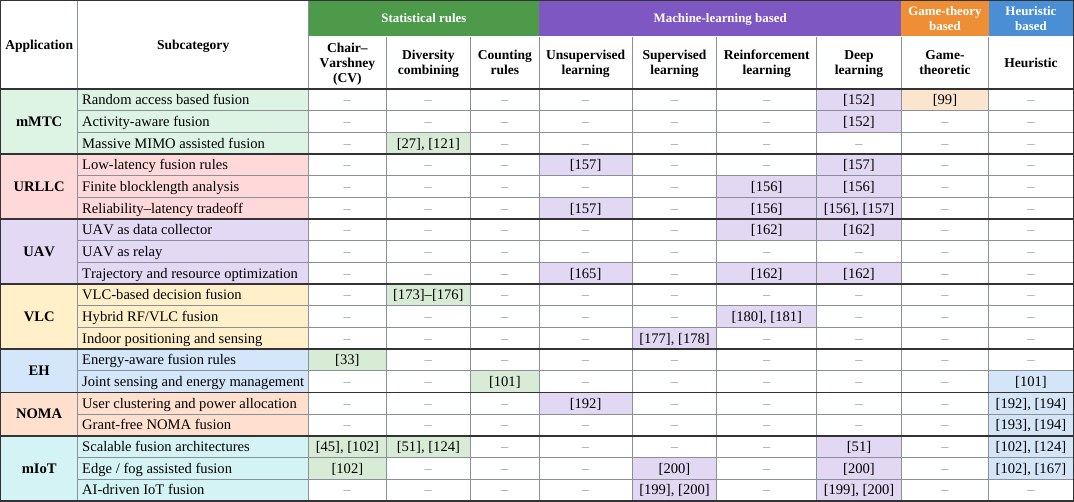}
\caption{Classification of Decision Fusion Literature by Emerging Application Area, Subcategory, and Fusion Approach}
\label{fig:emerging}
\vspace{-15pt} 
\end{figure*}

This section presents an overview of \gls{df} in emerging wireless technologies. Unlike conventional \glspl{wsn}, these advanced systems introduce operating conditions that directly affect the generation, transmission, and integration of local decisions at the \gls{fc}. Key factors, including massive connectivity, stringent reliability and latency requirements, node mobility, optical-channel constraints, intermittent energy availability, non-orthogonal access, and heterogeneous \gls{iot} deployments, complicate the independent design of sensing, reporting, and fusion stages. The following subsections analyze \gls{df} in \gls{mmtc}, \gls{urllc}, \gls{uav}-assisted communication, \gls{vlc}, \gls{eh} networks, \gls{noma}, \gls{miot}. Fig. \ref{fig:emerging} summarizes the existing \gls{df} techniques andd their subcategories and the corresponding fusion approaches used. For each technology, the analysis emphasizes system characteristics that affect report reliability and explores the resulting tradeoffs among detection accuracy, latency, energy consumption, communication overhead, complexity, and scalability.

\subsection{Decision Fusion in Massive Machine-Type Communication}
\label{subsec:decision_fusion_mmtc}

\Gls{mmtc} supports connectivity for a large number of low-power devices that typically transmit short and sporadic packets. In the context of \gls{df}, a key distinction from conventional \glspl{wsn} is that only a small and random subset of deployed devices may be active during any reporting interval. The \gls{fc} must infer the global hypothesis while addressing challenges such as grant-free access, activity uncertainty, packet collisions, \gls{csi} errors, and energy constraints. Transmitting raw observations from every device is generally impractical because it increases access load and rapidly depletes device energy. Therefore, hard decisions, coarsely quantized statistics, event-triggered reports, or compressed semantic representations are preferable when the monitored event is sparse in space or time \cite{bockelmann2016massive,chen2021massive,Tan2025GrantFreeActivity,Cao2025RISMassiveConnectivity}.

Reliable activity detection is therefore part of the fusion problem rather than a separate access-layer task. A false activity alarm can introduce a noise-generated symbol into the fusion statistic, whereas a missed activity event removes valid evidence. Recent work on fast maximum-likelihood and \gls{map} activity detection illustrates the importance of balancing detection quality against processing time in grant-free access
\cite{Tan2025GrantFreeActivity}. Similarly, tensor-based joint channel estimation and activity detection for \gls{ris}-assisted massive connectivity show that device identification and channel inference remain tightly coupled when additional propagation dimensions are introduced
\cite{Cao2025RISMassiveConnectivity}. In either case, a missing report should not automatically be interpreted as a local decision in favor of $\mathcal{H}_0$.

Large antenna arrays at the \gls{fc} are particularly attractive in \gls{mmtc} because they provide spatial degrees of freedom for separating simultaneous reports and suppressing multiuser interference. However, the corresponding gain depends on activity detection, channel estimation, pilot contamination, and large-scale fading variations. Sparse-aware, cluster-based, and low-complexity weighted rules may therefore be more practical than a fully centralized likelihood-ratio test in large deployments
\cite{ciuonzo2015massive,dey2020}. Communication-efficient aggregation provides another direction. Integrated semantic communication and over-the-air computation can reduce the amount of information transmitted by many sensing devices and directly aggregate task-relevant representations at the receiver
\cite{Yi2025SemanticAirCompCSS}. For \gls{df}, the resulting aggregation distortion should be reflected in the fusion weights or global threshold; otherwise, communication compression may be mistaken for sensing uncertainty. Overall, \gls{mmtc} fusion should be activity-aware, sparse-aware, and communication-efficient, while explicitly distinguishing inactivity, access failure, packet loss, and a legitimate negative decision.

\subsection{\gls{df} and Ultra-Reliable Low-Latency Communication}
\label{subsec:decision_fusion_urllc}

\gls{urllc} represents a different operating regime from \gls{mmtc}. Whereas \gls{mmtc} primarily emphasizes massive connectivity and energy-efficient access, \gls{urllc} requires a highly reliable decision within a strict latency deadline. This requirement is central to industrial automation, vehicular safety, remote control, smart grids, and mission-critical monitoring, where a delayed correct decision may be as harmful as an incorrect one. \gls{df} for \gls{urllc} should therefore be designed by jointly considering local sensing quality, short-packet reliability, channel-estimation uncertainty, processing delay, and the time required to collect a sufficient number of reports
\cite{popovski2018wireless,sutton2019enabling,
Zhang2023DRLURLLC,Zhang2024URLLCUnsupervised}.

For a serial sensing--reporting--fusion implementation, the end-to-end latency can be written as
\begin{equation}
T_{\mathrm{e2e}}=T_{\mathrm{sen}}+T_{\mathrm{rep}}+T_{\mathrm{fus}}
\leq T_{\max},
\label{eq:urllc_latency}
\end{equation}
where $T_{\mathrm{sen}}$, $T_{\mathrm{rep}}$, and $T_{\mathrm{fus}}$ denote the sensing, reporting, and fusion-processing durations. For parallel reporting, $T_{\mathrm{rep}}$ is the elapsed collection time, including any access, queueing, or retransmission delay, rather than the sum of individual packet durations. At the same time, the global probability of error should satisfy equation \eqref{eq-PrFusion-error}.
This set of equation shows that the problem is not simply to maximize detection probability, but to satisfy a joint reliability--latency requirement. Practical channel estimation further complicates this tradeoff because using more pilots can improve channel knowledge while leaving fewer symbols for data transmission. Recent learning-based \gls{urllc} designs explicitly account for imperfect \gls{csi} and for the tail probability of excessive packet error, rather than considering only the average reliability
\cite{Zhang2023DRLURLLC,Zhang2024URLLCUnsupervised}.

Information freshness is also important in \gls{urllc}. A report that arrives within the communication deadline may still be unhelpful if it describes an outdated system state. \gls{aoi}-aware scheduling in multi-server edge networks illustrates how server selection, computation, and finite-blocklength transmission jointly affect the timeliness of \gls{urllc} services
\cite{Wei2025AoIURLLC}. In a fusion system, the report age $\Delta_k(t)$ should therefore be incorporated into the reliability weight together with the short-packet error probability. Packet errors, bit errors in accepted reports, and local sensing errors are distinct quantities. Multi-connectivity, cooperative relaying, and multiple antennas can reduce the probability that all reporting links fail simultaneously, but excessive retransmissions and lengthy channel-estimation procedures can violate the latency budget. Consequently, \gls{urllc} fusion should favor simple, deadline-aware, freshness-aware, and reliability-weighted rules. Such rules may not reproduce the full likelihood-ratio test, but they provide a more credible operating point under finite-blocklength and strict-delay constraints
\cite{polyanskiy2010channel,Zhang2024URLLCUnsupervised,
Wei2025AoIURLLC}.
\subsection{Decision Fusion in UAV Communication}
\label{subsec:decision_fusion_uav}
\gls{uav} can participate in a decision-fusion network as mobile sensors, data collectors, relays, or aerial FCs. These configurations are useful when terrestrial infrastructure is unavailable, damaged, or expensive to deploy, such as in disaster response, agricultural monitoring, temporary surveillance, and remote-area inspection. Compared with a fixed \gls{fc}, a \gls{uav} can improve reporting links by adjusting its altitude and horizontal position. The same mobility, however, makes report reliability time varying and couples the fusion process with trajectory planning, sensor scheduling, communication power, and propulsion energy
\cite{Zeng2016UAV,Zeng2019Sky,Zhu2024UAVTrajectory}.

Consider $K$ ground sensors at $\boldsymbol x_k^{\mathrm{pos}}=[x_k,y_k,0]^{\mathsf T}$, while the \gls{uav} follows $\boldsymbol q_{\mathrm U}(t)=[x_{\mathrm U}(t),y_{\mathrm U}(t),z_{\mathrm U}(t)]^{\mathsf T}$. Here $t$ is continuous mission time and $z_{\mathrm U}(t)$ is altitude.
For a line-of-sight (LoS)-dominant sensor-to-\gls{uav} link, the distance is
\begin{equation}
d_k(t)=\|\boldsymbol q_{\mathrm U}(t)-\boldsymbol x_k^{\mathrm{pos}}\|_2,
\label{eq:uav_distance}
\end{equation}
and the reporting \gls{snr} can be approximated as
\begin{equation}
\gamma_k^{\mathrm{UAV}}(t)
=\frac{p_k(t)\beta_0}{\sigma_{\mathrm{rep},k}^{2}}
\left(\frac{d_0}{d_k(t)}\right)^{\kappa},
\label{eq:uav_reporting_snr}
\end{equation}
where $p_k(t)$ is the reporting power, $\beta_0$ is the channel power gain at reference distance $d_0$, $\kappa$ is the air-to-ground path-loss exponent, and $\sigma_{\mathrm{rep},k}^{2}$ is the reporting-noise power. The approximation is used over its valid propagation-distance range. The model in \eqref{eq:uav_reporting_snr} provides a useful first-order link between \gls{uav} position and reporting reliability, although practical air-to-ground channels may also experience \gls{nlos} propagation, shadowing, and blockage. Thus, two sensors with similar local detection probabilities may still deliver reports of substantially different quality.

The \gls{uav} trajectory directly affects the reporting SNRs and hence the fusion weights. Optimizing the trajectory only to maximize the sum \gls{snr} may repeatedly favor sensors that observe the same region and provide redundant evidence. A fusion-oriented design should also account for statistical informativeness and report complementarity. Let $\Theta_{\mathrm{UAV}}$ collect the complete trajectory, reporting and scheduling policies, fusion parameters, and mission duration. To make the trajectory-level objective explicit, consider the mission-average error
\begin{equation}    
\!\!\overline P_{\mathrm e,\mathrm{mission}}^{(g)}(\Theta_{\mathrm{UAV}})
=\frac{1}{T_{\mathrm{mission}}}
\int_0^{T_{\mathrm{mission}}}
P_{\mathrm e}^{(g)}(t;\Theta_{\mathrm{UAV}})\,\mathrm dt.
\end{equation}
The probability at each time is averaged over the random sensing and reporting states under the selected policy. A general design is

\begin{equation}
\label{eq:uav_joint_optimization}
\underset{\Theta _{\mathrm{UAV}}}{\min }\quad \overline{P}_{\mathrm{e},%
\mathrm{mission}}^{(g)}(\Theta _{\mathrm{UAV}})
\end{equation}%
subject to: \ 
\begin{tabular}{ll}
a) $\Vert $$\dot{\mathbf{q}}_{\mathrm{U}}$(t)$\Vert _{2}\leq V_{\max }$ & b) 
$E$$_{\mathrm{prop}}[\mathbf{q}$$_{\mathrm{U}}(\cdot )]\leq E_{\max }$ \\ 
c) $0<T$$_{\mathrm{mission}}\leq T_{\max }$ & d) $\mathbf{q}$$_{\mathrm{U}}$(%
$\cdot $)$\in \mathcal{Q}_{\mathrm{UAV}}$%
\end{tabular}
The set $\mathcal{Q}_{\mathrm{UAV}}$ specifies the mission endpoints and admissible flight region, and the speed bound holds throughout the mission. This mission-average formulation is a survey-level design template, rather than the exact objective of every cited \gls{uav} study. The propulsion-energy constraint is important because \gls{uav} motion may consume considerably more energy than communication and processing. Recent attention-reinforced trajectory optimization for large-scale, low-power data collection reinforces the need to coordinate data-acquisition quality with \gls{uav} and sensor energy expenditure
\cite{Zhu2024UAVTrajectory}. Likewise, studies of \gls{uav}-assisted cognitive-radio networks with \gls{noma} and \gls{eh} show that trajectory, spectrum access, power allocation, and available energy are strongly coupled, even when the immediate objective is throughput rather than global decision error
\cite{Bhowmick2022UAVCRNOMAEH}. These cross-layer results support joint rather than sequential design of mobility and fusion resources
\cite{Zeng2017Energy,Ghdiri2021UAVData}.

The role of an aerial \gls{fc} is also evolving toward task-oriented edge inference. Dynamic \gls{uav}-assisted cooperative edge-\gls{ai} systems aggregate information from multiple devices while jointly controlling trajectory and communication resources
\cite{Huang2025UAVEdgeAI}. Although some of these schemes operate at the feature level rather than on binary decisions, the underlying design lesson remains relevant: the \gls{uav} should move and allocate resources according to the value of the information for the final inference task, not merely according to received signal strength.

For multi-\gls{uav} networks, a hierarchical architecture is usually more scalable than forwarding every local decision to one aerial node. Each \gls{uav} can fuse reports from a local cluster and transmit a compact statistic to a leader \gls{uav} or terrestrial \gls{fc}. Reports from neighboring UAVs may nevertheless be correlated when their sensing regions overlap, and local statistics may have different scales when the clusters contain different sensor types. Collaborative perception and computing offloading in \gls{6g} air--ground integrated networks extends this idea by jointly considering vehicles, UAVs, roadside infrastructure, offloading delay, perception quality, and aerial energy consumption
\cite{Jiang2026AirGroundPerception}. Although such systems often fuse features or perception outputs, they provide a useful model for future hierarchical \gls{df}. Overall, \gls{uav}-assisted fusion should be mobility-aware, energy-aware, freshness-aware, and task-oriented.

\subsection{Decision Fusion in Visible Light Communication}
\label{subsec:decision_fusion_vlc}

\gls{vlc} uses light-emitting diodes or related optical sources for illumination and wireless communication. In a decision-fusion system, \gls{vlc} can serve as the reporting technology through which sensors transmit local decisions or soft statistics to an optical \gls{fc}. It is attractive in hospitals, aircraft cabins, industrial facilities, and electromagnetic-sensitive environments where radio-frequency communication may be restricted or undesirable. \gls{vlc} also offers unlicensed bandwidth and high spatial reuse. Its reliability, however, depends strongly on transmitter--receiver geometry, photodetector \gls{fov}, device orientation, ambient illumination, and blockage
\cite{Komine2004VLC,Karunatilaka2015VLC}. Cooperative sensing has also been investigated in cognitive \gls{vlc}, where sensing results from several nodes are exploited jointly rather than individually
\cite{V1_jiang2021collab}.

For an intensity-modulation/direct-detection link, let $\widetilde h_k^{\mathrm{LoS}}(t)$ denote the unblocked \gls{los} channel gain between sensor $k$ and the optical \gls{fc}. For front-hemisphere emission, $0\leq\phi_k^{\mathrm{irr}}(t)\leq\pi/2$, the FoV-limited LoS model is
\begin{equation}
\begin{aligned}
h_k^{\mathrm{VLC}}(t)&=
\begin{cases}
\widetilde h_k^{\mathrm{LoS}}(t),
&0\leq\psi_k^{\mathrm{inc}}(t)\leq\Psi_{\mathrm c},\\
0,&\text{otherwise},
\end{cases}\\
\widetilde h_k^{\mathrm{LoS}}(t)
&=\frac{(m_{\mathrm L}+1)A_{\mathrm{PD}}}{2\pi d_k^2(t)}
\cos^{m_{\mathrm L}}\!\bigl(\phi_k^{\mathrm{irr}}(t)\bigr)\\
&\quad\times T_{\mathrm{of}}\!\bigl(\psi_k^{\mathrm{inc}}(t)\bigr)
g_{\mathrm c}\!\bigl(\psi_k^{\mathrm{inc}}(t)\bigr)
\cos\!\bigl(\psi_k^{\mathrm{inc}}(t)\bigr),
\end{aligned}
\label{eq:vlc_channel_gain}
\end{equation}
where $A_{\mathrm{PD}}$ is the photodetector area, $d_k(t)$ is the link distance, $\phi_k^{\mathrm{irr}}(t)$ and $\psi_k^{\mathrm{inc}}(t)$ are the irradiance and incidence angles, $T_{\mathrm{of}}(\cdot)$ is the optical-filter gain, $g_{\mathrm c}(\cdot)$ is the concentrator gain, and $\Psi_{\mathrm c}$ is the \gls{fov} angle. The Lambertian order is
\begin{equation}
m_{\mathrm L}=-\frac{\ln2}{\ln(\cos\Phi_{1/2})},
\qquad 0<\Phi_{1/2}<\frac{\pi}{2},
\label{eq:lambertian_order}
\end{equation}
where $\Phi_{1/2}$ is the \gls{led} semi-angle at half optical power. This model describes the \gls{los} component; a blockage indicator or a separate reflection model is needed to represent blocked or \gls{nlos} operation. The \gls{los} channel can therefore disappear when the receiver moves outside the \gls{fov}, even when the physical distance remains small. Receiver orientation and blockage should consequently be incorporated into report reliability
\cite{Soltani2019Orientation}. Recent work also indicates that \gls{nlos} optical paths can be exploited for networking rather than treated only as unusable links, although their lower \gls{snr} and synchronization requirements must be accounted for explicitly
\cite{Wu2025NLOSVLC}.

Optical diversity can be obtained using several LEDs, photodetectors, or imaging receivers. Angle-diversity receivers, in particular, have been studied for improving the \gls{sinr} and for managing interference in indoor multi-cell \gls{vlc}, including the effect of the diversity-combining scheme and of overlapping receiver \glspl{fov}
\cite{V2_chen2014angle,V3_chen2014sinr,V4_adr2022,V5_alsakkaf2024}. Multiple optical branches carrying the same local decision should first be combined at the communication layer and then represented as one sensor report at the fusion layer. Otherwise, repeated copies of one decision may be incorrectly counted as independent evidence. Soft optical reporting can preserve an energy statistic, likelihood value, or confidence measure, but its potential detection gain must be balanced against optical bandwidth, illumination, dimming, peak-power, and implementation constraints. Information fusion has also been applied to visible-light indoor localization, where multiple classifiers or fusion-network layers are combined to improve accuracy
\cite{V8_guo2017fusion,V9_tlfn2019}; these works fuse learned features or classifier outputs rather than sensor decisions, but they illustrate the benefit of combining several optical observations. Recent power-constrained optical sensing work further illustrates the need to account for clipping and waveform limitations when communication and sensing functions share the same optical signal
\cite{Li2026OpticalPassiveSensing}.

Because continuous \gls{los} connectivity cannot always be guaranteed, hybrid \gls{rf}/\gls{vlc} operation is an important practical option. Sensors with a reliable optical path can report through \gls{vlc}, while blocked or poorly oriented sensors use \gls{rf}. Learning-based load balancing for hybrid LiFi/WiFi networks shows that the choice between optical and radio access can be adapted to the link state
\cite{V6_ahmad2020rl,V7_ahmad2022rl}, which is analogous to adapting the reliability weight in the fusion rule below. Since the two fusion statistics may have different scales, define their standardized forms by
\[
Z_a(t)=\frac{\Lambda_a(t)-\mu_{\Lambda,a,0}(t)}
{\sqrt{v_{\Lambda,a,0}(t)+\delta_a}},
\quad a\in\{\mathrm{VLC},\mathrm{RF}\},
\]
where $\mu_{\Lambda,a,0}$ and $v_{\Lambda,a,0}$ are the null-hypothesis mean and variance, and $\delta_a>0$ has variance units. The hybrid statistic is
\begin{equation}
\Lambda_{\mathrm{hyb}}(t)
=\eta_{\mathrm{hyb}}(t)Z_{\mathrm{VLC}}(t)
+[1-\eta_{\mathrm{hyb}}(t)]Z_{\mathrm{RF}}(t)
\underset{\mathcal{H}_0}{\overset{\mathcal{H}_1}{\gtrless}}\lambda_{\mathrm{hyb}}(t),
\label{eq:hybrid_rf_vlc_fusion}
\end{equation}
where $0\leq\eta_{\mathrm{hyb}}(t)\leq1$ reflects the reliability of the RF and \gls{vlc} links. Standardization does not make repeated or correlated reports independent; threshold calibration must account for the joint distribution of the two statistics. Aggregated \gls{rf}--\gls{vlc} designs for ultra-reliable and low-latency vehicular communication further support link-level aggregation when one technology alone cannot satisfy the required reliability
\cite{Kong2019HybridVLC,Singh2024AggregatedRFVLC}. Hence, \gls{vlc}-based fusion should be geometry-aware, blockage-aware, and power-aware, with hybrid reporting used when optical connectivity is intermittent.

\subsection{Decision Fusion and Energy Harvesting}
\label{subsec:decision_fusion_eh}

\gls{eh} enables a sensor to replenish part of its operating energy from solar radiation, vibration, thermal gradients, RF signals, or other ambient sources. For \gls{df}, the significance of \gls{eh} extends beyond increasing network lifetime. Random energy arrivals and finite battery capacities invalidate the conventional assumption that every sensor is continuously available for sensing and reporting. The participating set may change over time, which couples sensing quality, report reliability, and energy availability
\cite{Ulukus2015EHReview,Tarighati2017DHTEH}.

Let $B_k(t)$ denote the energy stored by sensor $k$ at the beginning of slot $t$, with $0\leq B_k(t)\leq B_k^{\max}$. Let $E_k^{\mathrm{har}}(t)$ denote the harvested energy and $a_k(t)\in\{0,1\}$ the activation variable. The energy consumed in one slot can be expressed as
\begin{equation}
E_k^{\mathrm{use}}(t)=a_k(t)\left[
E_k^{\mathrm{sen}}+E_k^{\mathrm{pro}}+E_k^{\mathrm{tx}}\bigl(p_k(t)\bigr)
\right],
\label{eq:EH_energy_consumption}
\end{equation}
where the three terms are sensing, processing, and reporting energies in joules. The function $E_k^{\mathrm{tx}}(p_k(t))$ includes transmission duration and any modeled circuit consumption, while $p_k(t)$ is power in watts. With harvested energy becoming usable in the next slot, the battery evolves according to
\begin{equation}
\begin{aligned}
B_k(t+1)&=\min\bigl\{B_k^{\max},\,B_k(t)
-E_k^{\mathrm{use}}(t)+E_k^{\mathrm{har}}(t)\bigr\},\\
&\qquad 0\leq E_k^{\mathrm{use}}(t)\leq B_k(t),
\end{aligned}
\label{eq:EH_battery_update}
\end{equation}
with $0\leq B_k(1)\leq B_k^{\max}$. The slot-wise inequality forbids consuming energy that is not in the battery. Every feasible policy also satisfies the cumulative condition
\begin{equation}
\sum\nolimits_{\tau=1}^{t}E_k^{\mathrm{use}}(\tau)
\leq B_k(1)+\sum\nolimits_{\tau=1}^{t-1}E_k^{\mathrm{har}}(\tau),
\qquad\forall k,t.
\label{eq:EH_energy_causality}
\end{equation}
The cumulative condition is necessary but is not sufficient for a finite battery, since it also counts energy that may have overflowed. The recursion and slot-wise expenditure bound in \eqref{eq:EH_battery_update} must therefore be enforced together. Energy arriving when the battery is full is lost, while aggressive consumption can prevent participation in later, more informative intervals. A transmission action that appears attractive in one slot may therefore be suboptimal over a longer horizon
\cite{Tutuncuoglu2012BatteryLimited,Ozel2012AWGN}.

The battery state should primarily determine whether the sensor is activated, how many samples it processes, and how much power it uses. Once a report is received, its fusion weight should be determined by sensing and communication reliability rather than by stored energy alone. A full battery does not make an inaccurate sensor statistically informative, although the battery state can indirectly influence reliability through the selected reporting power.

Energy availability also affects the choice among hard, quantized, soft, and censored reporting. Soft statistics preserve more information but generally require more transmission energy. Hard or quantized decisions are therefore attractive when the harvesting rate is low. Joint client scheduling and quantization in \gls{eh}-enabled learning networks demonstrates the broader value of coordinating participation, quantization accuracy, and energy causality instead of optimizing these variables independently \cite{Ni2024EHScheduling}. Content-based wake-up provides a related sensing-oriented principle: devices can be activated according to the expected value and timeliness of their information rather than according to energy state alone
\cite{Shiraishi2025ContentWakeUp}. For \gls{df}, such selective activation can reduce energy expenditure while retaining the most informative reports.

Censoring, battery depletion, and packet loss are nevertheless different events. A deliberately silent sensor may have an uninformative observation, while an energy-depleted sensor may be unable to report a highly informative observation. The \gls{fc} should receive an activity indicator or marginalize over the cause of a missing report. The same caution applies to energy-asymmetric signaling, where one local decision is represented by an active symbol and the other by a zero- or low-energy symbol. Without an explicit availability model, a legitimate zero-energy decision can be confused with a sensor that lacks sufficient energy, producing a systematic bias
\cite{Tarighati2017DHTEH}.

The activation policy, transmit powers, fusion weights, and global threshold can be jointly designed. Let $\Pi_{\mathrm{EH}}$ be a causal policy selecting these quantities from the information available at each slot. Write $P_{\mathrm e}^{(g)}(t;\Pi_{\mathrm{EH}})$ and $P_{\mathrm{fa}}^{(g)}(t;\Pi_{\mathrm{EH}})$ for probabilities already averaged over random sensing, channel, and energy states under that policy. A long-term problem is 
\begin{equation}
\label{eq:EH_joint_problem}
\underset{\Pi _{\mathrm{EH}}}{\min }\quad \limsup_{T\rightarrow \infty }%
\frac{1}{T}\sum\nolimits_{t=1}^{T}P_{\mathrm{e}}^{(g)}(t;\Pi _{\mathrm{EH}})
\end{equation}
\begin{tabular}{ll}
\multicolumn{2}{l}{subject to: \ \ \ \ a) $P_{\mathrm{fa}}^{(g)}(t;\Pi _{%
\mathrm{EH}})\leq \alpha _{\mathrm{FA}\text{ \ }}$} \\ 
\multicolumn{2}{l}{b) $\sum\limits_{\tau =1}^{t}E_{k}^{\mathrm{use}}(\tau
)\leq B_{k}(1)+\sum\limits_{\tau =1}^{t-1}E_{k}^{\mathrm{har}}(\tau )$} \\ 
c) $0\leq B$$_{k}(t)\leq B_{k}^{\max }$ & d) $a$$_{k}(t)\in \{0,1\}$ \\ 
e) $0\leq p_{k}(t)\leq a_{k}(t)P_{k}^{\max }$ & f) $0\leq E_{k}^{\mathrm{use}%
}(t)\leq B_{k}(t)$ \\ 
\multicolumn{2}{l}{g) $B_{k}(t+1)=\min \bigl\{B_{k}^{\max },$\thinspace $%
B_{k}(t)-E_{k}^{\mathrm{use}}(t)+E_{k}^{\mathrm{har}}(t)\bigr\}$}%
\end{tabular}\\
All sensor-wise constraints hold for every $k,t$ on each admissible energy-arrival realization, with a prescribed feasible initial battery state. The power bound sets the reporting power to zero when a sensor is inactive. Constraint~\eqref{eq:EH_joint_problem}-b is retained as the cumulative interpretation of energy causality, while \eqref{eq:EH_joint_problem}-d and \eqref{eq:EH_joint_problem}-e enforce the finite battery. The formulation emphasizes that \gls{eh}-aware fusion is sequential: present activation and energy-allocation decisions affect both current detection performance and future availability. When future energy arrivals are known, an offline policy can be used; in practice, online dynamic programming, stochastic optimization, or receding-horizon control is usually more appropriate. Save-then-report policies reduce the risk of energy shortage but increase delay, whereas best-effort reporting provides fresher information with more irregular participation. Joint scheduling and quantization techniques provide useful tools for managing this tradeoff under energy-causality constraints
\cite{Ozel2012AWGN,Ni2024EHScheduling}.

Sensor selection should account for both stored energy and information value. Selecting only the sensors with the largest batteries may yield highly correlated observations, while repeatedly selecting the strongest detectors can rapidly deplete a small subset of nodes. A balanced policy should favor complementary information while maintaining energy-neutral operation
\cite{Liu2020EH,Shiraishi2025ContentWakeUp}. Battery-state information at the \gls{fc} may itself be delayed or quantized because frequent battery reporting consumes energy. Status-updating policies under partial battery knowledge show that freshness and energy management remain coupled when the controller does not know the exact battery state
\cite{Hatami2025PartialBattery}. The \gls{fc} should therefore account for the age and uncertainty of battery reports before scheduling a sensor. Overall, \gls{eh}-based fusion should be battery-aware at the scheduling layer, channel-aware at the reporting layer, and statistically reliability-aware at the fusion layer.

\subsection{Decision Fusion with Non-Orthogonal Multiple Access}
\label{subsec:decision_fusion_noma}

\gls{noma} allows several sensors to share the same time--frequency resource by separating their reports in the power domain or through non-orthogonal signatures. This property is attractive for \gls{df} because local decisions are short, while orthogonal resource allocation becomes inefficient as the number of sensors grows. \gls{noma} deliberately retains multiuser interference and resolves it at the \gls{fc} through \gls{sic} or another multiuser detector. The spectral-efficiency gain is therefore accompanied by a tighter coupling between report decoding and the final fusion decision: an error at an early \gls{sic} stage can affect several reports decoded later
\cite{Islam2017PDNOMA,Liu2017NOMA}.

For an uplink power-domain \gls{noma} interval, let $\mathcal A_{\mathrm N}(t)$ be the true transmitting set on one resource. Each active sensor modulates its source decision $\mathbf{u}_k(t)\in\{-1,+1\}$, it does not transmit an \gls{fc}-decoded estimate. The received signal is
\begin{equation}
\mathbf{r}_{\mathrm N}(t)
=\sum\nolimits_{k\in\mathcal A_{\mathrm N}(t)}
\sqrt{p_k(t)}h_k(t)\mathbf{u}_k(t)+n_{\mathrm F}(t),
\label{eq:NOMA_received_signal}
\end{equation}
where $p_k(t)$ is the reporting power, $h_k(t)$ is the channel coefficient, and
$n_{\mathrm F}(t)\sim\mathcal{CN}(0,\sigma_{\mathrm F}^2)$. Define the instantaneous received power as
\begin{equation}
p_k^{\mathrm{rx}}(t)=p_k(t)|h_k(t)|^2.
\label{eq:NOMA_received_power}
\end{equation}
If $\varpi_t(i)$ is the sensor decoded at \gls{sic} stage $i$, a post-\gls{sic} \gls{sinr} approximation can be written as
\begin{equation}
\gamma_{\varpi_t(i)}^{\mathrm{SIC}}(t)
=\frac{p_{\varpi_t(i)}^{\mathrm{rx}}(t)}
{I_i^{\mathrm u}(t)+I_i^{\mathrm r}(t)+\sigma_{\mathrm F}^2}.
\label{eq:NOMA_SIC_SINR}
\end{equation}
\begin{subequations}
\addtocounter{parentequation}{-1}%
\def\theparentequation{\arabic{parentequation}}%
\renewcommand{\theequation}{\theparentequation-\alph{equation}}%
\begin{align}
I_i^{\mathrm u}(t)&=\sum\nolimits_{v=i+1}^{K_{\mathrm N}(t)}
p_{\varpi_t(v)}^{\mathrm{rx}}(t),
\label{eq:NOMA_uncancelled_interference}\\
I_i^{\mathrm r}(t)&=\sum\nolimits_{v=1}^{i-1}
\zeta_{\varpi_t(v)}(t)p_{\varpi_t(v)}^{\mathrm{rx}}(t).
\label{eq:NOMA_residual_interference}
\end{align}
\end{subequations}
Here, $K_{\mathrm N}(t)=|\mathcal A_{\mathrm N}(t)|$ and $\zeta_k(t)\in[0,1]$ is a phenomenological residual-power factor in the adopted partial-cancellation model. Perfect cancellation corresponds to zero. This approximation can represent residual channel-estimation and synchronization errors, but it does not specify the complete distribution of erroneous cancellation or the correlations among \gls{sic} decisions.

Recent cognitive-radio work integrates \gls{css} and sharing directly with power-domain \gls{noma}. In such a framework, local spectrum assessments are combined at an \gls{fc} and used together with interweave or underlay access decisions, while unreliable users should be prevented from dominating the global result
\cite{Machani2026PDNOMACSS}. This provides a direct example of why sensing reliability, user trust, spectrum-sharing mode, and \gls{noma} resource allocation should be jointly designed rather than optimized as separate layers.
Grant-free \gls{noma} affects information freshness because collisions, retransmission opportunities, and access probabilities determine the age of the delivered report. Recent analyses explicitly connect \gls{noma}-based grant-free transmission with \gls{aoi}
\cite{Ding2024NOMAAoI}. Time-critical and massive-MTC designs further combine scheduled uplink opportunities, grant-free \gls{noma}, \gls{sic}, reception probability, and both average and peak information age
\cite{Lu2026NOMAGrantFree}. These results support including $\Delta_k(t)$ and activity confidence in the fusion weight instead of considering only post-\gls{sic} \gls{sinr}
\cite{Shahab2020GFNOMA}.

Hard decisions remain attractive because they limit payload and energy, while soft reports preserve a likelihood or confidence measure at the cost of greater dynamic range and more demanding multiuser separation. Code-domain \gls{noma} can jointly infer activity and decisions through sparse signatures and message passing, although its complexity grows with the overload factor and the number of iterations. For a given admitted sensor set, the \gls{noma} group assignment, decoding order, powers, fusion weights, and threshold can be jointly optimized.

\subsection{Decision Fusion and Massive Internet of Things}
\label{subsec:decision_fusion_miot}

\gls{mmtc} mainly describes the radio-access problem created by a large population of sporadically active devices, whereas \gls{miot} refers to the broader end-to-end sensing environment. A \gls{miot} deployment may combine heterogeneous sensing modalities, several communication technologies, battery-powered and \gls{eh} devices, gateways, edge servers, \gls{uav}, and cloud platforms. The devices may use different sampling rates, local decision rules, confidence scales, and latency requirements. The central problem is therefore not merely to connect many sensors, but to preserve the statistical meaning of their reports as they are aggregated across a large and heterogeneous architecture
\cite{AlFuqaha2015IoT,Shi2016Edge}.

Forwarding every observation or local decision directly to one \gls{fc} is generally unattractive because access load, backhaul traffic, processing burden, and energy consumption grow with the number of devices. A hierarchical architecture is more scalable: sensors report to gateways, cluster heads, or edge nodes, which form local statistics and forward compressed results to a higher-level \gls{fc}
\cite{AlJarrah2020}. Recent M-ary distributed decision-fusion work in multihop relay sensor networks reinforces this architecture by jointly considering the \gls{dfr}, implementation framework, and end-to-end performance of multihop reporting
\cite{zhang2024mary}. Semantic communication and over-the-air aggregation provide complementary mechanisms for reducing the communication required from a large number of sensors \cite{Yi2025SemanticAirCompCSS}.

Asynchronous reporting creates a second source of unequal reliability. Sensors may sample at different rates, sleep for long periods, experience random-access delay, or transmit only after an event. A freshness-aware rule can multiply the statistical weight by a non-increasing function of
$\Delta_k(t)=t-t_k^{\mathrm{gen}}$. The decay rate should depend on the dynamics of the monitored phenomenon: an old report may remain useful for slow environmental processes but become dangerous in vehicular or industrial safety applications
\cite{quan2008,Yates2021AoI}.

Heterogeneous modalities also require calibration. A binary alarm from a low-cost sensor is not directly comparable with a confidence score from a camera or radar. Soft reports should therefore be transformed into calibrated likelihoods, posterior probabilities, or normalized confidence measures. Reliable federated multi-view learning provides a recent example of how heterogeneous views and local uncertainty can be combined at an edge server without assuming that every device produces the same type or quality of representation
\cite{Li2026FedRMVL}. Although the fusion may occur at the feature or model level, the same principle applies to \gls{df}: a numerically large score should not dominate unless it is statistically well calibrated.

Edge processing reduces backhaul traffic, but aggressive compression can discard information that the upper-level \gls{fc} cannot recover. Instead of forwarding one hard decision, an edge node can transmit a quantized log-likelihood, the local statistic, the number of participating sensors, and a compact quality indicator. Online vertical federated learning for \gls{css} further illustrates how distributed entities can learn from streaming, partitioned observations while limiting direct data exchange
\cite{Wang2025OVFLCSS}. Such learning-based mechanisms are promising for \gls{miot}, but their model outputs should be calibrated before they are used as fusion weights.

Air--ground integrated networks offer another hierarchical realization. Ground sensors or vehicles can perform local processing, while \gls{uav} and roadside units aggregate complementary information and offload computationally intensive tasks. Collaborative perception and computing offloading in \gls{6g} air--ground systems shows that perception quality, offloading delay, radio resources, and \gls{uav} energy are naturally coupled
\cite{Jiang2026AirGroundPerception}. This architecture is relevant to large-scale \gls{df} even when some intermediate nodes exchange features rather than final binary decisions. The openness and scale of \gls{miot} increase the likelihood of faulty, misconfigured, or compromised devices. Table~\ref{tab:section7_checklist} compares representative studies reviewed
in this section according to their information-fusion capabilities and
communication, timing, energy, and processing considerations. The articles
are grouped by their primary technology context. The checklist summarizes
the features explicitly attributed to each study in this section; general
design recommendations are not automatically assigned to every cited article.

\begin{table*}[!t]
\caption{Checklist Comparison of Representative Studies on \gls{df}
and Emerging Technologies.}
\label{tab:section7_checklist}
\centering
\begingroup
\footnotesize
\setlength{\tabcolsep}{2pt}
\renewcommand{\arraystretch}{1.06}
\renewcommand{\tabularxcolumn}[1]{m{#1}}

\newcommand{\svyes}{\ensuremath{\checkmark}}
\newcommand{\svns}{\textemdash}
\begin{tabularx}{\textwidth}{
@{}
c
c
>{\raggedright\arraybackslash}X
*{7}{>{\centering\arraybackslash}m{0.95cm}}
@{}
}
\toprule

\textbf{Ref.}
& \textbf{Year}
& \textbf{Main contribution reviewed}
& \textbf{IF}
& \textbf{AP}
& \textbf{CL}
& \textbf{TF}
& \textbf{EN}
& \textbf{RC}
& \textbf{EC}
\\

\midrule


\multicolumn{10}{@{}l}{
\textit{VII-A: Massive machine-type communication (\gls{mmtc})}}
\\
\specialrule{0.35pt}{0.8pt}{1.0pt}

\cite{Tan2025GrantFreeActivity}
& 2025
& Fast grant-free device-activity detection
& \svns & \svyes & \svns & \svns & \svns & \svns & \svns
\\
\specialrule{0.25pt}{0.9pt}{0.9pt}

\cite{Cao2025RISMassiveConnectivity}
& 2025
& \gls{ris}-assisted joint channel estimation and activity detection
& \svns & \svyes & \svyes & \svns & \svns & \svns & \svns
\\
\specialrule{0.25pt}{0.9pt}{0.9pt}

\cite{Yi2025SemanticAirCompCSS}
& 2025
& Semantic cooperative sensing with over-the-air aggregation
& \svyes & \svns & \svyes & \svns & \svns & \svyes & \svns
\\
\specialrule{0.35pt}{1.0pt}{1.2pt}


\multicolumn{10}{@{}l}{
\textit{VII-B: Ultra-reliable low-latency communication (\gls{urllc})}}
\\
\specialrule{0.35pt}{0.8pt}{1.0pt}

\cite{Zhang2023DRLURLLC}
& 2023
& Learning-based \gls{urllc} resource utilization with imperfect \gls{csi}
& \svns & \svns & \svyes & \svyes & \svns & \svns & \svns
\\
\specialrule{0.25pt}{0.9pt}{0.9pt}

\cite{Zhang2024URLLCUnsupervised}
& 2024
& \gls{urllc} learning with practical channel estimation
& \svns & \svns & \svyes & \svyes & \svns & \svns & \svns
\\
\specialrule{0.25pt}{0.9pt}{0.9pt}

\cite{Wei2025AoIURLLC}
& 2025
& Age-aware scheduling and server selection for edge \gls{urllc}
& \svns & \svyes & \svyes & \svyes & \svns & \svns & \svyes
\\
\specialrule{0.35pt}{1.0pt}{1.2pt}


\multicolumn{10}{@{}l}{
\textit{VII-C: \gls{uav}-assisted communication}}
\\
\specialrule{0.35pt}{0.8pt}{1.0pt}

\cite{Bhowmick2022UAVCRNOMAEH}
& 2022
& \gls{uav}--\gls{noma} throughput design with \gls{eh}
& \svns & \svns & \svns & \svns & \svyes & \svns & \svns
\\
\specialrule{0.25pt}{0.9pt}{0.9pt}

\cite{Zhu2024UAVTrajectory}
& 2024
& \gls{uav} trajectory design for low-power sensor-data collection
& \svns & \svns & \svyes & \svns & \svyes & \svns & \svns
\\
\specialrule{0.25pt}{0.9pt}{0.9pt}

\cite{Huang2025UAVEdgeAI}
& 2025
& Cooperative \gls{uav}-assisted edge-\gls{ai} inference
& \svyes & \svns & \svns & \svns & \svns & \svns & \svyes
\\
\specialrule{0.25pt}{0.9pt}{0.9pt}

\cite{Jiang2026AirGroundPerception}
& 2026
& Air--ground collaborative perception and computing offloading
& \svyes & \svns & \svns & \svyes & \svyes & \svns & \svyes
\\
\specialrule{0.35pt}{1.0pt}{1.2pt}


\multicolumn{10}{@{}l}{
\textit{VII-D: Visible light communication (\gls{vlc})}}
\\
\specialrule{0.35pt}{0.8pt}{1.0pt}

\cite{Singh2024AggregatedRFVLC}
& 2024
& Aggregated \gls{rf}--\gls{vlc} links for reliable, low-latency V2X
& \svns & \svns & \svyes & \svyes & \svns & \svns & \svns
\\
\specialrule{0.25pt}{0.9pt}{0.9pt}

\cite{Wu2025NLOSVLC}
& 2025
& \gls{nlos} optical networking under weak-link conditions
& \svns & \svns & \svyes & \svns & \svns & \svns & \svns
\\
\specialrule{0.25pt}{0.9pt}{0.9pt}

\cite{Li2026OpticalPassiveSensing}
& 2026
& Power-constrained passive optical sensing and waveform design
& \svns & \svns & \svyes & \svns & \svyes & \svns & \svns
\\
\specialrule{0.35pt}{1.0pt}{1.2pt}


\multicolumn{10}{@{}l}{
\textit{VII-E: Energy Harvesting \gls{eh}}}
\\
\specialrule{0.35pt}{0.8pt}{1.0pt}

\cite{Ni2024EHScheduling}
& 2024
& \gls{eh}-client scheduling and update quantization in federated learning
& \svns & \svyes & \svns & \svns & \svyes & \svyes & \svyes
\\
\specialrule{0.25pt}{0.9pt}{0.9pt}

\cite{Hatami2025PartialBattery}
& 2025
& Fresh status updating with partial battery-state knowledge
& \svns & \svyes & \svns & \svyes & \svyes & \svns & \svns
\\
\specialrule{0.25pt}{0.9pt}{0.9pt}

\cite{Shiraishi2025ContentWakeUp}
& 2025
& Content-based wake-up for energy-efficient, timely top-$k$ sensing
& \svns & \svyes & \svns & \svyes & \svyes & \svns & \svns
\\
\specialrule{0.35pt}{1.0pt}{1.2pt}


\multicolumn{10}{@{}l}{
\textit{VII-F: Non-orthogonal multiple access (\gls{noma})}}
\\
\specialrule{0.35pt}{0.8pt}{1.0pt}

\cite{Ding2024NOMAAoI}
& 2024
& Grant-free \gls{noma} access and age of information
& \svns & \svyes & \svyes & \svyes & \svns & \svns & \svns
\\
\specialrule{0.25pt}{0.9pt}{0.9pt}

\cite{Machani2026PDNOMACSS}
& 2026
& \gls{css} and sharing with PD-\gls{noma}
& \svyes & \svns & \svns & \svns & \svns & \svns & \svns
\\
\specialrule{0.25pt}{0.9pt}{0.9pt}

\cite{Lu2026NOMAGrantFree}
& 2026
& Scheduled/grant-free \gls{noma} for time-critical and \gls{mmtc}
& \svns & \svyes & \svyes & \svyes & \svns & \svns & \svns
\\
\specialrule{0.35pt}{1.0pt}{1.2pt}


\multicolumn{10}{@{}l}{
\textit{VII-G: Massive Internet of Things (\gls{miot})}}
\\
\specialrule{0.35pt}{0.8pt}{1.0pt}

\cite{zhang2024mary}
& 2024
& $M$-ary multihop \gls{df} and reporting implementation
& \svyes & \svns & \svyes & \svns & \svns & \svyes & \svns
\\
\specialrule{0.25pt}{0.9pt}{0.9pt}

\cite{Wang2025OVFLCSS}
& 2025
& Online vertical federated learning for cooperative sensing
& \svyes & \svns & \svns & \svns & \svns & \svyes & \svyes
\\
\specialrule{0.25pt}{0.9pt}{0.9pt}

\cite{Li2026FedRMVL}
& 2026
& Uncertainty-aware federated multi-view information fusion
& \svyes & \svns & \svns & \svns & \svns & \svns & \svyes
\\

\bottomrule
\end{tabularx}

\vspace{3pt}

\begin{minipage}{\textwidth}
\scriptsize
\raggedright

\textit{Legend:}
\svyes\ = explicitly described for the article in Section~VII;
\svns\ = not specified for that article in the section.
A dash does not establish that the complete article excludes the feature.
\par
\smallskip

\textit{Criteria:}
\textbf{IF}: fusion of sensing reports, features, or perception outputs.
\textbf{AP}: activity detection, participation/activation, or
scheduling/update policies;
\textbf{CL}: channel estimation, link reliability, or signal impairments;
\textbf{TF}: latency, deadlines, or information freshness;
\textbf{EN}: energy, battery availability, or transmit-power constraints;
\textbf{RC}: quantization, compressed/task-relevant reporting,
over-the-air aggregation, or reduced raw-data exchange;
\textbf{EC}: edge, federated, or offloaded computation.
\end{minipage}
\endgroup
\end{table*}
\section{Challenges and Future Directions}
\label{sec:future_directions}

The preceding sections indicate that \gls{df} can no longer be
treated as an isolated detector-design problem. In emerging wireless
systems, the quality of the global decision is jointly determined by the
sensing process, the reporting links, the availability of the
participating devices, and the computational resources used at the
\gls{fc} or edge nodes. These quantities may also change from
one sensing interval to another because of mobility, random access,
energy limitations, blockage, and nonstationary interference. Two
research directions are therefore particularly important. The first is
the development of end-to-end performance metrics and optimization
techniques that expose the tradeoffs among detection accuracy, reliability,
latency, energy, communication load, and implementation cost. The second
is the design of \gls{ml}-based fusion algorithms whose
convergence, generalization, and computational burden remain acceptable
when training and inference are distributed across heterogeneous
wireless devices.
\subsection{Performance-Metric Optimization}
\label{subsec:performance_metric_optimization}
The performance of a \gls{dfr} is commonly described using
the global probability of detection, global probability of false alarm,
probability of decision error, receiver \Gls{roc},
\gls{auroc}, or deflection coefficient. These
quantities remain fundamental, but a practical fusion system cannot
usually be assessed using one detection metric alone. Increasing the
number of participating sensors may improve the statistical separation
between the hypotheses, while simultaneously increasing access
contention, reporting delay, energy consumption, and \gls{fc} processing.
Likewise, a more conservative threshold can reduce false alarms at the
expense of additional missed detections. Recent work on multihop M-ary
\gls{df} and communication-efficient cooperative sensing further
shows that the implementation architecture and reporting mechanism can
materially change the achievable detection performance
\cite{varshney2012distributed,zhang2024mary,
Yi2025SemanticAirCompCSS}. Consequently, the relevant operating point is
application dependent rather than universally optimal.

A useful starting point is the application-dependent Bayesian detection
risk. Let $\boldsymbol{\theta}$ collect the local sensing thresholds,
sensor-activation variables, reporting parameters, fusion weights, and
global decision threshold. For binary \gls{df}, the detection
risk, with zero correct-decision costs, can be written as
\begin{equation}
\begin{aligned}
\mathcal R_{\mathrm{det}}(\boldsymbol\theta)
={}&\pi_0C_{10}P_{\mathrm{fa}}^{(g)}(\boldsymbol\theta)+\pi_1C_{01}\bigl[1-P_{\mathrm d}^{(g)}(\boldsymbol\theta)\bigr],
\end{aligned}
\label{eq:future_detection_risk}
\end{equation}
where $\pi_j=\Pr(\mathcal{H}_j)$, $\pi_0+\pi_1=1$, and the global events are
$P_{\mathrm{fa}}^{(g)}=\Pr(\widehat {\mathcal{H}}=\mathcal{H}_1\mid \mathcal{H}_0)$ and
$P_{\mathrm d}^{(g)}=\Pr(\widehat{\mathcal{H}}=\mathcal{H}_1\mid \mathcal{H}_1)$. Moreover, $C_{10}$ is the cost of deciding $\mathcal{H}_1$
when $\mathcal{H}_0$ is true, whereas $C_{01}$ is the cost of deciding $\mathcal{H}_0$ when
$\mathcal{H}_1$ is true. This formulation is more meaningful than an unweighted
error probability when the consequences of false alarms and missed
detections are different. For example, a missed alarm may dominate the
design of an industrial protection or healthcare system, while false
alarms may have a larger operational cost in other applications. Recent
adaptive-threshold decision-fusion techniques similarly combine event
detection with a minimum Bayesian decision loss, illustrating that the
threshold and the fusion cost should be selected together rather than
independently
\cite{Acharya2014NeymanPearson,Huang2024AdaptiveThresholdDF}.
When the priors or decision costs cannot be estimated reliably, a
\gls{np} formulation remains preferable because it maximizes
$P_{\mathrm d}^{(g)}$ subject to a prescribed constraint on $P_{\mathrm{fa}}^{(g)}$.

The evaluation metric should also include the complete reporting
process. Let $\mathcal K=\{1,\ldots,K\}$ denote the deployed sensors,
let $\mathsf D$ be the random set of accepted reports, and let
$A\subseteq\mathcal K$ be a realization. Define
\begin{equation}
  e_{j,A}(\boldsymbol\theta)
=\Pr(\widehat{\mathcal{H}}\ne \mathcal{H}_j\mid \mathcal{H}_j,\mathsf D=A;\boldsymbol\theta).  
\end{equation}
An explicit formulation that also permits hypothesis-dependent report
availability is
\begin{equation}
\begin{aligned}
P_{\mathrm e,\mathrm{e2e}}^{(g)}(\boldsymbol\theta)
={}&\sum_{j=0}^{1}\pi_j\sum_{A\subseteq\mathcal K}
\Pr(\mathsf D=A\mid \mathcal{H}_j;\boldsymbol\theta)\times e_{j,A}(\boldsymbol\theta),
\end{aligned}
\label{eq:future_e2e_error}
\end{equation}
This expression averages the conditional fusion error over both the
hypotheses and all available-report sets, including the empty set with
a specified fallback decision. It therefore includes scheduling,
packet loss, multihop forwarding errors, energy shortage, blockage,
and random-access failure through the availability and conditional-error
models. It is especially relevant to
multihop and over-the-air implementations, where the \gls{dfr} and the
communication mechanism are coupled
\cite{zhang2024mary,Yi2025SemanticAirCompCSS}. When activity depends on the true hypothesis, the two conditional
availability laws must be retained. When availability is independent of
the hypothesis, the expression reduces to the usual availability-weighted
error with the original priors. Evaluating only the intervals in which
all reports are received would instead give an optimistic view of the
deployed system.

For emerging wireless networks, a more complete assessment can be
organized through a vector of normalized objectives:
\begin{equation}
\!\!\!\!\boldsymbol J\!(\boldsymbol\theta)\!\!=\!\!\!\biggl[P_{\mathrm e,\mathrm{e2e}}^{(g)}(\boldsymbol\theta),
\frac{\overline E(\boldsymbol\theta)}{E_{\mathrm{ref}}},
\frac{\overline T_{\mathrm{e2e}}(\boldsymbol\theta)}{T_{\mathrm{ref}}},\frac{\overline\Delta(\boldsymbol\theta)}{\Delta_{\mathrm{ref}}},
\frac{\overline C_{\mathrm{com}}^{\mathrm{bit}}(\boldsymbol\theta)}
{C_{\mathrm{com,ref}}^{\mathrm{bit}}},
\frac{\mathcal C_{\mathrm{op}}(\boldsymbol\theta)}{C_{\mathrm{op,ref}}}
\biggr]^{\mathsf T}
\label{eq:future_metric_vector}
\end{equation}
where $\overline E$ is average sensing, processing, and reporting
energy per global decision; $\overline T_{\mathrm{e2e}}$ is average
end-to-end decision latency; $\overline\Delta$ is average information
age; $\overline C_{\mathrm{com}}^{\mathrm{bit}}$ is average communication
load in bits per decision; and $\mathcal C_{\mathrm{op}}$ is an
operation-count measure per decision. Averages refer to the operating
distribution or a stated long-term regime. Every reference quantity is
strictly positive and has the same units as its numerator, making the
components of \eqref{eq:future_metric_vector} dimensionless.
Without normalization, a weighted objective may be dominated by
the numerical range or physical unit of one component rather than by
its actual importance. This point becomes more significant when
semantic compression or over-the-air aggregation is employed, because
a reduction in transmitted information may alter both the communication
cost and the statistical quality of the fused result
\cite{Yi2025SemanticAirCompCSS}.

A scalar objective can be obtained as
\begin{equation}
J_{\boldsymbol\omega}(\boldsymbol\theta)
=\boldsymbol\omega^{\mathsf T}\boldsymbol J(\boldsymbol\theta),
\qquad\omega_i\geq0,\qquad\sum\nolimits_{i=1}^{6}\omega_i=1,
\label{eq:future_weighted_metric}
\end{equation}
where $\boldsymbol\omega$ contains application-dependent weights. For
an explicit mean-resource-budget interpretation, a constrained design is
\begin{equation}
\underset{\boldsymbol{\theta }}{\mathrm{min}}\quad J_{\boldsymbol{\omega }}(%
\boldsymbol{\theta })  \label{eq:future_metric_optimization}
\end{equation}%
subject to: 
\begin{tabular}{ll}
$\text{a) }P_{\mathrm{fa}}^{(g)}(\boldsymbol{\theta })\leq \alpha _{\mathrm{%
FA}}$ & $\text{b) }P_{\mathrm{d}}^{(g)}(\boldsymbol{\theta })\geq \beta _{%
\mathrm{D}}$ \\ 
$\text{c) }\overline{T}_{\mathrm{e2e}}(\boldsymbol{\theta })\leq T_{\max }$
& $\text{d) }\overline{E}(\boldsymbol{\theta })\leq E_{\max }$ \\ 
$\text{e) }\overline{C}_{\mathrm{com}}^{\mathrm{bit}}(\boldsymbol{\theta }%
)\leq C_{\mathrm{com,max}}^{\mathrm{bit}}$ & $\text{f) }\mathcal{C}_{\mathrm{%
op}}(\boldsymbol{\theta })\leq C_{\mathrm{op,max}}$%
\end{tabular}
Here, $\alpha_{\mathrm{FA}}$ and $\beta_{\mathrm D}$ specify the global
false-alarm and detection requirements. The remaining bounds are average
latency, average energy, average communicated bits, and computational
budgets. A bound on average latency does not guarantee a deadline for
every realization. Applications requiring that guarantee must use a
per-decision bound or an explicit tail constraint instead.
The difficulty in \eqref{eq:future_metric_optimization} is not limited
to solving the mathematical program. The weights and constraints must
also reflect the actual application. In an ultra-reliable low-latency
system, the decision deadline and the tail of the error distribution may
dominate the average energy cost. In an energy-harvesting network,
energy neutrality over many sensing intervals may be more important than
minimizing the error in a single interval. Environmental monitoring may
tolerate moderate delay but require unattended operation over a long
period. Therefore, a fixed set of weights should not be reported as
generally optimal without explaining the operating assumptions under
which it was selected.

A Pareto-based treatment is often more informative than a single
weighted solution. Rather than fixing $\boldsymbol{\omega}$ in advance,
the designer can determine a set of non-dominated operating points and
then select a solution according to the application requirements. This
makes it possible to quantify, for example, how much additional energy,
latency, or communication is required to achieve a given reduction in
global decision error. The approach is particularly useful for
mixed-integer and nonconvex designs involving sensor selection,
quantization levels, relay assignment, aggregation frequency, and
threshold adaptation. Recent hierarchical federated-learning work
illustrates how Pareto optimization can expose the joint tradeoff among
model accuracy, communication cost, and convergence time, which is
closely related to the resource-aware design of learned fusion systems
\cite{Deb2002NSGAII,Manjang2025ParetoHFL}. Nevertheless, the
computational cost and sensitivity of the Pareto-search procedure should
also be reported.

Another challenge is the absence of exact probability models during the design. Local detection probabilities, sensor correlations,
channel distributions, prior probabilities, and packet-error rates are
typically estimated from finite data and may drift after deployment. A
rule optimized for one nominal model can therefore perform poorly when
the operating environment changes. A robust alternative is to optimize
the worst-case risk over an uncertainty set $\mathcal{P}$:
\begin{equation}
\mathcal R_{\mathrm{rob}}(\boldsymbol\theta)
=\sup_{\mathbb Q\in\mathcal P}
\mathcal R_{\mathrm{det}}(\boldsymbol\theta;\mathbb Q),
\label{eq:future_robust_risk}
\end{equation}
where $\mathbb Q$ denotes a possible data-generating distribution. The protection offered by the criterion depends on how well the uncertainty set represents deployment conditions.
Depending on the available information, $\mathcal{P}$ may be defined
using bounds on moments, relative entropy, sensor operating
characteristics, or channel parameters. Robust optimization generally
sacrifices part of the nominal performance, but it can prevent a much
larger degradation under model mismatch \cite{Levy2009RobustHT}. A
complementary data-driven approach is to estimate predictive uncertainty
explicitly. Recent uncertainty-quantification results for wireless
signal classification show the value of evaluating predictive
distributions under low-\gls{snr} and out-of-distribution conditions rather
than reporting point predictions alone \cite{Yang2026UQAMC}.

The metric should also reflect the statistical uncertainty of the
reported performance. A very small estimated error probability obtained
from a limited test set may not provide sufficient evidence for a
safety-critical system. Confidence intervals, repeated-trial
variability, or statistically justified upper bounds should therefore
accompany the mean performance. This is particularly important for
rare-event detection and \gls{urllc}-like operation, where direct Monte Carlo
evaluation of extremely small error probabilities may require an
impractically large number of samples. Recent \gls{urllc} optimization work
distinguishes an average packet-error constraint from a constraint on
the probability that the packet error exceeds a prescribed level,
highlighting the importance of the tail rather than only the mean
\cite{Zhang2024URLLCUnsupervised}. Similar chance-constrained or
risk-sensitive criteria are promising for \gls{df} when
occasional large errors are unacceptable.

For \gls{ml}-based \glspl{dfr}, classification accuracy and \gls{auroc} should be
supplemented by probability-calibration and uncertainty measures. A
classifier may rank observations correctly while assigning confidence
values that are systematically larger than its true probability of
correctness. Since these confidence values are often used as soft fusion
inputs or reliability weights, poor calibration can make the \gls{fc}
unjustifiably confident. Reliability diagrams, negative log-likelihood,
the Brier score, expected calibration error, and uncertainty under
distribution shift are therefore useful complementary measures
\cite{Guo2017Calibration,Yang2026UQAMC}. Calibration should be examined
separately for different sensing \gls{snr}, reporting-channel qualities,
device populations, and event frequencies.

Fair comparison also requires a common experimental protocol. At a
minimum, the sensing \gls{snr}, reporting-channel model, channel-state
information assumptions, sensor correlation, device-activity process,
packet-loss model, energy budget, latency definition, and
threshold-selection method should be stated explicitly. For learned
techniques, the separation among training, validation, and test conditions
should be clear, and the test set should include operating points not
encountered during training. Recent multihop decision-fusion and
integrated semantic AirComp studies provide useful examples of
evaluating not only detection performance but also implementation
structure, reporting robustness, and scalability
\cite{zhang2024mary,Yi2025SemanticAirCompCSS}. Such end-to-end
reporting is more informative than presenting one accuracy value under
a single nominal scenario.

\subsection{Convergence and Complexity of ML Algorithms for \gls{df}}
\label{subsec:ml_convergence_complexity}
ML-based \gls{df} is attractive because it can learn nonlinear
relationships among heterogeneous sensor reports without requiring a
complete analytical description of the sensing and reporting channels.
However, the practical value of a learned \gls{dfr} depends on whether
the model can be trained reliably and executed within the resource
budget of the \gls{fc}, edge node, or sensor. Training accuracy alone does not
reveal the number of samples, iterations, communication rounds, or
energy units required to obtain the model. Similarly, a small training
loss does not guarantee that the \gls{dfr} will maintain its detection
performance under a different \gls{snr}, channel distribution, sensor
topology, or device-activity pattern. Recent wireless federated-learning
studies make this coupling explicit by relating convergence to
aggregation error, data heterogeneity, compression, and channel
resources
\cite{Wang2023GNNRISFL,Sami2024AirCluster,Oh2024FedVQCS}.

Let $\mathcal D_{\mathrm{tr}}=
\{(\boldsymbol y_n,\mathcal{H}_n^{\mathrm{lab}})\}_{n=1}^{N_{\mathrm{tr}}}$
denote the training set. Here $\boldsymbol y_n$ is a sensing or reporting
feature vector, and $\mathcal{H}_n^{\mathrm{lab}}$ is its hypothesis label. A
regularized empirical objective and stochastic update are
\begin{equation}
\label{eq:future_ml_training}
 L(\boldsymbol\vartheta)
=\frac{1}{N_{\mathrm{tr}}}\sum\nolimits_{n=1}^{N_{\mathrm{tr}}}
\ell\!\left(f_{\boldsymbol\vartheta}(\boldsymbol y_n),\mathcal{H}_n^{\mathrm{lab}}\right)
+\lambda_{\Omega}\Omega(\boldsymbol\vartheta),
\end{equation}
\begin{equation}
\boldsymbol\vartheta^{(i+1)}
=\boldsymbol\vartheta^{(i)}
-\eta_i\widehat{\nabla\mathcal L}\!
\left(\boldsymbol\vartheta^{(i)}\right).
\label{eq:future_ml_sgd}
\end{equation}
where $\boldsymbol\vartheta$ contains the model parameters,
$\ell(\cdot)$ is the sample loss, $\Omega(\cdot)$ is a regularizer, and
$\eta_i$ is the learning rate. For a nonempty uniformly sampled
minibatch $\mathcal I_i$, one gradient estimator is
\begin{multline}
\widehat{\nabla\mathcal L}(\boldsymbol\vartheta)
=\frac{1}{|\mathcal I_i|}\sum\nolimits_{n\in\mathcal I_i}
\nabla_{\boldsymbol\vartheta}
\ell\!\left(f_{\boldsymbol\vartheta}(\boldsymbol y_n),\mathcal{H}_n^{\mathrm{lab}}\right)+\lambda_{\Omega}\nabla\Omega(\boldsymbol\vartheta),
\end{multline}
under differentiability of the displayed objective.

For a nonconvex deep model, convergence to a global optimum is generally
not a realistic requirement. A more useful criterion is approximate
first-order stationarity:
\begin{equation}
\min_{0\leq i<I}\mathbb E\!\left[
\left\|\nabla\mathcal L(\boldsymbol\vartheta^{(i)})\right\|_2^2
\right]\leq\epsilon_{\mathrm{opt}},
\label{eq:future_ml_stationarity}
\end{equation}
where $I$ is the number of optimization iterations and
$\epsilon_{\mathrm{opt}}$ is the required tolerance. The expectation is
over optimization randomness. This is a stationarity criterion, not a
convergence theorem supplied by the formulation alone. Satisfying
\eqref{eq:future_ml_stationarity} only indicates that the optimizer has
reached an approximately stationary point. It does not guarantee a low
validation error, a calibrated confidence score, or compliance with the
required global false-alarm probability. Optimization convergence and
statistical generalization should therefore be reported separately.
This distinction is also important for communication-efficient
algorithms, for which an analytically valid convergence bound may
coexist with a practically unacceptable number of wireless rounds
\cite{Oh2024FedVQCS,Assaad2026CommEffFL}.

The convergence problem becomes more involved when training is
distributed across sensors or edge nodes. Let
$\mathcal D_k=\{(\boldsymbol y_{k,n},\mathcal{H}_{k,n}^{\mathrm{lab}})\}_{n=1}^{N_k}$
be node $k$'s local dataset, with $N_k\geq1$ for the nodes included in
the training objective. For compatible local models sharing one
parameter vector, a federated objective is
\begin{equation}
F(\boldsymbol\vartheta)
=\sum\nolimits_{k=1}^{K}\nu_k^{\mathrm{FL}}F_k(\boldsymbol\vartheta),
\end{equation}
\begin{equation}
F_k(\boldsymbol\vartheta)
=\frac{1}{N_k}\sum\nolimits_{n=1}^{N_k}
\ell\!\left(f_{\boldsymbol\vartheta}(\boldsymbol y_{k,n}),
\mathcal{H}_{k,n}^{\mathrm{lab}}\right),
\end{equation}
where $\nu_k^{\mathrm{FL}}\geq0$ and $\sum\nolimits_k\nu_k^{\mathrm{FL}}=1$.
The conventional choice $\nu_k^{\mathrm{FL}}=N_k/\sum\nolimits_jN_j$ weights
nodes according to dataset size.
For \gls{df}, this choice may be insufficient because a large
local dataset can contain highly correlated, outdated, or low-quality
observations. A smaller dataset may provide more complementary evidence
if it represents a different location, modality, or operating regime.
Reliability-aware, uncertainty-aware, or fairness-aware aggregation
weights are therefore promising. Recent federated multi-view learning
illustrates how local uncertainty and cross-view complementarity can be
incorporated into server-side information fusion when feature
dimensions and data quality differ across devices
\cite{Li2026FedRMVL}.

At communication round $\mathbf{r}$, let $\mathcal S_r$ denote participating
nodes and let $\boldsymbol\vartheta_k^{(r,I_k^{\mathrm{loc}})}$ be the
local model after $I_k^{\mathrm{loc}}$ updates. With
$s_r^{\mathrm{FL}}=\sum\nolimits_{k\in\mathcal S_r}\nu_k^{\mathrm{FL}}$, a
parameter-aggregation rule, including an empty-round fallback, is
\begin{equation}
\boldsymbol\vartheta^{(r+1)}=
\begin{cases}
\displaystyle\sum\nolimits_{k\in\mathcal S_r}
\frac{\nu_k^{\mathrm{FL}}}{s_r^{\mathrm{FL}}}
\boldsymbol\vartheta_k^{(r,I_k^{\mathrm{loc}})},
&s_r^{\mathrm{FL}}>0,\\[1ex]
\boldsymbol\vartheta^{(r)},&s_r^{\mathrm{FL}}=0
\end{cases},
\label{eq:future_federated_aggregation}
\end{equation}
This averaging requires equal-dimensional, aligned parameter vectors.
A zero participation weight retains the previous global model. Models
with different feature spaces or architectures, including general
vertical or multi-view schemes, require an explicitly compatible
aggregation mechanism rather than this parameter average.
The convergence of \eqref{eq:future_federated_aggregation} depends on the
learning-rate schedule, number of local updates, client participation,
data heterogeneity, and accuracy of the communicated updates. Increasing
$I_k^{\mathrm{loc}}$ reduces the frequency of global communication, but it can also
increase model drift when local data are \gls{noniid}. Clustered over-the-air
learning provides one recent response to this problem by training
several models for groups of devices with related data distributions and
by deriving convergence tradeoffs involving compression, channel
quality, and antenna resources \cite{Sami2024AirCluster}. Large-scale
experiments on real \gls{iot}--edge devices further show that computation and
communication costs can differ substantially across heterogeneous
hardware, which makes simulation-only complexity estimates incomplete
\cite{Wong2026IoTEdgeFL}.

Partial participation is not merely an implementation detail. Sensors
with depleted batteries, weak reporting links, limited memory, or long
processing times may participate less frequently than well-connected
nodes. If these devices observe a distinct part of the environment, the
learned model can become biased toward the conditions represented by
frequently selected clients. Asynchronous training can reduce the delay
caused by slow devices, but stale model updates may then degrade
convergence. \gls{uav}-enabled asynchronous federated learning has recently
quantified the influence of model asynchrony, device selection, and
communication errors on learning performance, illustrating why
scheduling and convergence analysis should be considered jointly
\cite{Zhai2025UAVAFL}. Future fusion systems should distinguish random
client subsampling from systematic participation bias and should account
for data novelty, class coverage, sensing reliability, energy
availability, and channel quality during client selection.

The wireless channel also becomes part of the learning algorithm when
gradients or model updates are transmitted over noisy links. A
simplified received-update model can be written as
\begin{equation}
\widetilde{\boldsymbol g}^{(r)}
=\sum\nolimits_{k\in\mathcal S_r}c_k^{\mathrm{agg},(r)}
\boldsymbol g_k^{\mathrm{grad},(r)}
+\boldsymbol n^{\mathrm{agg},(r)},
\label{eq:future_noisy_gradient}
\end{equation}
where $\boldsymbol g_k^{\mathrm{grad},(r)}$ is the local update,
$c_k^{\mathrm{agg},(r)}$ is an effective channel and power-control
coefficient, and $\boldsymbol n^{\mathrm{agg},(r)}$ is a
same-dimensional noise vector. The coefficients are not the binary
activation variables used in Section~\ref{sec:decision_fusion_emerging_technologies}.
For $s_r^{\mathrm{FL}}>0$, the intended aggregate is
\[
\boldsymbol g_{\mathrm{des}}^{(r)}
=\sum\nolimits_{k\in\mathcal S_r}
\frac{\nu_k^{\mathrm{FL}}}{s_r^{\mathrm{FL}}}
\boldsymbol g_k^{\mathrm{grad},(r)}.
\]
Unless the effective coefficients match these intended weights and the
noise has a suitable zero conditional mean,
\eqref{eq:future_noisy_gradient} can be biased relative to this target.
Gradient quantization,
sparsification, modulation errors, packet loss, and imperfect
synchronization introduce additional distortions. Recent work has
connected the accumulated aggregation error directly to a convergence
upper bound and used graph neural networks to reduce the complexity of
\gls{ris}-assisted resource optimization \cite{Wang2023GNNRISFL}. Other
studies show that useful convergence can still be achieved under
imperfect aggregation when device selection, power control,
quantization, and the wireless receiver are designed jointly
\cite{Sun2025RISFLImperfect}. Communication-efficient zero- and
first-order methods provide another direction by exchanging scalar
information and exploiting channel knowledge while retaining analytical
convergence guarantees under asynchronous operation
\cite{Assaad2026CommEffFL}. For sensing applications, semantic
communication and over-the-air computation can similarly reduce
reporting overhead while preserving the information needed for the
global detector \cite{Yi2025SemanticAirCompCSS}.

Computational complexity should be described using more than a single
parameter count. A useful implementation profile is
\begin{equation}
\boldsymbol c_{\mathrm{ML}}=\bigl[
N_{\mathrm{par}},N_{\mathrm{op}},M_{\mathrm{peak}},C_{\mathrm{tr}}^{\mathrm{bit}},T_{\mathrm{tr}},T_{\mathrm{inf}},E_{\mathrm{inf}}
\bigr]^{\mathsf T},
\label{eq:future_ml_complexity_vector}
\end{equation}
where $N_{\mathrm{par}}$ is the number of trainable parameters,
$N_{\mathrm{op}}$ is the number of arithmetic operations per inference,
$M_{\mathrm{peak}}$ is the peak memory requirement,
$C_{\mathrm{tr}}^{\mathrm{bit}}$ is the number of communicated training bits,
$T_{\mathrm{tr}}$ is the time required to reach a target validation
performance, $T_{\mathrm{inf}}$ is the inference latency, and
$E_{\mathrm{inf}}$ is the energy consumed per global decision. Parameters and operations are counts, memory is measured in bytes, communication in bits, time in seconds, and energy in joules. Reporting
only the number of parameters or floating-point operations can be
misleading because memory movement, communication, hardware parallelism,
and numerical precision also influence the actual execution time and
energy. Measurements on heterogeneous \gls{iot}--edge platforms are therefore
valuable for validating analytical complexity claims
\cite{Sze2017EfficientDNN,Wong2026IoTEdgeFL}.

This distinction is important when comparing centralized, edge-based,
and fully distributed learning. A centralized model may require fewer
total arithmetic operations than a collection of local models, while
incurring a much larger communication cost. Conversely, model
compression may reduce storage and reporting load but perturb the
distribution of the output scores and shift the operating threshold of
the detector. \gls{fedvqcs} demonstrates that dimensionality reduction and
vector quantization can substantially reduce update communication,
while its convergence still depends on the reconstruction error of the
global update \cite{Oh2024FedVQCS}. Imperfect-aggregation results also
indicate that increasing the number of quantization bits is not
automatically beneficial when the corresponding symbol-error rate is
considered \cite{Sun2025RISFLImperfect}. Consequently, a compressed
fusion model should be recalibrated and retested under the same
false-alarm, detection, channel, and latency conditions as the
uncompressed model.

Adaptive inference offers another promising direction. A small model or
a limited subset of sensors can first process an easy observation, while
additional sensors, deeper network layers, or a more complex \gls{dfr}
are activated only when the initial confidence is insufficient. This
conditional-computation strategy can reduce average energy and latency
without requiring every decision to use the worst-case resource budget.
Its success, however, depends on trustworthy uncertainty estimates. An
overconfident early classifier may terminate the process before
sufficient evidence has been collected. Deep-ensemble uncertainty
quantification under low-\gls{snr} and out-of-distribution conditions provides
a useful basis for designing such confidence-aware stopping mechanisms
\cite{Yang2026UQAMC}.

Calibration is equally important when the output of an \gls{ml} classifier is
interpreted as a likelihood, posterior probability, or sensor
reliability weight. A low classification error does not imply that the
reported confidence is statistically meaningful. Post-training
calibration, uncertainty-aware aggregation, validation over several
channel conditions, and adaptive threshold selection should therefore
be included in the deployment procedure
\cite{Guo2017Calibration,Yang2026UQAMC,Li2026FedRMVL}. For
\gls{np} applications, the final threshold should be determined
using an independent validation set and then verified under the expected
reporting-channel and sensor-availability conditions.

Nonstationarity introduces a further convergence problem after
deployment. The sensing environment, interference statistics, device
population, and event prevalence may change over time. Periodic
retraining can track these changes, but frequent updates consume energy
and communication resources and may overwrite information learned from
earlier conditions. Online and continual-learning techniques should
therefore balance adaptation speed against model stability. Asynchronous
training, adaptive client selection, and dynamically selected
aggregation intervals provide useful mechanisms, but they should be
evaluated in terms of both convergence and end-to-end decision risk
\cite{Zhai2025UAVAFL,Manjang2025ParetoHFL}.

Security and privacy also affect convergence. Faulty or malicious
devices may inject abnormal gradients, mislabeled observations, or
fabricated local models. A conventional average can then move the global
model toward an unreliable solution even when only a limited number of
clients are compromised. At the same time, \gls{noniid} data can make a
benign update appear statistically unusual. Recent Byzantine-robust and
privacy-preserving federated learning addresses this tension by
combining heterogeneous-gradient comparison with encrypted aggregation
\cite{Zhu2026BPFLH}. Nevertheless, distinguishing a malicious update
from a rare but valid sensing condition remains difficult, and an overly
aggressive defense may discard precisely the observations needed to
detect an uncommon event.

The learning algorithm and resource-allocation variables can be
optimized jointly. Let $\boldsymbol{z}$ collect the model architecture,
participating nodes, number of local updates, quantization resolution,
aggregation interval, and compression parameters. Detector calibration,
including the final threshold, is among the design variables.
A general formulation is
\begin{equation}
\label{eq:future_joint_ml_design}
\underset{\boldsymbol{\vartheta },\boldsymbol{z}}{\min}\quad 
\mathcal{L}_{\mathrm{val}}(\boldsymbol{\vartheta },\boldsymbol{z})+\lambda _{%
\mathrm{bit}}C_{\mathrm{tr}}^{\mathrm{bit}}(\boldsymbol{z})+\lambda _{%
\mathrm{en}}E_{\mathrm{tr}}(\boldsymbol{z})
\end{equation}
\begin{tabular}{ll}
subject to: & a) $P_{\mathrm{fa}}^{(g)}(\boldsymbol{\vartheta },\boldsymbol{z%
})\leq \alpha _{\mathrm{FA}}$ \\ 
b) $T_{\mathrm{inf}}(\boldsymbol{\vartheta },\boldsymbol{z})\leq T_{\max }$
& c) $M_{\mathrm{peak}}(\boldsymbol{\vartheta },\boldsymbol{z})\leq M_{\max }
$ \\ 
d) $N_{\mathrm{rnd}}(\boldsymbol{z})\leq N_{\mathrm{rnd}}^{\max }$ & e) $E_{%
\mathrm{inf}}(\boldsymbol{\vartheta },\boldsymbol{z})\leq E_{\max }$%
\end{tabular}
where $\mathcal L_{\mathrm{val}}$ is the validation loss,
$C_{\mathrm{tr}}^{\mathrm{bit}}$ counts communicated training bits,
$E_{\mathrm{tr}}$ is training energy, and $N_{\mathrm{rnd}}$ is the
number of communication rounds. These costs are evaluated for a stated
training protocol and stopping criterion. The weights
$\lambda_{\mathrm{bit}}$ and $\lambda_{\mathrm{en}}$ provide the units
or normalization needed to combine loss, bits, and joules. A false-alarm
constraint alone does not guarantee an application-specific detection
floor; such a floor must be imposed separately when required.
Some constraints in \eqref{eq:future_joint_ml_design}, particularly the
false-alarm requirement, may not be differentiable. A smooth surrogate
can be used during training, but the final operating point should be
checked using independent data and the actual fusion threshold. Recent
Pareto-based hierarchical learning, imperfect wireless aggregation, and
communication-efficient FL demonstrate that the architecture,
aggregation schedule, device selection, and physical-layer resources
should be optimized together rather than sequentially
\cite{Manjang2025ParetoHFL,Sun2025RISFLImperfect,
Assaad2026CommEffFL}.

In summary, the evaluation of an \gls{ml}-based \gls{dfr} should separate
optimization convergence, statistical generalization, confidence
calibration, robustness, and implementation complexity. Future
algorithms should be communication-aware, hardware-aware, \gls{noniid}-aware,
and resilient to partial participation, channel errors, and malicious
updates. The most useful model is not necessarily the one that achieves
the smallest unconstrained training loss, but the one that reaches a
reliable and calibrated operating point within the available sensing,
communication, energy, memory, and latency budgets.

\section{Conclusion}
\label{sec:conclusion}
This survey has reviewed decision-fusion techniques for wireless sensor networks and emerging communication systems, covering their models, architectures, reporting strategies, performance measures, and representative applications. Statistical, learning-based, game-theoretic, and heuristic approaches were discussed alongside advanced combining and likelihood-approximation methods. Their roles in \gls{mmtc}, \gls{urllc}, \gls{uav}-assisted communication, \gls{vlc}, \gls{eh}, \gls{noma}, and \gls{miot} were also examined. The reviewed studies highlight that the choice of a fusion strategy depends not only on detection performance but also on the information, communication, and computational resources required to achieve it. A central observation is that sensing quality, reporting reliability, and fusion processing should be considered jointly rather than optimized independently. \gls{ml} techniques offer useful alternatives to full likelihood-based processing, but their benefits depend on the operating assumptions, available data, and implementation constraints. Future work should therefore emphasize resource-aware design, uncertainty and confidence calibration, and robustness to heterogeneous sensors, missing reports, and changing wireless conditions. For learning-based fusion, convergence and generalization must also be assessed alongside training and inference costs. Common evaluation conditions and validation on realistic devices remain important for establishing whether algorithmic gains translate into practical benefits. Ultimately, an effective fusion system should deliver reliable and timely decisions within the resources available to the application.

\appendices

\section*{Appendix I: List of Acronyms}
\printnoidxglossary[type=\acronymtype, title={}]
\Nomenclature{\ensuremath{K}}{Number of sensors}
\Nomenclature{\ensuremath{k}}{Sensor index}
\Nomenclature{\ensuremath{T}}{Number of observations (samples) processed by each sensor}
\Nomenclature{\ensuremath{t}}{Time slot (sample) index}
\Nomenclature{\ensuremath{n_k}}{Additive white Gaussian noise at sensor $k$}
\Nomenclature{\ensuremath{\tau_k}}{Predefined local decision threshold of sensor $k$}
\Nomenclature{\ensuremath{\mathcal{N}(0,\sigma^2)}}{Real Gaussian distribution with zero mean and variance $\sigma^2$}
\Nomenclature{\ensuremath{\mathcal{CN}(0,\sigma^2)}}{Circularly symmetric complex Gaussian distribution with zero mean and variance $\sigma^2$}
\Nomenclature{\ensuremath{\sigma_n^2}}{Variance of the sensor noise $n_k$}
\Nomenclature{\ensuremath{\gamma_k}}{Signal-to-noise ratio (SNR) of sensor $k$}
\Nomenclature{\ensuremath{h_k}}{Fading coefficient of the channel from sensor $k$ to the FC}
\Nomenclature{\ensuremath{\phi_k}}{Additive noise at the FC input for sensor $k$}
\Nomenclature{\ensuremath{P_k}}{Average transmit power of the signal from sensor $k$}
\Nomenclature{\ensuremath{r_k}}{Signal received at the FC from sensor $k$}
\Nomenclature{\ensuremath{r}}{Vector of the signals received at the FC from all $K$ sensors}
\Nomenclature{\ensuremath{M}}{Number of relayed (multihop) paths per sensor}
\Nomenclature{\ensuremath{L}}{Number of relay groups; each relaying path has $L+1$ hops}
\Nomenclature{\ensuremath{m}}{Relayed path index}
\Nomenclature{\ensuremath{p}}{Branch index of the received signals at the FC}
\Nomenclature{\ensuremath{f(r|H_1)}}{Conditional joint probability density function (PDF) of $r$ given $H_1$}
\Nomenclature{\ensuremath{f(r|H_0)}}{Conditional joint probability density function (PDF) of $r$ given $H_0$}
\Nomenclature{\ensuremath{\tau_0}}{System-level threshold chosen to satisfy a certain $P_{FA}$}
\Nomenclature{\ensuremath{P_{FA}}}{Prescribed system-level false alarm probability}
\Nomenclature{\ensuremath{\lambda_0}}{Threshold of the log-likelihood ratio, $\lambda_0 = \log(\tau_0)$}

\Nomenclature{\ensuremath{\Lambda}}{Real fusion statistic}
\Nomenclature{\ensuremath{\lambda}}{Method-specific decision threshold}
\Nomenclature{\ensuremath{\Lambda_{CV}}}{Chair--Varshney (CV) log-likelihood ratio statistic}
\Nomenclature{\ensuremath{p_k}}{Reporting (transmit) power of sensor $k$}
\Nomenclature{\ensuremath{\Lambda_{MRC}}}{Sensing-weighted maximal-ratio combining (MRC) statistic}
\Nomenclature{\ensuremath{\Lambda_{EGC}}}{Equal-gain combining (EGC) statistic}
\Nomenclature{\ensuremath{\Lambda_{SC}}}{Selection combining (SC) statistic}
\Nomenclature{\ensuremath{\angle h_k}}{Phase of the channel coefficient $h_k$}
\Nomenclature{\ensuremath{y}}{Feature vector of sensing or reporting features}
\Nomenclature{\ensuremath{\vartheta}}{Parameters of the learning model (neural network)}
\Nomenclature{\ensuremath{\hat{H}}}{Decided (estimated) hypothesis}
\Nomenclature{\ensuremath{w_k}}{Fusion weight of sensor $k$}

\Nomenclature{\ensuremath{\Lambda_{ILS}}}{Ideal local sensors (ILS) statistic}
\Nomenclature{\ensuremath{\Lambda_{CMRC}}}{Cooperative MRC (CMRC) statistic}
\Nomenclature{\ensuremath{\Lambda_{LAR}}}{Link-adaptive regeneration (LAR) combiner statistic}
\Nomenclature{\ensuremath{\Lambda_{EMRC}}}{Enhanced MRC (EMRC) statistic}
\Nomenclature{\ensuremath{\Lambda_{CEGC}}}{Cooperative EGC (CEGC) phase-only statistic}
\Nomenclature{\ensuremath{\Lambda_{SMRC}}}{Selection MRC (SMRC) statistic}
\Nomenclature{\ensuremath{\Lambda_{MLLA}}}{Max-log with linear approximation (MLLA) statistic}
\Nomenclature{\ensuremath{\Lambda_{IR}}}{Ideal relays (IR) statistic}
\Nomenclature{\ensuremath{\Lambda_{CA}}}{Channel-aware plug-in log-likelihood ratio statistic}
\Nomenclature{\ensuremath{\hat{\theta}_{stat}}}{Estimated sensor, channel and transition parameters}
\Nomenclature{\ensuremath{\Phi}}{Configurable front end (reconfigurable surface) of the receiver}
\Nomenclature{\ensuremath{w}}{Fusion weight vector}
\Nomenclature{\ensuremath{\alpha_{FA}}}{Target (required) global false alarm probability}

\Nomenclature{\ensuremath{P_d}}{Probability of detection}
\Nomenclature{\ensuremath{P_{fa}}}{Probability of false alarm}
\Nomenclature{\ensuremath{P_e}}{Probability of fusion error}
\Nomenclature{\ensuremath{\Pr(H_0)}}{Prior probability of $H_0$}
\Nomenclature{\ensuremath{\Pr(H_1)}}{Prior probability of $H_1$}
\Nomenclature{\ensuremath{y_c}}{Linear combination of the sensor decisions at the FC}
\Nomenclature{\ensuremath{D(w)}}{Deflection coefficient}
\Nomenclature{\ensuremath{A(\gamma)}}{Area under the ROC curve (AUROC)}
\Nomenclature{\ensuremath{KL(P\|Q)}}{Kullback--Leibler divergence between distributions $P$ and $Q$}
\Nomenclature{\ensuremath{F_1}}{F1 score (harmonic mean of precision and recall)}

\Nomenclature{\ensuremath{\Lambda_{mMTC}(t)}}{mMTC reliability-aware fusion statistic}
\Nomenclature{\ensuremath{\Lambda_U(t)}}{URLLC partial fusion statistic at time $t$}
\Nomenclature{\ensuremath{\Lambda_{UAV}(t)}}{Aerial fusion statistic}
\Nomenclature{\ensuremath{\Lambda_{VLC}(t)}}{VLC reliability-aware fusion statistic}
\Nomenclature{\ensuremath{\Lambda_{EH}(t)}}{EH-aware fusion statistic}
\Nomenclature{\ensuremath{\Lambda_N(t)}}{Post-SIC reliability-aware fusion statistic}
\Nomenclature{\ensuremath{\Lambda_{mIoT}(t)}}{mIoT global fusion statistic}

\Nomenclature{\ensuremath{T_{e2e}}}{End-to-end decision latency}
\Nomenclature{\ensuremath{T_{max}}}{Maximum allowed end-to-end latency (deadline)}

\Nomenclature{\ensuremath{\beta_0}}{Channel power gain at the reference distance}
\Nomenclature{\ensuremath{d_0}}{Reference distance}
\Nomenclature{\ensuremath{\kappa}}{Air-to-ground path-loss exponent}

\Nomenclature{\ensuremath{\Psi_c}}{Field-of-view (FoV) angle of the photodetector}
\Nomenclature{\ensuremath{m_L}}{Lambertian order}

\Nomenclature{\ensuremath{Q}}{Number of clusters (edge nodes) in the mIoT network}
\Nomenclature{\ensuremath{\mathcal{C}_q}}{Set of sensors associated with edge node $q$}

\Nomenclature{\ensuremath{\theta}}{Design variables (local thresholds, activation, reporting parameters, weights and global threshold)}
\Nomenclature{\ensuremath{C_{10}}}{Cost of deciding $H_1$ when $H_0$ is true}
\Nomenclature{\ensuremath{C_{01}}}{Cost of deciding $H_0$ when $H_1$ is true}
\Nomenclature{\ensuremath{J(\theta)}}{Vector of normalized performance objectives}
\Nomenclature{\ensuremath{J_\omega(\theta)}}{Weighted scalar objective}
\Nomenclature{\ensuremath{\omega}}{Application-dependent weight vector of the objectives}
\Nomenclature{\ensuremath{\mathcal{P}}}{Uncertainty set of data-generating distributions}
\Nomenclature{\ensuremath{\mathcal{D}_{tr}}}{Training set}
\Nomenclature{\ensuremath{\mathcal{L}(\vartheta)}}{Regularized empirical training objective}
\Nomenclature{\ensuremath{c_{ML}}}{Implementation (complexity) profile vector of the ML model}

\vspace{5mm}
\renewcommand{\nomname}{Appendix II: List of Symbols}
\printnomenclature

\bibliographystyle{IEEEtran}
\bibliography{IEEE_abrev, references_all}

@STRING{IEEE_J_AES        = "{IEEE} Trans. Aerosp. Electron. Syst."}

@STRING{IEEE_J_ITS        = "{IEEE} Trans. Intell. Transp. Syst."}

@STRING{IEEE_J_VT         = "{IEEE} Trans. Veh. Technol."}

@STRING{IEEE_J_SPL        = "{IEEE} Signal Process. Lett."}

@STRING{IEEE_J_SP         = "{IEEE} Trans. Signal Process."}

@STRING{IEEE_J_COML       = "{IEEE} Commun. Lett."}

@STRING{IEEE_J_JSAC       = "{IEEE} J. Sel. Areas Commun."}

@STRING{IEEE_J_COM        = "{IEEE} Trans. Commun."}

@STRING{IEEE_J_WCOM       = "{IEEE} Trans. Wireless Commun."}

@STRING{IEEE_J_WCOML       = "{IEEE} Wireless Commun. Lett."}

@STRING{IEEE_J_IT         = "{IEEE} Trans. Inf. Theory"}

@STRING{IEEE_J_EVC        = "{IEEE} Trans. Evol. Comput."}

@STRING{IEEE_J_MC          = "{IEEE} Trans. Mobile Comput."}

@STRING{IEEE_J_NET        = "{IEEE/ACM} Trans. Netw."}

@STRING{IEEE_J_IP         = "{IEEE} Trans. Image Process."}

@STRING{IEEE_J_MM         = "{IEEE} Trans. Multimedia"}

@STRING{IEEE_J_SMCA       = "{IEEE} Trans. Syst., Man, Cybern. {A}, Syst. Humans"}

@STRING{IEEE_J_GRS        = "{IEEE} Trans. Geosci. Remote Sens."}

@STRING{IEEE_J_PROC       = "Proc. {IEEE}"}

@STRING{IEEE_J_CE         = "{IEEE} Trans. Consum. Electron."}

@STRING{IEEE_J_IM         = "{IEEE} Trans. Instrum. Meas."}

@STRING{IEEE_J_NS         = "{IEEE} Trans. Nucl. Sci."}

@STRING{IEEE_J_SENSOR     = "{IEEE} Sensors J."}

@STRING{IEEE_M_AES        = "{IEEE} Aerosp. Electron. Syst. Mag."}

@STRING{IEEE_M_COM        = "{IEEE} Commun. Mag."}

@STRING{IEEE_M_MM         = "{IEEE} Multimedia"}

@STRING{IEEE_M_NET        = "{IEEE} Netw."}

@STRING{IEEE_M_SP         = "{IEEE} Signal Process. Mag."}

@STRING{IEEE_J_CASII_EB                            = "{IEEE} Trans. Circuits Syst. {II}, Exp. Briefs"}

@STRING{IEEE_J_SYS                                 = "{IEEE} Syst. J."}

@STRING{IEEE_J_SIPN                                = "{IEEE} Trans. Signal Inf. Process. Netw."}

@STRING{IEEE_J_TASLP                               = "{IEEE/ACM} Trans. Audio, Speech, Language Process."}

@STRING{IEEE_J_ACCESS                              = "{IEEE} Access"}

@STRING{IEEE_J_JSTSP                               = "{IEEE} J. Sel. Topics Signal Process."}

@STRING{IEEE_J_JBHI                                = "{IEEE} J. Biomed. Health Inform."}

@STRING{IEEE_J_ICPS                                = "{IEEE} Trans. Ind. Cyber-Phys. Syst."}

@STRING{IEEE_J_II                                  = "{IEEE} Trans. Ind. Informat."}

@STRING{IEEE_J_TCCN                                = "{IEEE} Trans. Cogn. Commun. Netw."}

@STRING{IEEE_J_TIFS                                = "{IEEE} Trans. Inf. Forensics Security"}

@STRING{IEEE_J_IOT                                 = "{IEEE} Internet Things J."}

@STRING{IEEE_J_COMST                               = "{IEEE} Commun. Surveys Tuts."}

@STRING{IEEE_J_TGCN                                = "{IEEE} Trans. Green Commun. Netw."}

@STRING{IEEE_J_TNNLS                               = "{IEEE} Trans. Neural Netw. Learn. Syst."}

@STRING{IEEE_J_TDSC                                = "{IEEE} Trans. Dependable Secure Comput."}

@STRING{IEEE_M_COMSTD                              = "{IEEE} Commun. Stand. Mag."}

@STRING{IET_J_SP                                   = "{IET} Signal Process."}

@STRING{IET_J_COM                                  = "{IET} Commun."}

@STRING{J_COMPNET                                  = "Comput. Netw."}

@STRING{J_FUTURE_TRANSP                            = "Future Transp."}

@STRING{J_BIGDATA_RES                              = "Big Data Res."}

@STRING{J_APPL_EARTH_OBS                           = "Int. J. Appl. Earth Obs. Geoinf."}

@STRING{J_INF_FUSION                               = "Inf. Fusion"}

@STRING{J_WIRELESS_PERS_COM                        = "Wireless Pers. Commun."}

@STRING{J_EURASIP_ASP                              = "{EURASIP} J. Adv. Signal Process."}

@STRING{J_SIGNAL_PROCESS                           = "Signal Process."}

@STRING{J_SENSORS                                  = "Sensors"}

@STRING{J_WIRELESS_COM_NET                         = "J. Wireless Commun. Netw."}

@STRING{J_PHYS_COM                                 = "Phys. Commun."}

@STRING{J_COMPUTER_VISION                          = "Int. J. Comput. Vis."}

@STRING{J_CMC                                      = "Comput. Mater. Contin."}

@STRING{IEEE_C_ICC                                 = "Proc. {IEEE} Int. Conf. Commun. ({ICC})"}

@STRING{IEEE_C_GLOBECOM                            = "Proc. {IEEE} Global Commun. Conf. ({GLOBECOM})"}

@STRING{C_ISCIT                                    = "Proc. Int. Symp. Commun. Inf. Technol. ({ISCIT})"}

@STRING{IEEE_C_FUSION_24                           = "Proc. {IEEE} 24th Int. Conf. Inf. Fusion ({FUSION})"}

@STRING{C_ASET                                     = "Proc. Adv. Sci. Eng. Technol. Int. Conf. ({ASET})"}

@STRING{C_ICSI_5                                   = "Proc. 5th Int. Conf. Sensors Inf. Technol. ({ICSI})"}

@STRING{IEEE_C_DASC_PICOM_CBDCOM_CYBERSCITECH      = "Proc. {IEEE} {DASC/PiCom/CBDCom/CyberSciTech}"}

@STRING{C_FUSION_17                                = "Proc. 17th Int. Conf. Inf. Fusion ({FUSION})"}

@STRING{C_ICML_34                                  = "Proc. 34th Int. Conf. Mach. Learn. ({ICML})"}

@STRING{IEEE_C_IV                                  = "Proc. {IEEE} Intell. Veh. Symp. ({IV})"}

@ARTICLE{javadi2016,
  author={Javadi, S. Hamed},
  journal=IEEE_M_AES, 
  title={Detection over sensor networks: a tutorial}, 
  year={2016},
  volume={31},
  number={3},
  pages={2-18},
  doi={10.1109/MAES.2016.140128}}

@book{callaway2003wireless,
  title={Wireless sensor networks: architectures and protocols},
  author={Callaway, Jr., Edgar H.},
  year={2003},
  publisher={CRC press},
  isbn={9780849318238}
}

@book{shorey2005mobile,
  title={Mobile, wireless, and sensor networks: technology, applications, and future directions},
  editor={Shorey, Rajeev and Ananda, Akkihebbal L. and Chan, Mun Choon and Ooi, Wei Tsang},
  year={2005},
  publisher={John Wiley \& Sons}
}

@book{karl2007protocols,
  title={Protocols and architectures for wireless sensor networks},
  author={Karl, Holger and Willig, Andreas},
  year={2007},
  publisher={John Wiley \& Sons},
  isbn={9780470519233}
}

@book{dargie2011fundamentals,
  title={Fundamentals of wireless sensor networks: theory and practice},
  author={Dargie, Waltenegus and Poellabauer, Christian},
  year={2011},
  publisher={John Wiley \& Sons}
}

@article{hamdi2008bandwidth,
  title={Bandwidth-effective design of a satellite-based hybrid wireless sensor network for mobile target detection and tracking},
  author={Hamdi, Mohamed and Boudriga, Noureddine and Obaidat, Mohammad S},
  journal=IEEE_J_SYS,
  volume={2},
  number={1},
  pages={74--82},
  year={2008},
  publisher={IEEE}
}

@article{tan2010exploiting,
  author={Tan, Rui and Xing, Guoliang and Wang, Jianping and So, Hing Cheung},
  journal={IEEE_J_MC}, 
  title={Exploiting Reactive Mobility for Collaborative Target Detection in Wireless Sensor Networks}, 
  year={2010},
  volume={9},
  number={3},
  pages={317-332},
  doi={10.1109/TMC.2009.125}}

@article{amaldi2012design,
  author={Amaldi, Edoardo and Capone, Antonio and Cesana, Matteo and Filippini, Ilario},
  journal={IEEE_J_NET}, 
  title={Design of Wireless Sensor Networks for Mobile Target Detection}, 
  year={2012},
  volume={20},
  number={3},
  pages={784-797},
  doi={10.1109/TNET.2011.2175746}}

@article{you2014new,
  title={New method for target identification in a foliage environment using selected bispectra and chaos particle swarm optimisation-based support vector machine},
  author={You, Minglei and Jiang, Ting},
  journal=IET_J_SP,
  volume={8},
  number={1},
  pages={76--84},
  year={2014},
  publisher={IET}
}

@article{yu2016consensus,
   author={Yu, Yihua},
  journal={IEEE_J_VT}, 
  title={Consensus-Based Distributed Mixture Kalman Filter for Maneuvering Target Tracking in Wireless Sensor Networks}, 
  year={2016},
  volume={65},
  number={10},
  pages={8669-8681},
  doi={10.1109/TVT.2015.2508456}}

@article{yu2017distributed,
  title={Distributed target tracking in wireless sensor networks with data association uncertainty},
  author={Yu, Yihua},
  journal=IEEE_J_COML,
  volume={21},
  number={6},
  pages={1281--1284},
  year={2017},
  publisher={IEEE}
}

@article{larios2012energy,
  title={Energy efficient wireless sensor network communications based on computational intelligent data fusion for environmental monitoring},
  author={Larios, Diego Francisco and Barbancho, Julio and Rodr{\'\i}guez, Gustavo and Sevillano, Jos{\'e} Luis and Molina, Francisco Javier and Le{\'o}n, Carlos},
  journal=IET_J_COM,
  volume={6},
  number={14},
  pages={2189--2197},
  year={2012},
  publisher={IET}
}

@article{chen2017cooperative,
  title={Cooperative spectrum sensing with M-ary quantized data in cognitive radio networks under SSDF attacks},
  author={Chen, Huifang and Zhou, Ming and Xie, Lei and Li, Jie},
 journal=IEEE_J_WCOM,
  volume={16},
  number={8},
  pages={5244--5257},
  year={2017},
  publisher={IEEE}
}

@article{tenney1981detection,  title={Detection with distributed sensors},
  author={Tenney, Robert R. and Sandell, Jr., Nils R.},
  journal=IEEE_J_AES, volume={AES-17}, number={4},
  pages={501--510}, month=jul, year={1981}}

@article{chair1986optimal,title={Optimal data fusion in multiple sensor detection systems},
  author={Chair, Z. and Varshney, Pramod K.},
  journal=IEEE_J_AES, volume={AES-22}, number={1},
  pages={98--101}, month=jan, year={1986}}

@book{varshney2012distributed,
  title={Distributed detection and data fusion},
  author={Varshney, Pramod K},
  year={2012},
  publisher={Springer Science \& Business Media},
  isbn={9781461273332}
}

@article{maleki2013censored,
  title={Censored truncated sequential spectrum sensing for cognitive radio networks},
  author={Maleki, Sina and Leus, Geert},
  journal=IEEE_J_JSAC,
  volume={31},
  number={3},
  pages={364--378},
  year={2013},
  publisher={IEEE}
}

@article{falco2026,
title = {The evolution of Dynamic Spectrum Sensing: A two-decade survey from foundations to frontiers},
journal = J_COMPNET,
volume = {278},
pages = {112095},
year = {2026},
issn = {1389-1286},
doi = {https://doi.org/10.1016/j.comnet.2026.112095},
author = {Mariana Falco and Antonio Scarvaglieri and Fabio Busacca and Farzam Nosrati and Daniele Croce}}

@article{niu2006fusion,
  title={Fusion of decisions transmitted over Rayleigh fading channels in wireless sensor networks},
  author={Niu, Ruixin and Chen, Biao and Varshney, Pramod K},
  journal=IEEE_J_SP,
  volume={54},
  number={3},
  pages={1018--1027},
  year={2006},
  publisher={IEEE}
}

@article{zhang2009optimization,
  title={Optimization of cooperative spectrum sensing with energy detection in cognitive radio networks},
  author={Zhang, Wei and Mallik, Ranjan K and Letaief, Khaled Ben},
  journal=IEEE_J_WCOM,
  volume={8},
  number={12},
  pages={5761--5766},
  year={2009},
  publisher={IEEE}
}

@article{lei2010multiple,
  title={Multiple-symbol differential decision fusion for mobile wireless sensor networks},
  author={Lei, Andre and Schober, Robert},
  journal=IEEE_J_WCOM,
  volume={9},
  number={2},
  pages={778--790},
  year={2010},
  publisher={IEEE}
}

@article{poh2010quality,
  title={Quality-based score normalization with device qualitative information for multimodal biometric fusion},
  author={Poh, Norman and Kittler, Josef and Bourlai, Thirimachos},
  journal=IEEE_J_SMCA,
  volume={40},
  number={3},
  pages={539--554},
  year={2010},
  publisher={IEEE}
}

@article{lai2010fusion,
  title={Fusion of decisions transmitted over flat fading channels via maximizing the deflection coefficient},
  author={Lai, Kuei-Chiang and Yang, Yun-Lin and Jia, Jiun-Jie},
  journal=IEEE_J_VT,
  volume={59},
  number={7},
  pages={3634--3640},
  year={2010},
  publisher={IEEE}
}

@article{chen2011maximum,
  title={A maximum likelihood approach to joint image registration and fusion},
  author={Chen, Siyue and Guo, Qing and Leung, Henry and Bosse, Eloi},
  journal=IEEE_J_IP,
  volume={20},
  number={5},
  pages={1363--1372},
  year={2011},
  publisher={IEEE}
}

@article{ahmadi2011distributed,
  title={Distributed detection with adaptive topology and nonideal communication channels},
  author={Ahmadi, Hamid R and Vosoughi, Azadeh},
  journal=IEEE_J_SP,
  volume={59},
  number={6},
  pages={2857--2874},
  year={2011},
  publisher={IEEE}
}

@article{cheng2012performance,
  title={Performance analysis of distributed decision fusion using a multilevel censoring scheme in wireless sensor networks},
  author={Cheng, Victor W and Wang, Tsang-Yi},
  journal=IEEE_J_VT,
  volume={61},
  number={4},
  pages={1610--1619},
  year={2012},
  publisher={IEEE}
}

@article{Ciuonzo2014UnknownPd,
  author  = {Ciuonzo, D. and Salvo Rossi, Pierluigi},
  title   = {Decision Fusion With Unknown Sensor Detection
             Probability},
  journal = IEEE_J_SPL,
  volume  = {21},
  number  = {2},
  pages   = {208--212},
  month   = feb,
  year    = {2014},
  doi     = {10.1109/LSP.2013.2295054}
}

@article{minotto2014simultaneous,
  title={Simultaneous-speaker voice activity detection and localization using mid-fusion of SVM and HMMs},
  author={Minotto, Vicente P and Jung, Claudio R and Lee, Bowon},
  journal=IEEE_J_MM,
  volume={16},
  number={4},
  pages={1032--1044},
  year={2014},
  publisher={IEEE}
}

@article{rossi2015hmm,
  title={HMM-based decision fusion in wireless sensor networks with noncoherent multiple access},
  author={Salvo Rossi, Pierluigi and Ciuonzo, Domenico and Ekman, Torbj{\"o}rn},
  journal=IEEE_J_COML,
  volume={19},
  number={5},
  pages={871--874},
  year={2015},
  publisher={IEEE}
}

@article{goel2016selection,
  title={Selection relaying in decode-and-forward multi-hop cognitive radio systems using energy detection},
  author={Goel, Dhanesh and Krishna, Vemuri Sai and Bhatnagar, Manav},
  journal=IET_J_COM,
  volume={10},
  number={7},
  pages={753--760},
  year={2016},
  publisher={IET}
}

@article{peng2016multimodal,
  title={Multimodal ensemble fusion for disambiguation and retrieval},
  author={Peng, Yang and Zhou, Xiaofeng and Wang, Daisy Zhe and Patwa, Ishan and Gong, Dihong and Fang, Chunsheng},
  journal=IEEE_M_MM,
  year={2016},
  publisher={IEEE}
}

@article{nurellari2016distributed,
  title={Distributed two-step quantized fusion rules via consensus algorithm for distributed detection in wireless sensor networks},
  author={Nurellari, Edmond and McLernon, Des and Ghogho, Mounir},
  journal=IEEE_J_SIPN,
  volume={2},
  number={3},
  pages={321--335},
  year={2016},
  publisher={IEEE}
}

@article{rossi2016performance,
  title={Performance analysis of energy detection for MIMO decision fusion in wireless sensor networks over arbitrary fading channels},
  author={Salvo Rossi, Pierluigi and Ciuonzo, Domenico and Kansanen, Kimmo and Ekman, Torbj{\"o}rn},
 journal=IEEE_J_WCOM,
  volume={15},
  number={11},
  pages={7794--7806},
  year={2016},
  publisher={IEEE}
}

@article{stefanakis2017perpendicular,
  title={Perpendicular cross-spectra fusion for sound source localization with a planar microphone array},
  author={Stefanakis, Nikolaos and Pavlidi, Despoina and Mouchtaris, Athanasios},
  journal=IEEE_J_TASLP,
  volume={25},
  number={9},
  pages={1821--1835},
  year={2017},
  publisher={IEEE}
}

@article{zhou2017hyperspectral,
  title={Hyperspectral and multispectral image fusion based on local low rank and coupled spectral unmixing},
  author={Zhou, Yuan and Feng, Liyang and Hou, Chunping and Kung, Sun-Yuan},
  journal=IEEE_J_GRS,
  volume={55},
  number={10},
  pages={5997--6009},
  year={2017},
  publisher={IEEE}
}

@article{galluccio2004analytical,
  title={Analytical evaluation of a tradeoff between energy efficiency and responsiveness of neighbor discovery in self-organizing ad hoc networks},
  author={Galluccio, Laura and Morabito, Giacomo and Palazzo, Sergio},
  journal=IEEE_J_JSAC,
  volume={22},
  number={7},
  pages={1167--1182},
  year={2004},
  publisher={IEEE}
}

@article{saleem2014empirical,
  title={Empirical studies of bio-inspired self-organized secure autonomous routing protocol},
  author={Saleem, Kashif and Fisal, Norsheila and Al-Muhtadi, Jalal},
  journal=IEEE_J_SENSOR,
  volume={14},
  number={7},
  pages={2232--2239},
  year={2014},
  publisher={IEEE}
}

@article{wang2014self,
  title={Self-organizing map neural network-based nearest neighbor position estimation scheme for continuous crystal PET detectors},
  author={Wang, Yonggang and Li, Deng and Lu, Xiaoming and Cheng, Xinyi and Wang, Liwei},
  journal=IEEE_J_NS,
  volume={61},
  number={5},
  pages={2446--2455},
  year={2014},
  publisher={IEEE}
}

@ARTICLE{park2015multiband,
  author={Park, Hyunhee and Kim, Yongsun and Song, Taewon and Pack, Sangheon},
  journal=IEEE_J_VT, 
  title={Multiband Directional Neighbor Discovery in Self-Organized mmWave Ad Hoc Networks}, 
  year={2015},
  volume={64},
  number={3},
  pages={1143-1155},
  doi={10.1109/TVT.2014.2329303}}

@inproceedings{da2009performance,
  title={Performance of cooperative diversity networks: Analysis of amplify-and-forward relaying under equal-gain and maximal-ratio combining},
  author={da Costa, Daniel Benevides and A{\"\i}ssa, Sonia},
  booktitle=IEEE_C_ICC,
  pages={1--5},
  year={2009},
  organization={IEEE}
}

@article{ionita2014detectors,
  title={Detectors for Estimate-and-Forward Wireless Relay Networks},
  author={Ionita, Corina I and Lilleberg, Jorma and Aazhang, Behnaam},
 journal=IEEE_J_WCOM,
  volume={13},
  number={11},
  pages={6342--6355},
  year={2014},
  publisher={IEEE}
}

@article{bai2015performance,
  title={Performance analysis of SNR-based incremental hybrid decode-amplify-forward cooperative relaying protocol},
  author={Bai, Zhiquan and Jia, Jianlan and Wang, Cheng-Xiang and Yuan, Dongfeng},
  journal=IEEE_J_COM,
  volume={63},
  number={6},
  pages={2094--2106},
  year={2015},
  publisher={IEEE}
}

@article{feng2017tas,
  title={TAS-based incremental hybrid decode--amplify--forward relaying for physical layer security enhancement},
  author={Feng, Youhong and Yan, Shihao and Yang, Zhen and Yang, Nan and Zhu, Wei-Ping},
  journal=IEEE_J_COM,
  volume={65},
  number={9},
  pages={3876--3891},
  year={2017},
  publisher={IEEE}
}

@inproceedings{sharma2009performance,
  title={Performance analysis of maximum likelihood decode and forward cooperative systems in rayleigh fading},
  author={Sharma, G. V. V. and Ganwani, Vijay and Desai, Uday B and Merchant, S. N.},
  booktitle=IEEE_C_ICC,
  pages={1--5},
  year={2009},
  organization={IEEE}
}

@article{sharma2010performance,
  title={Performance analysis of maximum likelihood detection for decode and forward MIMO relay channels in Rayleigh fading},
  author={Sharma, G. V. V. and Ganwani, Vijay and Desai, Uday B and Merchant, Shabbir N},
 journal=IEEE_J_WCOM,
  volume={9},
  number={9},
  pages={2880--2889},
  year={2010},
  publisher={IEEE}
}

@ARTICLE{sharma2011conditionally,
  author={Sharma, G. V. V. and Desai, Uday B. and Merchant, S. N.},
  journal=IEEE_J_WCOM, 
  title={Conditionally Gaussian Distributions and their Application in the Performance of Maximum Likelihood Decode and Forward Cooperative Systems}, 
  year={2011},
  volume={10},
  number={1},
  pages={27-32},
  doi={10.1109/TWC.2010.111910.091690}}

@article{jain2011exact,
  title={Exact analysis of the piecewise linear combiner for decode and forward cooperation with three relays},
  author={Jain, Abhijeet and Sharma, G. V. V. and Desai, Uday B and Merchant, S. N.},
 journal=IEEE_J_WCOM,
  volume={10},
  number={8},
  pages={2461--2467},
  year={2011},
  publisher={IEEE}
}

@article{ngo2011linear,
  title={Linear multihop amplify-and-forward relay channels: Error exponent and optimal number of hops},
  author={Ngo, Hien Quoc and Larsson, Erik G},
 journal=IEEE_J_WCOM,
  volume={10},
  number={11},
  pages={3834--3842},
  year={2011},
  publisher={IEEE}
}

@article{wang2013dynamic,
  title={Dynamic programming for route selection in multihop fixed gain amplify-and-forward relay networks},
  author={Wang, Yue and Bocus, Mohammud Z and Coon, Justin P},
  journal=IEEE_J_COML,
  volume={17},
  number={5},
  pages={932--935},
  year={2013},
  publisher={IEEE}
}

@article{lopez2015average,
  title={Average fade duration for amplify-and-forward relay networks in fading channels},
  author={Lopez-Martinez, F Javier and Kurniawan, Ernest and Islam, Russ and Goldsmith, Andrea},
 journal=IEEE_J_WCOM,
  volume={14},
  number={10},
  pages={5454--5467},
  year={2015},
  publisher={IEEE}
}

@book{goldsmith2005wireless,
  title={Wireless communications},
  author={Goldsmith, Andrea},
  year={2005},
  publisher={Cambridge university press}
}

@ARTICLE{chen2004,
  author={Biao Chen and Ruixiang Jiang and Kasetkasem, T. and Varshney, P.K.},
   journal= IEEE_J_SP,
  title={Channel aware decision fusion in wireless sensor networks}, 
  year={2004},
  volume={52},
  number={12},
  pages={3454-3458},
  doi={10.1109/TSP.2004.837404}}

@article{chen2006channel,
  title={Channel-aware distributed detection in wireless sensor networks},
  author={Chen, Biao and Tong, Lang and Varshney, Pramod K},
  journal=IEEE_M_SP,
  volume={23},
  number={4},
  pages={16--26},
  year={2006},
  publisher={IEEE}
}

@article{ciuonzo2012channel,
  title={Channel-aware decision fusion in distributed MIMO wireless sensor networks: Decode-and-fuse vs. decode-then-fuse},
  author={Ciuonzo, Domenico and Romano, Gianmarco and Salvo Rossi, Pierluigi},
  journal=IEEE_J_WCOM,
  volume={11},
  number={8},
  pages={2976--2985},
  year={2012},
  publisher={IEEE}
}

@article{picinbono1995,
  author={Picinbono, Bernard},
  title={On Deflection as a Performance Criterion in Detection},
  journal=IEEE_J_AES, volume={31}, number={3},
  pages={1072--1081}, month=jul, year={1995},
  doi={10.1109/7.395235}}

@ARTICLE{golvaei2021,
  author={Golvaei, Mehran and Fakharzadeh, Mohammad},
  journal=IEEE_J_CASII_EB, 
  title={A Fast Soft Decision Algorithm for Cooperative Spectrum Sensing}, 
  year={2021},
  volume={68},
  number={1},
  pages={241-245},}

@ARTICLE{liu2019,
  author={Liu, Mingxia and You, Xiaoming and Yu, Xingxing and Liu, Sheng},
  journal=IEEE_J_ACCESS, 
  title={KL Divergence-Based Pheromone Fusion for Heterogeneous Multi-Colony Ant Optimization}, 
  year={2019},
  volume={7},
  number={},
  pages={152646-152657},
  doi={10.1109/ACCESS.2019.2948395}}

@ARTICLE{dey2020,
  author={Dey, I. and Ciuonzo, D. and Salvo Rossi, Pierluigi},
  journal=IEEE_J_WCOM, 
  title={Wideband Collaborative Spectrum Sensing Using Massive MIMO Decision Fusion}, 
  year={2020},
  volume={19},
  number={8},
  pages={5246-5260},
  doi={10.1109/TWC.2020.2991113}}

@ARTICLE{jayakrishan2008,
  author={Unnikrishnan, Jayakrishnan and Veeravalli, Venugopal V.},
  journal=IEEE_J_JSTSP, 
  title={Cooperative Sensing for Primary Detection in Cognitive Radio}, 
  year={2008},
  volume={2},
  number={1},
  pages={18-27},
  doi={10.1109/JSTSP.2007.914880}}

@ARTICLE{quan2008,
  author={Quan, Zhi and Cui, Shuguang and Sayed, Ali H.},
  journal=IEEE_J_JSTSP, 
  title={Optimal Linear Cooperation for Spectrum Sensing in Cognitive Radio Networks}, 
  year={2008},
  volume={2},
  number={1},
  pages={28-40},
  doi={10.1109/JSTSP.2007.914882}}

@ARTICLE{sarikhani2020,
  author={Sarikhani, Rahil and Keynia, Farshid},
  journal=IEEE_J_COML, 
  title={Cooperative Spectrum Sensing Meets Machine Learning: Deep Reinforcement Learning Approach}, 
  year={2020},
  volume={24},
  number={7},
  pages={1459-1462},
  doi={10.1109/LCOMM.2020.2984430}}

@ARTICLE{atapattu2010,
  author={Atapattu, Saman and Tellambura, Chintha and Jiang, Hai},
  journal=IEEE_J_WCOM, 
  title={Analysis of area under the ROC curve of energy detection}, 
  year={2010},
  volume={9},
  number={3},
  pages={1216-1225},
  doi={10.1109/TWC.2010.03.091085}}

@ARTICLE{attapattu2011,
  author={Atapattu, Saman and Tellambura, Chintha and Jiang, Hai},
  journal=IEEE_J_COML, 
  title={MGF Based Analysis of Area under the ROC Curve in Energy Detection}, 
  year={2011},
  volume={15},
  number={12},
  pages={1301-1303},
  doi={10.1109/LCOMM.2011.103111.111420}}

@ARTICLE{vladeanu2022,
  author={Vladeanu, Calin and Martian, Alexandru and Popescu, Dimitrie C.},
  journal=IEEE_J_ACCESS, 
  title={Spectrum Sensing With Energy Detection in Multiple Alternating Time Slots}, 
  year={2022},
  volume={10},
  number={},
  pages={38565-38574},
  doi={10.1109/ACCESS.2022.3165556}}

@ARTICLE{gahane2018,
  author={Gahane, Lokesh and Sharma, Prabhat Kumar and Varshney, Neeraj and Tsiftsis, Theodoros A. and Kumar, Preetam},
  journal=IEEE_J_COM, 
  title={An Improved Energy Detector for Mobile Cognitive Users Over Generalized Fading Channels}, 
  year={2018},
  volume={66},
  number={2},
  pages={534-545},
  doi={10.1109/TCOMM.2017.2754250}}

@ARTICLE{chin2016,
  author={Chin, Wen-Long and Li, Jia-Ming and Chen, Hsiao-Hwa},
  journal=IEEE_J_VT, 
  title={Low-Complexity Energy Detection for Spectrum Sensing With Random Arrivals of Primary Users}, 
  year={2016},
  volume={65},
  number={2},
  pages={947-952},
  doi={10.1109/TVT.2015.2405558}}

@INPROCEEDINGS{khalek2020,
  author={Khalek, Nada Abdel and Hamouda, Walaa},
  booktitle=IEEE_C_GLOBECOM, 
  title={Learning-Based Cooperative Spectrum Sensing in Hybrid Underlay-Interweave Secondary Networks}, 
  year={2020},
  volume={},
  number={},
  pages={1-6},
  doi={10.1109/GLOBECOM42002.2020.9322133}}

@ARTICLE{thilina2013,
  author={Thilina, Karaputugala Madushan and Choi, Kae Won and Saquib, Nazmus and Hossain, Ekram},
  journal=IEEE_J_JSAC, 
  title={Machine Learning Techniques for Cooperative Spectrum Sensing in Cognitive Radio Networks}, 
  year={2013},
  volume={31},
  number={11},
  pages={2209-2221},
  doi={10.1109/JSAC.2013.131120}}

@INPROCEEDINGS{chen2016,
  author={Chen, Xin and Hou, Fukang and Huang, Hai and Jing, Xiaojun},
  booktitle=C_ISCIT, 
  title={Principle component analysis based cooperative spectrum sensing in cognitive radio}, 
  year={2016},
  volume={},
  number={},
  pages={602-605},
  doi={10.1109/ISCIT.2016.7751703}}

@Article{futuretransp6030103,
AUTHOR = {Ji, Ang and Ju, Xilu and Liu, Nieyangzi and Chen, Junxian and Dai, Zhe},
TITLE = {Cooperative Connected and Automated Mobility: A Survey},
JOURNAL = J_FUTURE_TRANSP,
VOLUME = {6},
YEAR = {2026},
NUMBER = {3},
ARTICLE-NUMBER = {103},
ISSN = {2673-7590},
DOI = {10.3390/futuretransp6030103}
}

@ARTICLE{baccari2024,
  author={Baccari, Sihem and Hadded, Mohamed and Ghazzai, Hakim and Touati, Haifa and Elhadef, Mourad},
  journal=IEEE_J_ACCESS, 
  title={Anomaly Detection in Connected and Autonomous Vehicles: A Survey, Analysis, and Research Challenges}, 
  year={2024},
  volume={12},
  number={},
  pages={19250-19276},
  doi={10.1109/ACCESS.2024.3361829}}

@ARTICLE{Liu2022,
  author={Liu, Ze and Cai, Yingfeng and Wang, Hai and Chen, Long and Gao, Hongbo and Jia, Yunyi and Li, Yicheng},
  journal=IEEE_J_ITS, 
  title={Robust Target Recognition and Tracking of Self-Driving Cars With Radar and Camera Information Fusion Under Severe Weather Conditions}, 
  year={2022},
  volume={23},
  number={7},
  pages={6640-6653},
  doi={10.1109/TITS.2021.3059674}}

@ARTICLE{yaldiz2023,
  author={Yaldiz, Cem O. and Buller, Mark J. and Richardson, Kristine L. and An, Sungtae and Lin, David J. and Satish, Aprameya and Driver, Kyla and Atkinson, Emma and Mesite, Timothy and King, Christopher and Bursey, Max and Galer, Meghan and Millard-Stafford, Mindy and Sawka, Michael N. and Medda, Alessio and Inan, Omer T.},
  journal=IEEE_J_JBHI, 
  title={Early Prediction of Impending Exertional Heat Stroke With Wearable Multimodal Sensing and Anomaly Detection}, 
  year={2023},
  volume={27},
  number={12},
  pages={5803-5814},
  doi={10.1109/JBHI.2023.3323014}}

@article{gumaei2022,
author = {Gumaei, Abdu and Ismail, Walaa N. and Rafiul Hassan, Md. and Hassan, Mohammad Mehedi and Mohamed, Ebtsam and Alelaiwi, Abdullah and Fortino, Giancarlo},
title = {A Decision-Level Fusion Method for COVID-19 Patient Health Prediction},
year = {2022},
issue_date = {Feb 2022},
publisher = {Elsevier Science Publishers B. V.},
address = {NLD},
volume = {27},
number = {C},
issn = {2214-5796},
doi = {10.1016/j.bdr.2021.100287},
journal = J_BIGDATA_RES,
month = feb,
numpages = {11}}

@article{XU2024,
title = {Detecting forest fire omission error based on data fusion at subpixel scale},
journal = J_APPL_EARTH_OBS,
volume = {128},
pages = {103737},
year = {2024},
issn = {1569-8432},
author = {Haizhou Xu and Gui Zhang and Rong Chu and Juan Zhang and Zhigao Yang and Xin Wu and Huashun Xiao},
}

@article{zWANG2024,
title = {Historical information fusion of dense multi-source satellite image time series for flood extent mapping},
journal = J_INF_FUSION,
volume = {109},
pages = {102445},
year = {2024},
issn = {1566-2535},
author = {Zhihao Wang and Xueqian Wang and Gang Li and Wei Wu and Yu Liu and Zhaohui Song and Huina Song},
}

@ARTICLE{Xue2024,
  author={Xue, Yawen and Pan, Jie and Geng, Yangyang and Yang, Zeyu and Liu, Mengxiang and Deng, Ruilong},
  journal=IEEE_J_ICPS, 
  title={Real-Time Intrusion Detection Based on Decision Fusion in Industrial Control Systems}, 
  year={2024},
  volume={2},
  number={},
  pages={143-153},
  doi={10.1109/TICPS.2024.3406505}}

@ARTICLE{Liu2021,
  author={Liu, Mingqian and Yang, Ke and Zhao, Nan and Chen, Yunfei and Song, Hao and Gong, Fengkui},
  journal=IEEE_J_II, 
  title={Intelligent Signal Classification in Industrial Distributed Wireless Sensor Networks Based Industrial Internet of Things}, 
  year={2021},
  volume={17},
  number={7},
  pages={4946-4956},
  doi={10.1109/TII.2020.3016958}}

@INPROCEEDINGS{Tabella2021,
  author={Tabella, Gianluca and Ciuonzo, Domenico and Paltrinieri, Nicola and Salvo Rossi, Pierluigi},
  booktitle=IEEE_C_FUSION_24, 
  title={Spatio-Temporal Decision Fusion for Quickest Fault Detection Within Industrial Plants: The Oil and Gas Scenario}, 
  year={2021},
  volume={},
  number={},
  pages={1-8},
  doi={10.23919/FUSION49465.2021.9626941}}

@INPROCEEDINGS{nair2024,
  author={Nair, Mahesh and Voudouris, Konstantinos N},
  booktitle=C_ASET, 
  title={Distributed Fusion to Improve the Efficiency of Industrial Wireless Sensor Network}, 
  year={2024},
  volume={},
  number={},
  pages={1-4},
  doi={10.1109/ASET60340.2024.10708710}}

@INPROCEEDINGS{Qiu2025,
  author={Qiu, Guangping and Li, Jincan and Zhang, Jiaheng and Deng, Jizhong and Wang, Weixing},
  booktitle=C_ICSI_5, 
  title={Multi-Sensor Fusion for Environmental Perception and Decision-Making in Smart Fertilization Robots}, 
  year={2025},
  volume={},
  number={},
  pages={669-673},
  doi={10.1109/ICSI64877.2025.11009313}}

@INPROCEEDINGS{Lambrinos2019,
  author={Lambrinos, Lambros},
  booktitle=IEEE_C_DASC_PICOM_CBDCOM_CYBERSCITECH, 
  title={Internet of Things in Agriculture: A Decision Support System for Precision Farming}, 
  year={2019},
  volume={},
  number={},
  pages={889-892},
  doi={10.1109/DASC/PiCom/CBDCom/CyberSciTech.2019.00163}}

@ARTICLE{saki2025,
  author={Saki, Mahdi and Keshavarz, Rasool and Franklin, Daniel and Abolhasan, Mehran and Lipman, Justin and Shariati, Negin},
  journal=IEEE_J_ACCESS, 
  title={A Data-Driven Review of Remote Sensing-Based Data Fusion in Precision Agriculture From Foundational to Transformer-Based Techniques}, 
  year={2025},
  volume={13},
  number={},
  pages={166188-166209},
  doi={10.1109/ACCESS.2025.3610649}}

@article{Lee2019DeepCooperative,
  author={W. Lee and M. Kim and D. H. Cho},
  journal=IEEE_J_VT,
  title={Deep Cooperative Sensing: Cooperative Spectrum Sensing Based on Convolutional Neural Networks},
  volume={68},
  number={3},
  pages={3005--3009},
  year={2019}
}

@article{Gao2019DeepLearning,
  author={J. Gao and X. Yi and C. Zhong and X. Chen and Z. Zhang},
  journal=IEEE_J_WCOML,
  title={Deep Learning for Spectrum Sensing},
  volume={8},
  number={6},
  pages={1727--1730},
  year={2019}
}

@article{Rajendran2018,
  author={S. Rajendran and W. Meert and D. Giustiniano and V. Lenders and S. Pollin},
  journal=IEEE_J_TCCN,
  title={Deep Learning Models for Wireless Signal Classification With Distributed Low-Cost Spectrum Sensors},
  volume={4},
  number={3},
  pages={433--445},
  year={2018}
}

@article{Abrardo2016,
  author={A. Abrardo and M. Barni and K. Kallas and B. Tondi},
  journal=IEEE_J_TIFS,
  title={A Game-Theoretic Framework for Optimum Decision Fusion in the Presence of Byzantines},
  volume={11},
  number={6},
  pages={1333--1345},
  year={2016}
}

@article{Wang2010,
  author={B. Wang and K. J. R. Liu and T. C. Clancy},
  journal=IEEE_J_COM,
  title={Evolutionary Cooperative Spectrum Sensing Game: How to Collaborate?},
  volume={58},
  number={3},
  pages={890--900},
  year={2010}
}

@article{Hao2012,
  author={X. Hao and M. H. Cheung and V. W. S. Wong and V. C. M. Leung},
  journal=IEEE_J_WCOM,
  title={Hedonic Coalition Formation Game for Cooperative Spectrum Sensing and Channel Access in Cognitive Radio Networks},
  volume={11},
  number={11},
  pages={3968--3979},
  year={2012}
}

@ARTICLE{AlJarrah2020,
  author={Al-Jarrah, Mohammad A. and Yaseen, Maysa A. and Al-Dweik, Arafat and Dobre, Octavia A. and Alsusa, Emad},
  journal=IEEE_J_IOT, 
  title={Decision Fusion for IoT-Based Wireless Sensor Networks}, 
  year={2020},
  volume={7},
  number={2},
  pages={1313-1326},
  doi={10.1109/JIOT.2019.2954720}}

@ARTICLE{Liu2020EH,
  author={Liu, Xiaoying and Zheng, Kechen and Chi, Kaikai and Zhu, Yi-Hua},
  journal=IEEE_J_WCOM, 
  title={Cooperative Spectrum Sensing Optimization in Energy-Harvesting Cognitive Radio Networks}, 
  year={2020},
  volume={19},
  number={11},
  pages={7663-7676},
  doi={10.1109/TWC.2020.3015260}}

@ARTICLE{AlJarrah2019DFCoopWSN,
  author={Al-Jarrah, Mohammad A. and Al-Dweik, Arafat and Kalil, Mohamad and Ikki, Salama S.},
  journal=IEEE_J_VT, 
  title={Decision Fusion in Distributed Cooperative Wireless Sensor Networks}, 
  year={2019},
  volume={68},
  number={1},
  pages={797-811},
  doi={10.1109/TVT.2018.2879413}}

@article{chamberland2003decentralized,
  author  = {J.-F. Chamberland and V. V. Veeravalli},
  title   = {Decentralized Detection in Sensor Networks},
  journal = IEEE_J_SP,
  volume  = {51},
  number  = {2},
  pages   = {407--416},
  month   = feb,
  year    = {2003},
  doi     = {10.1109/TSP.2002.806982}
}

@article{chamberland2004asymptotic,
  author  = {J.-F. Chamberland and V. V. Veeravalli},
  title   = {Asymptotic Results for Decentralized Detection in Power Constrained Wireless Sensor Networks},
  journal = IEEE_J_JSAC,
  volume  = {22},
  number  = {6},
  pages   = {1007--1015},
  month   = aug,
  year    = {2004},
  doi     = {10.1109/JSAC.2004.830894}
}

@article{chamberland2007wireless,
  author  = {J.-F. Chamberland and V. V. Veeravalli},
  title   = {Wireless Sensors in Distributed Detection Applications},
  journal = IEEE_M_SP,
  volume  = {24},
  number  = {3},
  pages   = {16--25},
  month   = may,
  year    = {2007},
  doi     = {10.1109/MSP.2007.361598}
}

@article{cattivelli2011distributed,
  author  = {F. S. Cattivelli and A. H. Sayed},
  title   = {Distributed Detection Over Adaptive Networks Using Diffusion Adaptation},
  journal = IEEE_J_SP,
  volume  = {59},
  number  = {5},
  pages   = {1917--1932},
  month   = may,
  year    = {2011}
}

@article{aldalahmeh2019fusion,
  author  = {S. A. Aldalahmeh and S. O. Al-Jazzar and D. McLernon and S. A. R. Zaidi and M. Ghogho},
  title   = {Fusion Rules for Distributed Detection in Clustered Wireless Sensor Networks With Imperfect Channels},
  journal = IEEE_J_SIPN,
  volume  = {5},
  number  = {3},
  pages   = {585--597},
  month   = sep,
  year    = {2019},
  doi     = {10.1109/TSIPN.2019.2901198}
}

@article{ma2008soft,
  author  = {J. Ma and G. Zhao and Y. Li},
  title   = {Soft Combination and Detection for Cooperative Spectrum Sensing in Cognitive Radio Networks},
  journal = IEEE_J_WCOM,
  volume  = {7},
  number  = {11},
  pages   = {4502--4507},
  month   = nov,
  year    = {2008},
  doi     = {10.1109/T-WC.2008.070941}
}

@article{nevat2014distributed,
  author  = {I. Nevat and G. W. Peters and I. B. Collings},
  title   = {Distributed Detection in Sensor Networks Over Fading Channels With Multiple Antennas at the Fusion Centre},
  journal = IEEE_J_SP,
  volume  = {62},
  number  = {3},
  pages   = {671--683},
  month   = feb,
  year    = {2014},
  doi     = {10.1109/TSP.2013.2293970}
}

@article{appadwedula2008censoring,
  author  = {S. Appadwedula and V. V. Veeravalli and D. L. Jones},
  title   = {Decentralized Detection With Censoring Sensors},
  journal = IEEE_J_SP,
  volume  = {56},
  number  = {4},
  pages   = {1362--1373},
  month   = apr,
  year    = {2008},
  doi     = {10.1109/TSP.2007.909355}
}

@article{nguyen2015bits,
  author  = {N. Nguyen-Thanh and P. Ciblat and S. Maleki and V.-T. Nguyen},
  title   = {How Many Bits Should Be Reported in Quantized Cooperative Spectrum Sensing?},
  journal = IEEE_J_WCOML,
  volume  = {4},
  number  = {5},
  pages   = {465--468},
  month   = oct,
  year    = {2015},
  doi     = {10.1109/LWC.2015.2437879}
}

@article{bockelmann2016massive,
  author    = {C. Bockelmann and N. Pratas and H. Nikopour and K. Au and T. Svensson and C. Stefanovic and P. Popovski and A. Dekorsy},
  title     = {Massive Machine-Type Communications in {5G}: Physical and {MAC}-Layer Solutions},
  journal   = IEEE_M_COM,
  volume    = {54},
  number    = {9},
  pages     = {59--65},
  month     = sep,
  year      = {2016},
  doi       = {10.1109/MCOM.2016.7565189}
}

@article{chen2021massive,
  author    = {X. Chen and D. W. K. Ng and W. Yu and E. G. Larsson and N. Al-Dhahir and R. Schober},
  title     = {Massive Access for {5G} and Beyond},
  journal   = IEEE_J_JSAC,
  volume    = {39},
  number    = {3},
  pages     = {615--637},
  month     = mar,
  year      = {2021},
  doi       = {10.1109/JSAC.2020.3019724}
}

@article{ciuonzo2015massive,
  author    = {D. Ciuonzo and Salvo Rossi, Pierluigi and S. Dey},
  title     = {Massive MIMO Channel-Aware Decision Fusion},
  journal   = IEEE_J_SP,
  volume    = {63},
  number    = {3},
  pages     = {604--619},
  month     = feb,
  year      = {2015},
  doi       = {10.1109/TSP.2014.2376886}
}

@article{popovski2018wireless,
  author    = {P. Popovski and J. J. Nielsen and C. Stefanovic and E. de Carvalho and E. Strom and K. F. Trillingsgaard and A.-S. Bana and D. M. Kim and R. Kotaba and J. Park and R. B. Sorensen},
  title     = {Wireless Access for Ultra-Reliable Low-Latency Communication: Principles and Building Blocks},
  journal   = IEEE_M_NET,
  volume    = {32},
  number    = {2},
  pages     = {16--23},
  month     = mar,
  year      = {2018},
  doi       = {10.1109/MNET.2018.1700258}
}

@article{sutton2019enabling,
  author    = {G. J. Sutton and J. Zeng and R. P. Liu and W. Ni and D. N. Nguyen and B. A. Jayawickrama and X. Huang and M. Abolhasan and Z. Zhang and E. Dutkiewicz and T. Lv},
  title     = {Enabling Technologies for Ultra-Reliable and Low Latency Communications: From PHY and MAC Layer Perspectives},
  journal   = IEEE_J_COMST,
  volume    = {21},
  number    = {3},
  pages     = {2488--2524},
  year      = {2019},
  doi       = {10.1109/COMST.2019.2897800}
}

@article{polyanskiy2010channel,
  author    = {Y. Polyanskiy and H. V. Poor and S. Verd{\'u}},
  title     = {Channel Coding Rate in the Finite Blocklength Regime},
  journal   = IEEE_J_IT,
  volume    = {56},
  number    = {5},
  pages     = {2307--2359},
  month     = may,
  year      = {2010},
  doi       = {10.1109/TIT.2010.2043769}
}

@article{Zeng2016UAV,
  author  = {Zeng, Yong and Zhang, Rui and Lim, Teng Joon},
  title   = {Wireless Communications with Unmanned Aerial Vehicles:
             Opportunities and Challenges},
  journal = IEEE_M_COM,
  volume  = {54},
  number  = {5},
  pages   = {36--42},
  month   = may,
  year    = {2016},
  doi     = {10.1109/MCOM.2016.7470933}
}

@article{Zeng2017Energy,
  author  = {Zeng, Yong and Zhang, Rui},
  title   = {Energy-Efficient {UAV} Communication With Trajectory
             Optimization},
  journal = IEEE_J_WCOM,
  volume  = {16},
  number  = {6},
  pages   = {3747--3760},
  month   = jun,
  year    = {2017},
  doi     = {10.1109/TWC.2017.2688328}
}

@article{Zeng2019Sky,
  author  = {Zeng, Yong and Wu, Qingqing and Zhang, Rui},
  title   = {Accessing From the Sky: A Tutorial on {UAV}
             Communications for {5G} and Beyond},
  journal = IEEE_J_PROC,
  volume  = {107},
  number  = {12},
  pages   = {2327--2375},
  month   = dec,
  year    = {2019},
  doi     = {10.1109/JPROC.2019.2952892}
}

@article{Ghdiri2021UAVData,
  author  = {Ghdiri, Oussama and Jaafar, Wael and Alfattani, Safwan
             and Ben Abderrazak, Jihene and Yanikomeroglu, Halim},
  title   = {Offline and Online {UAV}-Enabled Data Collection in
             Time-Constrained {IoT} Networks},
  journal = IEEE_J_TGCN,
  volume  = {5},
  number  = {4},
  pages   = {1918--1933},
  year    = {2021},
  doi     = {10.1109/TGCN.2021.3104801}
}

@article{Komine2004VLC,
  author  = {Komine, Toshihiko and Nakagawa, Masao},
  title   = {Fundamental Analysis for Visible-Light Communication
             System Using {LED} Lights},
  journal = IEEE_J_CE,
  volume  = {50},
  number  = {1},
  pages   = {100--107},
  month   = feb,
  year    = {2004},
  doi     = {10.1109/TCE.2004.1277847}
}

@article{Karunatilaka2015VLC,
  author  = {Karunatilaka, Dilukshan A. and Zafar, Fahad and
             Kalavally, Vineetha and Parthiban, Rajendran},
  title   = {{LED}-Based Indoor Visible Light Communications:
             State of the Art},
  journal = IEEE_J_COMST,
  volume  = {17},
  number  = {3},
  pages   = {1649--1678},
  year    = {2015},
  doi     = {10.1109/COMST.2015.2417576}
}

@article{Soltani2019Orientation,
  author  = {Dehghani Soltani, Mohammad and Purwita, Ardimas Andi
             and Zeng, Zhihong and Haas, Harald and Safari, Majid},
  title   = {Modeling the Random Orientation of Mobile Devices:
             Measurement, Analysis and {LiFi} Use Case},
  journal = IEEE_J_COM,
  volume  = {67},
  number  = {3},
  pages   = {2157--2172},
  month   = mar,
  year    = {2019},
  doi     = {10.1109/TCOMM.2018.2882213}
}

@article{Kong2019HybridVLC,
  author  = {Kong, Justin and Ismail, Muhammad and Serpedin, Erchin
             and Qaraqe, Khalid A.},
  title   = {Energy Efficient Optimization of Base Station
             Intensities for Hybrid {RF/VLC} Networks},
  journal = IEEE_J_WCOM,
  volume  = {18},
  number  = {8},
  pages   = {4171--4183},
  month   = aug,
  year    = {2019},
  doi     = {10.1109/TWC.2019.2922611}
}

@article{Ulukus2015EHReview,
  author  = {S. Ulukus and A. Yener and E. Erkip and O. Simeone
             and M. Zorzi and P. Grover and K. Huang},
  title   = {Energy Harvesting Wireless Communications:
             A Review of Recent Advances},
  journal = IEEE_J_JSAC,
  volume  = {33},
  number  = {3},
  pages   = {360--381},
  month   = mar,
  year    = {2015}
}

@article{Tutuncuoglu2012BatteryLimited,
  author  = {K. Tutuncuoglu and A. Yener},
  title   = {Optimum Transmission Policies for Battery Limited
             Energy Harvesting Nodes},
  journal = IEEE_J_WCOM,
  volume  = {11},
  number  = {3},
  pages   = {1180--1189},
  month   = mar,
  year    = {2012}
}

@article{Ozel2012AWGN,
  author  = {O. Ozel and S. Ulukus},
  title   = {Achieving {AWGN} Capacity Under Stochastic
             Energy Harvesting},
  journal = IEEE_J_IT,
  volume  = {58},
  number  = {10},
  pages   = {6471--6483},
  month   = oct,
  year    = {2012}
}

@article{Tarighati2017DHTEH,
  author  = {A. Tarighati and J. Gross and J. Jald{\'e}n},
  title   = {Decentralized Hypothesis Testing in Energy Harvesting
             Wireless Sensor Networks},
  journal = IEEE_J_SP,
  volume  = {65},
  number  = {18},
  pages   = {4862--4873},
  month   = sep,
  year    = {2017}
}

@article{Islam2017PDNOMA, author = {S. M. R. Islam and N. Avazov and O. A. Dobre and K. S. Kwak}, title = {Power-Domain Non-Orthogonal Multiple Access ({NOMA}) in {5G} Systems: Potentials and Challenges}, journal = IEEE_J_COMST, volume = {19}, number = {2}, pages = {721--742}, year = {2017} }

@article{Liu2017NOMA, author = {Y. Liu and Z. Qin and M. Elkashlan and Z. Ding and A. Nallanathan and L. Hanzo}, title = {Nonorthogonal Multiple Access for {5G} and Beyond}, journal = IEEE_J_PROC, volume = {105}, number = {12}, pages = {2347--2381}, month = dec, year = {2017} }

@article{Shahab2020GFNOMA, author = {M. B. Shahab and R. Abbas and M. Shirvanimoghaddam and S. J. Johnson}, title = {Grant-Free Non-Orthogonal Multiple Access for {IoT}: A Survey}, journal = IEEE_J_COMST, volume = {22}, number = {3}, pages = {1805--1838}, year = {2020} }

@article{AlFuqaha2015IoT, author = {A. Al-Fuqaha and M. Guizani and M. Mohammadi and M. Aledhari and M. Ayyash}, title = {Internet of Things: A Survey on Enabling Technologies, Protocols, and Applications}, journal = IEEE_J_COMST, volume = {17}, number = {4}, pages = {2347--2376}, year = {2015} }

@article{Shi2016Edge, author = {W. Shi and J. Cao and Q. Zhang and Y. Li and L. Xu}, title = {Edge Computing: Vision and Challenges}, journal = IEEE_J_IOT, volume = {3}, number = {5}, pages = {637--646}, month = oct, year = {2016} }

@article{Yates2021AoI, author = {R. D. Yates and Y. Sun and Brown, III, D. Richard and S. K. Kaul and E. Modiano and S. Ulukus}, title = {Age of Information: An Introduction and Survey}, journal = IEEE_J_JSAC, volume = {39}, number = {5}, pages = {1183--1210}, month = may, year = {2021} }

@INPROCEEDINGS{Acharya2014NeymanPearson,
  author={Acharya, Sayandeep and Wang, Ji and Kam, Moshe},
  booktitle=C_FUSION_17, 
  title={Distributed decision fusion using the Neyman-Pearson criterion}, 
  year={2014},
  volume={},
  number={},
  pages={1-7},
  doi={}}

@article{Deb2002NSGAII,
  author  = {K. Deb and A. Pratap and S. Agarwal and T. Meyarivan},
  title   = {A Fast and Elitist Multiobjective Genetic Algorithm:
             {NSGA-II}},
  journal = IEEE_J_EVC,
  volume  = {6},
  number  = {2},
  pages   = {182--197},
  month   = apr,
  year    = {2002}
}

@article{Levy2009RobustHT,
  author  = {B. C. Levy},
  title   = {Robust Hypothesis Testing With a Relative
             Entropy Tolerance},
  journal = IEEE_J_IT,
  volume  = {55},
  number  = {1},
  pages   = {413--421},
  month   = jan,
  year    = {2009}
}

@inproceedings{Guo2017Calibration,
  author    = {C. Guo and G. Pleiss and Y. Sun
               and K. Q. Weinberger},
  title     = {On Calibration of Modern Neural Networks},
  booktitle = C_ICML_34,
  series    = {Proc. Mach. Learn. Res.},
  volume    = {70},
  pages     = {1321--1330},
  publisher = {PMLR},
  month     = aug,
  year      = {2017}
}

@article{Sze2017EfficientDNN,
  author  = {V. Sze and Y.-H. Chen
             and T.-J. Yang and J. S. Emer},
  title   = {Efficient Processing of Deep Neural Networks:
             A Tutorial and Survey},
  journal = IEEE_J_PROC,
  volume  = {105},
  number  = {12},
  pages   = {2295--2329},
  month   = dec,
  year    = {2017}
}

@article{Luong2019DRLSurvey,
  author  = {N. C. Luong and D. T. Hoang and S. Gong and
             D. Niyato and P. Wang and Y.-C. Liang and D. I. Kim},
  title   = {Applications of Deep Reinforcement Learning in
             Communications and Networking: A Survey},
  journal = IEEE_J_COMST,
  volume  = {21},
  number  = {4},
  pages   = {3133--3174},
  year    = {2019},
  doi     = {10.1109/COMST.2019.2916583}
}

@article{Gao2021MARLCSS,
  author  = {A. Gao and C. Du and S. X. Ng and W. Liang},
  title   = {A Cooperative Spectrum Sensing With Multi-Agent
             Reinforcement Learning Approach in Cognitive
             Radio Networks},
  journal = IEEE_J_COML,
  volume  = {25},
  number  = {8},
  pages   = {2604--2608},
  month   = aug,
  year    = {2021},
  doi     = {10.1109/LCOMM.2021.3078442}
}

@article{Ciuonzo2013MRC,
  author  = {D. Ciuonzo and G. Romano and Salvo Rossi, Pierluigi},
  title   = {Performance Analysis and Design of Maximum-Ratio
             Combining in Channel-Aware {MIMO} Decision Fusion},
  journal = IEEE_J_WCOM,
  volume  = {12},
  number  = {9},
  pages   = {4716--4728},
  month   = sep,
  year    = {2013},
  doi     = {10.1109/TWC.2013.071913.130269}
}

@article{Wang2007CooperativeDemodulation,
  author  = {T. Wang and A. Cano and G. B. Giannakis
             and J. N. Laneman},
  title   = {High-Performance Cooperative Demodulation With
             Decode-and-Forward Relays},
  journal = IEEE_J_COM,
  volume  = {55},
  number  = {7},
  pages   = {1427--1438},
  month   = jul,
  year    = {2007},
  doi     = {10.1109/TCOMM.2007.900631}
}

@article{Wang2008LAR,
  author  = {T. Wang and G. B. Giannakis and R. Wang},
  title   = {Smart Regenerative Relays for Link-Adaptive
             Cooperative Communications},
  journal = IEEE_J_COM,
  volume  = {56},
  number  = {11},
  pages   = {1950--1960},
  month   = nov,
  year    = {2008},
  doi     = {10.1109/TCOMM.2008.060688}
}

@article{Vien2009SmartRelaying,
  author  = {N. H. Vien and H. H. Nguyen and T. Le-Ngoc},
  title   = {Diversity Analysis of Smart Relaying},
  journal = IEEE_J_VT,
  volume  = {58},
  number  = {6},
  pages   = {2849--2862},
  month   = jul,
  year    = {2009},
  doi     = {10.1109/TVT.2008.2005991}
}

@article{Nasri2012EMRC,
  author  = {A. Nasri and R. Schober and M. Uysal},
  title   = {Enhanced MRC for Decode-and-Forward Cooperative Diversity Systems},
  journal = IEEE_J_WCOM,
  volume  = {11},
  number  = {10},
  pages   = {3418--3423},
  month   = oct,
  year    = {2012},
  doi={10.1109/TWC.2012.083112.111074}}

@article{Vien2012SmartEGC,
  author  = {N. H. Vien and H. H. Nguyen and T. Le-Ngoc},
  title   = {Performance Analysis of Smart Relaying With Equal
             Gain Combining},
  journal = J_WIRELESS_PERS_COM,
  volume  = {65},
  number  = {2},
  pages   = {273--292},
  year    = {2012},
  doi     = {10.1007/s11277-011-0249-z}
}

@article{Ikki2008BestRelaySelection,
  author  = {S. S. Ikki and M. H. Ahmed},
  title   = {Performance of Multiple-Relay Cooperative Diversity
             Systems With Best Relay Selection Over Rayleigh
             Fading Channels},
  journal = J_EURASIP_ASP,
  volume  = {2008},
  pages   = {1--7},
  year    = {2008},
  note    = {Art. no. 580368},
  doi     = {10.1155/2008/580368}
}

@article{Yi2008RelaySelectionDiversity,
  author  = {Z. Yi and I.-M. Kim},
  title   = {Diversity Order Analysis of the Decode-and-Forward
             Cooperative Networks With Relay Selection},
  journal = IEEE_J_WCOM,
  volume  = {7},
  number  = {5},
  pages   = {1792--1799},
  month   = may,
  year    = {2008},
  doi     = {10.1109/TWC.2008.061041}
}

@article{Lei2010MaxLog,
  author  = {A. Lei and R. Schober},
  title   = {Coherent Max-Log Decision Fusion in Wireless
             Sensor Networks},
  journal = IEEE_J_COM,
  volume  = {58},
  number  = {5},
  pages   = {1327--1332},
  month   = may,
  year    = {2010},
  doi     = {10.1109/TCOMM.2010.05.090100}
}

@article{Chaudhari2012HardSoft,
  author  = {S. Chaudhari and J. Lund{\'e}n and V. Koivunen
             and H. V. Poor},
  title   = {Cooperative Sensing With Imperfect Reporting Channels:
             Hard Decisions or Soft Decisions?},
  journal = IEEE_J_SP,
  volume  = {60},
  number  = {1},
  pages   = {18--28},
  month   = jan,
  year    = {2012},
  doi     = {10.1109/TSP.2011.2170978}
}

@article{Huang2024AdaptiveThresholdDF,
  author  = {Y. Huang and Y. Zhu and J. Pan and Y. Gao and F. Peng and Y. Sun},
  title   = {Decision Fusion-Based Nonintrusive Load Identification
             Involving Adaptive Threshold Event Detection},
  journal = IEEE_J_IM,
  volume  = {73},
  pages   = {1--11},
  month   = jun,
  year    = {2024},
  note    = {Art. no. 9004111},
  doi     = {10.1109/TIM.2024.3413152}
}

@article{Zhang2024URLLCUnsupervised,
  author  = {L. Zhang and C. She and K. Ying and Y. Li and B. Vucetic},
  title   = {Unsupervised Learning for Ultra-Reliable and Low-Latency
             Communications With Practical Channel Estimation},
  journal = IEEE_J_WCOM,
  volume  = {23},
  number  = {4},
  pages   = {3633--3647},
  month   = apr,
  year    = {2024},
  doi     = {10.1109/TWC.2023.3309900}
}

@article{Manjang2025ParetoHFL,
  author  = {O. Manjang and Y. Zhai and J. Shen and A. Sarwar and
             X. He and H. Wang and L. Zhu},
  title   = {Efficient Hierarchical Federated Learning With
             Pareto-Optimal Bi-Level Reinforcement Learning},
  journal = IEEE_J_IOT,
  volume  = {12},
  number  = {15},
  pages   = {30710--30724},
  month   = aug,
  year    = {2025},
  doi     = {10.1109/JIOT.2025.3572900}
}

@article{Yang2026UQAMC,
  author  = {H. Yang and R. Sahay},
  title   = {An Uncertainty Quantification Framework for
             Deep Learning-Based Automatic Modulation Classification},
  journal = IEEE_J_IOT,
  volume  = {13},
  number  = {3},
  pages   = {5087--5096},
  month   = feb,
  year    = {2026},
  doi     = {10.1109/JIOT.2025.3640108}
}

@article{Wang2023GNNRISFL,
  author  = {Z. Wang and Y. Zhou and Y. Zou and Q. An and
             Y. Shi and M. Bennis},
  title   = {A Graph Neural Network Learning Approach to Optimize
             {RIS}-Assisted Federated Learning},
  journal = IEEE_J_WCOM,
  volume  = {22},
  number  = {9},
  pages   = {6092--6106},
  month   = sep,
  year    = {2023},
  doi     = {10.1109/TWC.2023.3239400}
}

@article{Sami2024AirCluster,
  author  = {H. U. Sami and B. G{\"u}ler},
  title   = {Over-the-Air Clustered Federated Learning},
  journal = IEEE_J_WCOM,
  volume  = {23},
  number  = {7},
  pages   = {7877--7893},
  month   = jul,
  year    = {2024},
  doi     = {10.1109/TWC.2023.3345730}
}

@article{Oh2024FedVQCS,
  author  = {Y. Oh and Y.-S. Jeon and M. Chen and W. Saad},
  title   = {{FedVQCS}: Federated Learning via Vector Quantized
             Compressed Sensing},
  journal = IEEE_J_WCOM,
  volume  = {23},
  number  = {3},
  pages   = {1755--1770},
  month   = mar,
  year    = {2024},
  doi     = {10.1109/TWC.2023.3291877}
}

@article{Zhai2025UAVAFL,
  author  = {Z. Zhai and X. Yuan and X. Wang and H. Yang},
  title   = {{UAV}-Enabled Asynchronous Federated Learning},
  journal = IEEE_J_WCOM,
  volume  = {24},
  number  = {3},
  pages   = {2358--2372},
  month   = mar,
  year    = {2025},
  doi     = {10.1109/TWC.2024.3520501}
}

@article{Sun2025RISFLImperfect,
  author  = {P. Sun and E. Liu and W. Ni and R. Wang and Z. Xing and
             B. Li and A. Jamalipour},
  title   = {Reconfigurable Intelligent Surface-Assisted Wireless
             Federated Learning With Imperfect Aggregation},
  journal = IEEE_J_COM,
  volume  = {73},
  number  = {2},
  pages   = {1058--1071},
  month   = feb,
  year    = {2025},
  doi     = {10.1109/TCOMM.2024.3450605}
}

@article{Assaad2026CommEffFL,
  author  = {M. Assaad and Z. Nehme and M. Debbah},
  title   = {Communication-Efficient Zero-Order and First-Order
             Federated Learning Methods Over Wireless Networks},
  journal = IEEE_J_SP,
  volume  = {74},
  pages   = {2706--2723},
  month   = may,
  year    = {2026},
  doi     = {10.1109/TSP.2026.3697131}
}

@article{Wong2026IoTEdgeFL,
  author  = {K.-S. Wong and D.-M. Nguyen and K. Le and L. Ho and
             C. Do and D. Le-Phuoc},
  title   = {An Empirical Study of Federated Learning on
             {IoT}--Edge Devices: Resource Allocation and Heterogeneity},
  journal = IEEE_J_TNNLS,
  volume  = {37},
  number  = {2},
  pages   = {753--765},
  month   = feb,
  year    = {2026},
  doi     = {10.1109/TNNLS.2025.3611415}
}

@article{Li2026FedRMVL,
  author  = {D. Li and Z. Yang and J. Kang and Z. Xiong and
             D. Niyato and S. Xie},
  title   = {Reliable Federated Multi-View Learning for Heterogeneous
             Information Fusion in Mobile Edge Computing},
  journal = IEEE_J_MC,
  volume  = {25},
  number  = {5},
  pages   = {7062--7076},
  month   = may,
  year    = {2026},
  doi     = {10.1109/TMC.2025.3640705}
}

@article{Zhu2026BPFLH,
  author  = {G. Zhu and W. Shen and Z. Liu and J. Qin and J. Ma},
  title   = {{BPFLH}: Byzantine-Robust Privacy-Preserving Federated
             Learning for Heterogeneous Data},
  journal = IEEE_J_TDSC,
  volume  = {23},
  number  = {3},
  pages   = {6103--6118},
  month   = may,
  year    = {2026},
  doi     = {10.1109/TDSC.2026.3661522}
}

@article{Machani2026PDNOMACSS,
  author  = {G. W. Machani and T. Walingo},
  title   = {Cooperative Spectrum Sensing and Sharing in
             {PD-NOMA} Cognitive Radio Networks},
  journal = IEEE_J_ACCESS,
  volume  = {14},
  pages   = {2592--2606},
  month   = jan,
  year    = {2026},
  doi     = {10.1109/ACCESS.2025.3650426}
}

@article{Jiang2026AirGroundPerception,
  author  = {H. Jiang and J. Guo and Z. Xiao and K. Yang
             and T. Li and B. Li},
  title   = {Collaborative Perception and Computing Offloading
             in {6G} Air-Ground Integrated Networks},
  journal = IEEE_J_JSAC,
  volume  = {44},
  pages   = {2821--2838},
  year    = {2026},
  doi     = {10.1109/JSAC.2025.3645021}
}

@article{Tan2025GrantFreeActivity,
  author  = {B. Tan and Y. Cui},
  title   = {Fast {MLE} and {MAPE}-Based Device Activity Detection for
             Grant-Free Access via {PSCA} and {PSCA-Net}},
  journal = IEEE_J_WCOM,
  volume  = {24},
  number  = {6},
  pages   = {5066--5083},
  month   = jun,
  year    = {2025},
  doi     = {10.1109/TWC.2025.3545649}
}

@article{Cao2025RISMassiveConnectivity,
  author  = {Y. Cao and C. Xing and N. Wei and S. Gong and N. Zhao
             and D. Niyato},
  title   = {Tensor-Based Joint Channel Estimation and Activity Detection
             for Reconfigurable Intelligent Surface-Assisted Massive
             Connectivity},
  journal = IEEE_J_COM,
  volume  = {73},
  number  = {11},
  pages   = {11870--11885},
  month   = nov,
  year    = {2025},
  doi     = {10.1109/TCOMM.2025.3578806}
}

@ARTICLE{Yi2025SemanticAirCompCSS,
  author={Yi, Peng and Cao, Yang and Kang, Xin and Liang, Ying-Chang},
  journal=IEEE_J_COM, 
  title={Integrated Distributed Semantic Communication and Over-the-Air Computation for Cooperative Spectrum Sensing}, 
  year={2025},
  volume={73},
  number={4},
  pages={2416-2430},
  doi={10.1109/TCOMM.2024.3468215}}

@article{Zhang2023DRLURLLC,
  author  = {L. Zhang and C. She and K. Ying and Y. Li and B. Vucetic},
  title   = {Deep Reinforcement Learning for Improving Resource Utilization
             Efficiency of {URLLC} With Imperfect Channel State Information},
  journal = IEEE_J_WCOML,
  volume  = {12},
  number  = {10},
  pages   = {1796--1800},
  month   = oct,
  year    = {2023},
  doi     = {10.1109/LWC.2023.3294910}
}

@article{Wei2025AoIURLLC,
  author  = {J. Wei and C. Shi and Y. Zhu and Y. Hu and A. Schmeink},
  title   = {Age-of-Information Minimization in Multi-Server Edge Computing
             Networks for {URLLC} Services},
  journal = IEEE_J_VT,
  volume  = {74},
  number  = {9},
  pages   = {14317--14330},
  month   = sep,
  year    = {2025},
  doi     = {10.1109/TVT.2025.3566137}
}

@article{Zhu2024UAVTrajectory,
  author  = {Y. Zhu and B. Yang and M. Liu and Z. Li},
  title   = {{UAV} Trajectory Optimization for Large-Scale and Low-Power
             Data Collection: An Attention-Reinforced Learning Scheme},
  journal = IEEE_J_WCOM,
  volume  = {23},
  number  = {4},
  pages   = {3009--3024},
  month   = apr,
  year    = {2024},
  doi     = {10.1109/TWC.2023.3304900}
}

@article{Huang2025UAVEdgeAI,
  author  = {J. Huang and L. Lian and D. Wen and Y. Zhou and F. Wang
             and W. Wang and Y. Shi},
  title   = {Dynamic {UAV}-Assisted Cooperative Edge {AI} Inference},
  journal = IEEE_J_WCOM,
  volume  = {24},
  number  = {1},
  pages   = {615--628},
  month   = jan,
  year    = {2025},
  doi     = {10.1109/TWC.2024.3497593}
}

@article{Bhowmick2022UAVCRNOMAEH,
  author  = {A. Bhowmick and S. D. Roy and S. Kundu},
  title   = {Throughput Maximization of a {UAV} Assisted {CR} Network
             With {NOMA}-Based Communication and Energy-Harvesting},
  journal = IEEE_J_VT,
  volume  = {71},
  number  = {1},
  pages   = {362--374},
  month   = jan,
  year    = {2022},
  doi     = {10.1109/TVT.2021.3123183}
}

@article{Singh2024AggregatedRFVLC,
  author  = {G. Singh and A. Srivastava and V. A. Bohara
             and M. Noor-A-Rahim and Z. Liu and D. Pesch},
  title   = {Toward {6G-V2X}: Aggregated {RF-VLC} for Ultra-Reliable
             and Low-Latency Autonomous Driving},
  journal = IEEE_M_COMSTD,
  volume  = {8},
  number  = {4},
  pages   = {80--87},
  month   = dec,
  year    = {2024},
  doi     = {10.1109/MCOMSTD.0001.2300005}
}

@article{Wu2025NLOSVLC,
  author  = {J. Wu and P. Zhang and Y. Huang and C. Liu and Y. Sun
             and C. Chen and Y. Yang and J. Luo},
  title   = {Exploring the Potential of Utilizing Nonline-of-Sight
             Channels for Networking in Visible Light Communication},
  journal = IEEE_J_IOT,
  volume  = {12},
  number  = {15},
  pages   = {31723--31736},
  month   = aug,
  year    = {2025},
  doi     = {10.1109/JIOT.2025.3574349}
}

@article{Li2026OpticalPassiveSensing,
  author  = {B. Li and J. Wang and D. Lian and Y. Gao and J. Lian
             and C. Tang and B. Lian},
  title   = {Passive Sensing Technique Using Cyclic Prefix in
             Power-Constrained Optical Wireless Communication Systems},
  journal = IEEE_J_VT,
  volume  = {75},
  number  = {6},
  pages   = {10426--10437},
  month   = jun,
  year    = {2026},
  doi     = {10.1109/TVT.2025.3640107}
}

@article{Ni2024EHScheduling,
  author  = {Z. Ni and Z. Zhang and N. C. Luong and D. Niyato
             and D. I. Kim and S. Feng},
  title   = {Joint Client Scheduling and Quantization Optimization in
             Energy Harvesting-Enabled Federated Learning Networks},
  journal = IEEE_J_WCOM,
  volume  = {23},
  number  = {8},
  pages   = {9566--9582},
  month   = aug,
  year    = {2024},
  doi     = {10.1109/TWC.2024.3363706}
}

@article{Hatami2025PartialBattery,
  author  = {M. Hatami and M. Leinonen and M. Codreanu},
  title   = {Status Updating Under Partial Battery Knowledge in
             Energy Harvesting {IoT} Networks},
  journal = IEEE_J_TGCN,
  volume  = {9},
  number  = {3},
  pages   = {1003--1020},
  month   = sep,
  year    = {2025},
  doi     = {10.1109/TGCN.2024.3484132}
}

@article{Shiraishi2025ContentWakeUp,
  author  = {J. Shiraishi and A. E. Kal{\o}r and I. Leyva-Mayorga
             and F. Chiariotti and P. Popovski and H. Yomo},
  title   = {Content-Based Wake-Up for Energy-Efficient and Timely
             Top-$k$ {IoT} Sensing Data Retrieval},
  journal = IEEE_J_COM,
  volume  = {73},
  number  = {12},
  pages   = {13983--13998},
  month   = dec,
  year    = {2025},
  doi     = {10.1109/TCOMM.2025.3605477}
}

@article{Ding2024NOMAAoI,
  author  = {Z. Ding and R. Schober and H. V. Poor},
  title   = {Impact of {NOMA} on Age of Information: A Grant-Free
             Transmission Perspective},
  journal = IEEE_J_WCOM,
  volume  = {23},
  number  = {5},
  pages   = {3975--3989},
  month   = may,
  year    = {2024},
  doi     = {10.1109/TWC.2023.3313612}
}

@article{Lu2026NOMAGrantFree,
  author  = {Lan Lu and {\v{C}}edomir Stefanovi{\'c} and Bo Ai
             and Yan Kyaw Tun and Yuxuan Sun and Wei Chen},
  title   = {{NOMA} Grant-Free Scheme in Periodically Scheduled Uplink
             for Time-Critical- and Massive-{MTC} in Railways},
  journal = IEEE_J_VT,
  volume  = {75},
  number  = {8},
  pages   = {17407--17422},
  month   = aug,
  year    = {2026},
  doi     = {10.1109/TVT.2026.3677464}
}

@article{Wang2025OVFLCSS,
  author  = {H. Wang and J. Xu},
  title   = {Online Vertical Federated Learning for Cooperative
             Spectrum Sensing},
  journal = IEEE_J_TCCN,
  volume  = {11},
  number  = {4},
  pages   = {2317--2326},
  month   = aug,
  year    = {2025},
  doi     = {10.1109/TCCN.2024.3519354}
}

@article{Tabella2023TimeAware,
  author  = {G. Tabella and D. Ciuonzo and Y. Yilmaz and
             X. Wang and Salvo Rossi, Pierluigi},
  title   = {Time-Aware Distributed Sequential Detection of Gas
             Dispersion via Wireless Sensor Networks},
  journal = IEEE_J_SIPN,
  volume  = {9},
  pages   = {721--735},
  year    = {2023},
  doi     = {10.1109/TSIPN.2023.3324586}
}

@article{Mao2024MultiBitSparse,
  author  = {L. Mao and S. Yan and Z. Sui and H. Li},
  title   = {Multi-Bit Distributed Detection of Sparse Stochastic
             Signals Over Error-Prone Reporting Channels},
  journal = IEEE_J_SIPN,
  volume  = {10},
  pages   = {881--893},
  year    = {2024},
  doi     = {10.1109/TSIPN.2024.3496253}
}

@article{Aziz2025MultiBitFusion,
  author  = {A. M. Aziz},
  title   = {From Centralized Data Fusion to Hard- and Soft-Decision
             Fusion in Multisensor Distributed Detection Systems:
             A New Multiple-Bit Decision Fusion Approach},
  journal = IEEE_J_ACCESS,
  volume  = {13},
  pages   = {167178--167195},
  month   = sep,
  year    = {2025},
  doi     = {10.1109/ACCESS.2025.3611786}
}

@article{Mao2026HybridBandwidth,
  author  = {L. Mao and Z. Sui and M. Matthaiou and H. Li},
  title   = {Distributed Detection and Bandwidth Allocation With
             Hybrid Quantized and Full-Precision Observations Over
             Multiplicative Fading Channels},
  journal = IEEE_J_VT,
  volume  = {75},
  number  = {6},
  pages   = {11710--11715},
  month   = jun,
  year    = {2026},
  doi     = {10.1109/TVT.2025.3644870}
}

@ARTICLE{Zhang2025MultiBitDDF,
  author={Zhang, Gaoyuan and Si, Ruisong and Chen, Kai and Wang, Weiguang and Mu, Yu and Li, Yiwei and Ji, Baofeng and Mumtaz, Shahid},
  journal=IEEE_J_SP, 
  title={Multi-bit Distributed Decision Fusion Rules for Multi-hop Multi-relay Wireless Sensor Networks}, 
  year={2025},
  volume={},
  number={},
  pages={1-14},
  doi={10.1109/TSP.2025.3608188}}

@article{Perez2026BlindFusion,
  author  = {J. P{\'e}rez and I. Santamar{\'i}a and
             A. Pag{\`e}s-Zamora},
  title   = {Blind Learning of the Optimal Fusion Rule in Wireless
             Sensor Networks},
  journal = J_SIGNAL_PROCESS,
  volume  = {239},
  pages   = {110238},
  month   = feb,
  year    = {2026},
  doi     = {10.1016/j.sigpro.2025.110238}
}

@ARTICLE{Ciuonzo2026HolographicTarget,
  author={Ciuonzo, Domenico and Zappone, A. and Di Renzo, Marco and d'Elia, Ciro},
  journal=IEEE_J_AES, 
  title={Holographic \& Channel-Aware Distributed Detection of a Non-Cooperative Target}, 
  year={2026},
  volume={},
  number={},
  pages={1-17},
  doi={10.1109/TAES.2026.3659081}}

@article{yan2015decision,
  title={Decision fusion with channel errors in distributed decode-then-fuse sensor networks},
  author={Yan, Yongsheng and Wang, Haiyan and Shen, Xiaohong and Zhong, Xionghu},
  journal=J_SENSORS,
  volume={15},
  number={8},
  pages={19157--19180},
  year={2015},
  publisher={MDPI}
}

@ARTICLE{Arafat2020-SJ,
  author={Al-Jarrah, Mohammad A. and Al-Dweik, Arafat and Ikki, Salama S. and Alsusa, Emad},
  journal=IEEE_J_SYS, 
  title={Spectrum-Occupancy Aware Cooperative Spectrum Sensing Using Adaptive Detection}, 
  year={2020},
  month={Jun.},
  volume={14},
  number={2},
  pages={2225-2236},
  doi={10.1109/JSYST.2019.2922773}}

@ARTICLE{Berber2026-LSP,
  author={Berber, Alperen and Dulek, Berkan},
  journal=IEEE_J_SPL, 
  title={Optimal Binary Hypothesis Testing Based on the Behavioral Kullback–Leibler Divergence Criterion}, 
  year={2026},
  volume={33},
  number={},
  pages={161-165},
  doi={10.1109/LSP.2025.3640513}}

@article{viswanathan1997distributed,
  author  = {Viswanathan, Ramanarayanan and Varshney, Pramod K.},
  title   = {Distributed Detection with Multiple Sensors: Part {I}---Fundamentals},
  journal = IEEE_J_PROC,
  year    = {1997},
  volume  = {85},
  number  = {1},
  pages   = {54--63},
  month   = jan,
  doi     = {10.1109/5.554208},
}

@article{zhang2024mary,
  author  = {Zhang, Gaoyuan and Li, Yongen and Ji, Baofeng and Chen, Kai and Mu, Yu and Wang, Weiguang},
  title   = {{M}-ary Distributed Decision Fusion for Multihop Relay Wireless Sensor Networks: Decision Fusion Rule, Implementation Framework, and Performance Analysis},
  journal = IEEE_J_IOT,
  year    = {2024},
  volume  = {11},
  number  = {18},
  pages   = {29569--29587},
}

@article{song2024energy,
  author  = {Song, Yu and Zhang, Shilong and Wang, Shubin},
  title   = {An energy efficient fusing data gathering protocol in wireless sensor networks},
  journal = J_COMPNET,
  year    = {2024},
  volume  = {243},
  pages   = {110305},
  doi     = {10.1016/j.comnet.2024.110305},
}

@article{lilhore2026securefusion,
  author  = {Lilhore, Umesh Kumar and Simaiya, Sarita and Rahoof, P. P. and Alroobaea, Roobaea and Baqasah, Abdullah M. and Alsafyani, Majed and Alhazmi, Afnan and Tekeste, Lidia Gosy},
  title   = {Advanced threat detection for smart cities through {IoT} sensor and network data integration with {IoT-securefusion}},
  journal = J_WIRELESS_COM_NET,
  year    = {2026},
  volume  = {2026},
  pages   = {20},
  doi     = {10.1186/s13638-025-02554-w},
}

@misc{zhang2025byzantine,
  author        = {Zhang, Gaoyuan and Song, Gaolei and Li, Boyuan and Li, Zijian and Ji, Baofeng and Zheng, Ruijuan and Zheng, Guoqiang and Quek, Tony Q. S.},
  title         = {Byzantine Attacks in {RIS}-Enhanced Cooperative Spectrum Sensing: A Decision Fusion Perspective},
  year          = {2025},
  howpublished  = {arXiv preprint arXiv:2510.27175},
  eprint        = {2510.27175},
  archivePrefix = {arXiv},
  primaryClass  = {cs.IT},
  doi           = {10.48550/arXiv.2510.27175},
}

@article{ciuonzo2025holographic,
  author  = {Ciuonzo, Domenico and Zappone, Alessio and Di Renzo, Marco},
  title   = {Channel-Aware Holographic Decision Fusion},
  journal = IEEE_J_IOT,
  year    = {2025},
  volume  = {12},
  number  = {16},
  pages   = {33011--33025},
  doi     = {10.1109/JIOT.2025.3574794},
}

@article{ciuonzo2025score,
  author  = {Ciuonzo, Domenico},
  title   = {Score-Based Fading-Aware Decision Fusion},
  journal = IEEE_J_SPL,
  year    = {2025},
  volume  = {32},
  pages   = {2952--2956},
  doi     = {10.1109/LSP.2025.3591002},
}

@article{AKYILDIZ201140,
title = {Cooperative spectrum sensing in cognitive radio networks: A survey},
journal = J_PHYS_COM,
volume = {4},
number = {1},
pages = {40-62},
year = {2011},
issn = {1874-4907},
author = {Ian F. Akyildiz and Brandon F. Lo and Ravikumar Balakrishnan}
}

@ARTICLE{cichon2016,
  author={Cichoń, Krzysztof and Kliks, Adrian and Bogucka, Hanna},
  journal=IEEE_J_COMST, 
  title={Energy-Efficient Cooperative Spectrum Sensing: A Survey}, 
  year={2016},
  volume={18},
  number={3},
  pages={1861-1886},
  doi={10.1109/COMST.2016.2553178}}

@Article{Arjoune2019,
AUTHOR = {Arjoune, Youness and Kaabouch, Naima},
TITLE = {A Comprehensive Survey on Spectrum Sensing in Cognitive Radio Networks: Recent Advances, New Challenges, and Future Research Directions},
JOURNAL = J_SENSORS,
VOLUME = {19},
YEAR = {2019},
NUMBER = {1},
ARTICLE-NUMBER = {126},
PubMedID = {30609719},
ISSN = {1424-8220},
DOI = {10.3390/s19010126}
}

@article{Li2026MultimodalAlignment,
  author  = {Songtao Li and Hao Tang},
  title   = {Multimodal Alignment and Fusion: A Survey},
  journal = J_COMPUTER_VISION,
  volume  = {134},
  month   = feb,
  year    = {2026},
  note    = {Art. no. 103},
  doi     = {10.1007/s11263-025-02667-1}
}

@inproceedings{Wei2025IntegratingSensors,
  author    = {Chuheng Wei and Ziye Qin and Ziyan Zhang
               and Guoyuan Wu and Matthew J. Barth},
  title     = {Integrating Multi-Modal Sensors: A Review of Fusion
               Techniques for Intelligent Vehicles},
  booktitle = IEEE_C_IV,
  pages     = {1817--1824},
  year      = {2025},
  doi       = {10.1109/IV64158.2025.11097396}
}

@article{Qian2025MultiSensorFusion,
  author  = {Hui Qian and Mingchen Wang and Maotao Zhu and Hai Wang},
  title   = {A Review of Multi-Sensor Fusion in Autonomous Driving},
  journal = J_SENSORS,
  volume  = {25},
  number  = {19},
  month   = oct,
  year    = {2025},
  note    = {Art. no. 6033},
  doi     = {10.3390/s25196033}
}

@article{Mittal2026SensorFusionModels,
  author  = {Sangeeta Mittal and Chetna Gupta and Varun Gupta},
  title   = {Sensor Fusion Models in Autonomous Systems: A Review},
  journal = J_CMC,
  volume  = {87},
  number  = {1},
  month   = feb,
  year    = {2026},
  note    = {Art. no. 6},
  doi     = {10.32604/cmc.2025.071599}
}

@article{agri2026,
    author = {Wang, Qiufen and Wang, Tianhao},
    title = {Machine learning-driven sensor fusion for precision agriculture: current applications and future perspectives},
    journal = {Sensor Review},
    volume = {46},
    number = {6},
    pages = {1029-1054},
    year = {2026},
    month = {02},
    issn = {0260-2288},
    doi = {10.1108/SR-08-2025-0613}
}

@ARTICLE{varshney2002,
  author={Biao Chen and Ruixiang Jiang and Kasetkasem, T. and Varshney, P.K.},
  journal={IEEE Transactions on Signal Processing}, 
  title={Channel aware decision fusion in wireless sensor networks}, 
  year={2004},
  volume={52},
  number={12},
  pages={3454-3458},
  doi={10.1109/TSP.2004.837404}}

@INPROCEEDINGS{supervised2012,
  author={Thilina, Karaputugala Madushan and Choi, Kae Won and Saquib, Nazmus and Hossain, Ekram},
  booktitle={2012 IEEE Global Communications Conference (GLOBECOM)}, 
  title={Pattern classification techniques for cooperative spectrum sensing in cognitive radio networks: SVM and W-KNN approaches}, 
  year={2012},
  volume={},
  number={},
  pages={1260-1265},
  doi={10.1109/GLOCOM.2012.6503286}}

@INPROCEEDINGS{game2008,
  author={Wang, Beibei and Liu, K.J. Ray and Clancy, T. Charles},
  booktitle={IEEE GLOBECOM 2008 - 2008 IEEE Global Telecommunications Conference}, 
  title={Evolutionary Game Framework for Behavior Dynamics in Cooperative Spectrum Sensing}, 
  year={2008},
  volume={},
  number={},
  pages={1-5},
  doi={10.1109/GLOCOM.2008.ECP.599}}

@ARTICLE{deep2017,
  author={Lee, Woongsup and Kim, Minhoe and Cho, Dong-Ho},
  journal={IEEE Transactions on Vehicular Technology}, 
  title={Deep Cooperative Sensing: Cooperative Spectrum Sensing Based on Convolutional Neural Networks}, 
  year={2019},
  volume={68},
  number={3},
  pages={3005-3009},
  doi={10.1109/TVT.2019.2891291}}

@article{heuristic2000,
title = {Threshold optimization algorithm for weak signal in distributed-sensor fusion system},
journal = {Signal Processing},
volume = {82},
number = {10},
pages = {1329-1336},
year = {2002},
issn = {0165-1684},
doi = {https://doi.org/10.1016/S0165-1684(02)00298-0},
url = {https://www.sciencedirect.com/science/article/pii/S0165168402002980},
author = {Weixian Liu and Yilong Lu and Jeffrey S. Fu},
}

@INPROCEEDINGS{rl2010,
  author={Lo, Brandon F. and Akyildiz, Ian F.},
  booktitle={21st Annual IEEE International Symposium on Personal, Indoor and Mobile Radio Communications}, 
  title={Reinforcement learning-based cooperative sensing in cognitive radio ad hoc networks}, 
  year={2010},
  volume={},
  number={},
  pages={2244-2249},
  doi={10.1109/PIMRC.2010.5671686}}

@article{V1_jiang2021collab,
  author  = {Zile Jiang and Xiaodi You and Chaoran Xiong and Gangxiang Shen and Biswanath Mukherjee},
  title   = {Collaborative spectrum sensing for cognitive visible light communications},
  journal = {Opt. Express},
  volume  = {29},
  number  = {11},
  pages   = {17683--17699},
  year    = {2021},
  doi     = {10.1364/OE.427776}
}

@inproceedings{V2_chen2014angle,
  author    = {Zhe Chen and Nikola Serafimovski and Harald Haas},
  title     = {Angle diversity for an indoor cellular visible light communication system},
  booktitle = {Proc. IEEE 79th Veh. Technol. Conf. (VTC Spring)},
  address   = {Seoul, South Korea},
  month     = may,
  year      = {2014},
  pages     = {1--5},
  doi       = {10.1109/VTCSpring.2014.7022832}
}

@inproceedings{V3_chen2014sinr,
  author    = {Zhe Chen and Dobroslav Tsonev and Harald Haas},
  title     = {Improving {SINR} in indoor cellular visible light communication networks},
  booktitle = {Proc. IEEE Int. Conf. Commun. (ICC)},
  address   = {Sydney, Australia},
  month     = jun,
  year      = {2014},
  pages     = {3383--3388},
  doi       = {10.1109/ICC.2014.6883844}
}

@article{V4_adr2022,
  author  = {Milton Rom{\'a}n Ca{\~n}izares and Pablo Palacios J{\'a}tiva and Cesar A. Azurdia-Meza and Samuel Montejo-S{\'a}nchez and Sandra C{\'e}spedes},
  title   = {Impact of diversity combining schemes in a multi-cell {VLC} system with angle diversity receivers},
  journal = {Photonic Netw. Commun.},
  volume  = {43},
  number  = {1},
  pages   = {13--22},
  year    = {2022},
  doi     = {10.1007/s11107-022-00966-y}
}

@article{V5_alsakkaf2024,
  author  = {Ahmed Gaafar Ahmed Al-Sakkaf and M{\'a}ximo Morales-C{\'e}spedes},
  title   = {Interference management for {VLC} indoor systems based on overlapping field-of-view angle diversity receivers},
  journal = {IEEE Access},
  volume  = {12},
  pages   = {51431--51449},
  year    = {2024},
  doi     = {10.1109/ACCESS.2024.3381968}
}

@article{V6_ahmad2020rl,
  author  = {Rizwana Ahmad and Mohammad Dehghani Soltani and Majid Safari and Anand Srivastava and Abir Das},
  title   = {Reinforcement learning based load balancing for hybrid {LiFi} {WiFi} networks},
  journal = {IEEE Access},
  volume  = {8},
  pages   = {132273--132284},
  year    = {2020},
  doi     = {10.1109/ACCESS.2020.3007871}
}

@article{V7_ahmad2022rl,
  author  = {Rizwana Ahmad and Mohammad Dehghani Soltani and Majid Safari and Anand Srivastava},
  title   = {Reinforcement learning-based near-optimal load balancing for heterogeneous {LiFi} {WiFi} network},
  journal = {IEEE Syst. J.},
  volume  = {16},
  number  = {2},
  pages   = {3084--3095},
  month   = jun,
  year    = {2022},
  doi     = {10.1109/JSYST.2021.3088302}
}

@article{V8_guo2017fusion,
  author  = {Xiansheng Guo and Sihua Shao and Nirwan Ansari and Abdallah Khreishah},
  title   = {Indoor localization using visible light via fusion of multiple classifiers},
  journal = {IEEE Photon. J.},
  volume  = {9},
  number  = {6},
  year    = {2017},
  doi     = {10.1109/JPHOT.2017.2767576}
}

@article{V9_tlfn2019,
  author  = {Xiansheng Guo and Fangzi Hu and Nkrow Raphael Elikplim and Lin Li},
  title   = {Indoor localization using visible light via two-layer fusion network},
  journal = {IEEE Access},
  volume  = {7},
  pages   = {16421--16430},
  year    = {2019},
  doi     = {10.1109/ACCESS.2019.2895131}
}

@ARTICLE{covert2023,
  author={Arghavani, Abbas and Dey, Subhrakanti and Ahlén, Anders},
  journal={IEEE Transactions on Signal Processing}, 
  title={Covert Outage Minimization in the Presence of Multiple Wardens}, 
  year={2023},
  volume={71},
  number={},
  pages={686-700},
  doi={10.1109/TSP.2023.3248869}}
\begin{IEEEbiography}[{\includegraphics[width=1in,height=1.25in,clip,keepaspectratio]{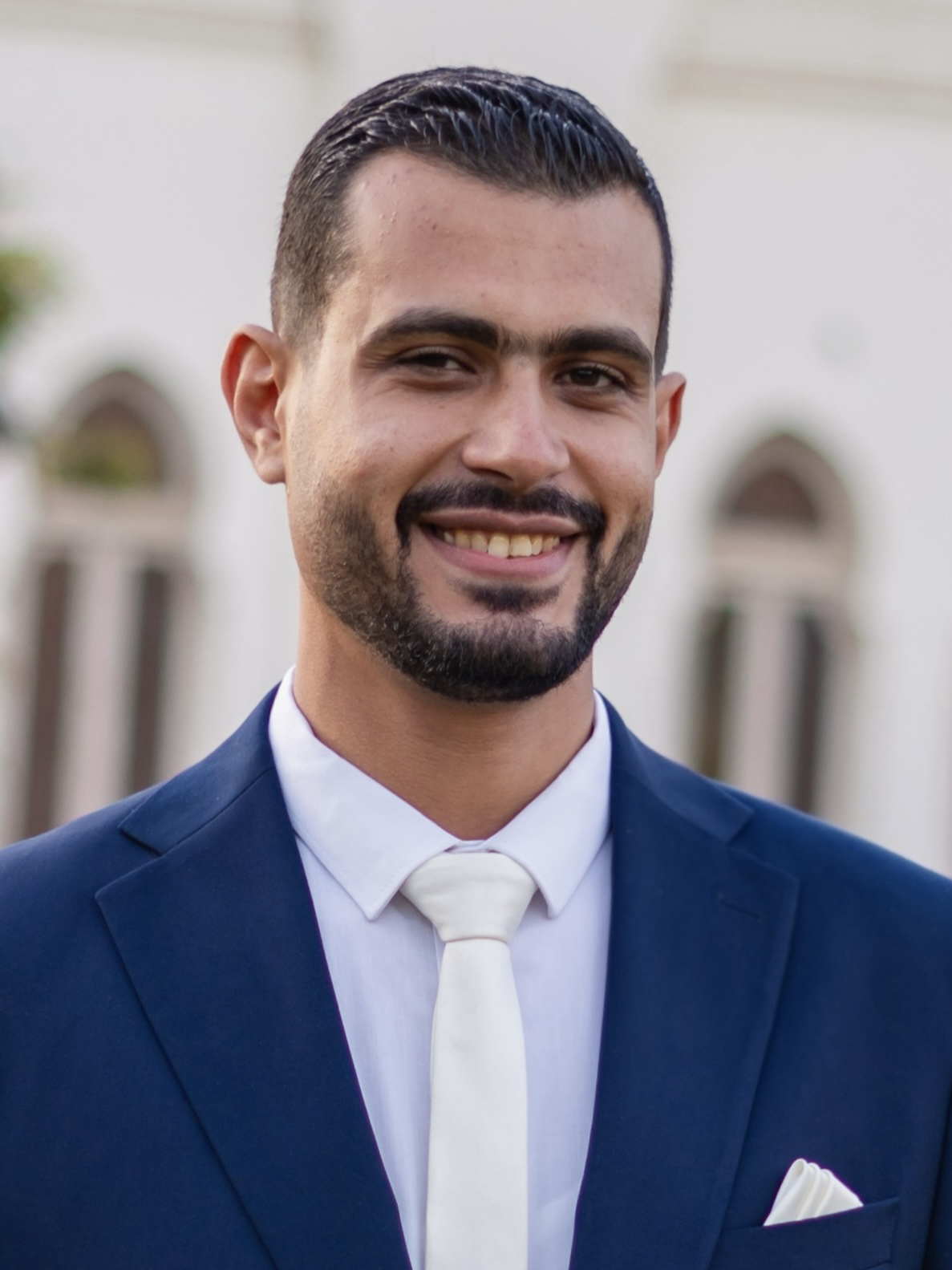}}]{Mohamed Abdelkader} received the B.Sc. and M.Sc. degrees in communication and electronics engineering from Alexandria University, Egypt, in 2019 and 2025, respectively. He is currently pursuing the Ph.D. degree in communications and computer engineering with Khalifa University, UAE. He previously served as a Teaching and Research Assistant with Alexandria University, The American University in Cairo, and Alamein International University, Egypt. His research interests include communication channel modeling, data fusion, spectrum sensing, wireless communication systems, UAV communications and machine learning.
\end{IEEEbiography}
\begin{IEEEbiography}[{\includegraphics[width=1in,height=1.25in,clip,keepaspectratio]{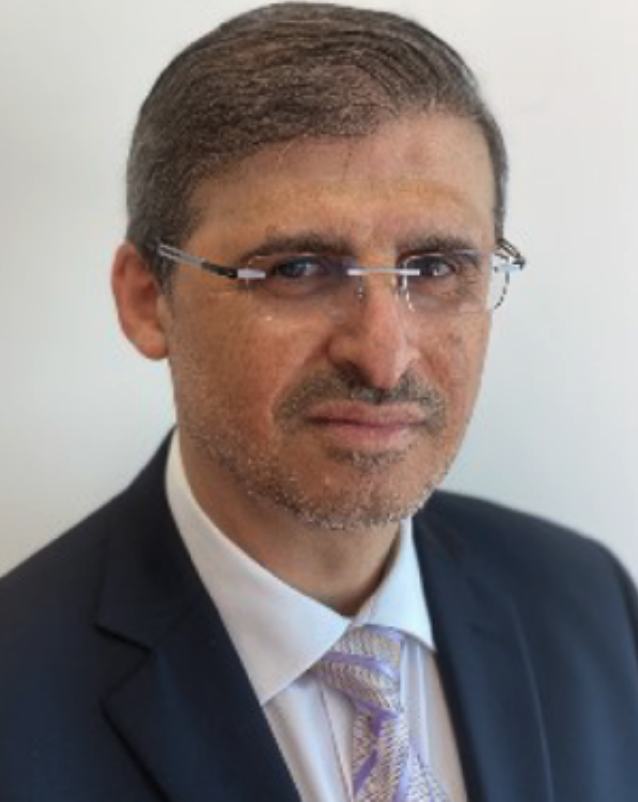}}]{Arafat Al-Dweik}(Senior Member, IEEE) received the M.S. (summa cum laude) and Ph.D. (magna cum laude) degrees in electrical engineering from Cleveland State University, Cleveland, OH, USA, in 1998 and 2001, respectively. He was with Efficient Channel Coding Inc., Cleveland; the Department of Information Technology, Arab American University, Jenin, Palestine, and the University of Guelph, Guelph, ON, Canada. He is currently with the Department of Electrical Engineering and Computer Science, Khalifa University, Abu Dhabi, United Arab Emirates. He was a Visiting Research Fellow with the School of Electrical, Electronic, and Computer Engineering, Newcastle University, Newcastle upon Tyne, U.K., and a Research Professor with Western University, London, ON, Canada, and the University of Guelph. His research focuses on various areas of wireless communications and signal processing, such as multiple access, signal design, transceiver design, the IoT, WSNs, and AI for wireless communications. He received the UAE Leader-Founder Award 2019, Dubai Award for Sustainable Transportation 2016, the Hijjawi Award for Applied Sciences 2003, and the Fulbright Alumni Development Grants 2003 and 2005. He was awarded the Fulbright Graduate Student Scholarship in 1997. He has been an Editor of IEEE TRANSACTIONS ON VEHICULAR TECHNOLOGY since 2012 and an Associate Editor of Frontiers in Communications and Networks and IET Communications from 2015 to 2020. He is a registered Professional Engineer in Province of Ontario, Canada. He is a Distinguished Lecturer of IEEE from 2023 to 2025.\end{IEEEbiography}
\begin{IEEEbiography}[{\includegraphics[width=1in,height=1.25in,clip,keepaspectratio]{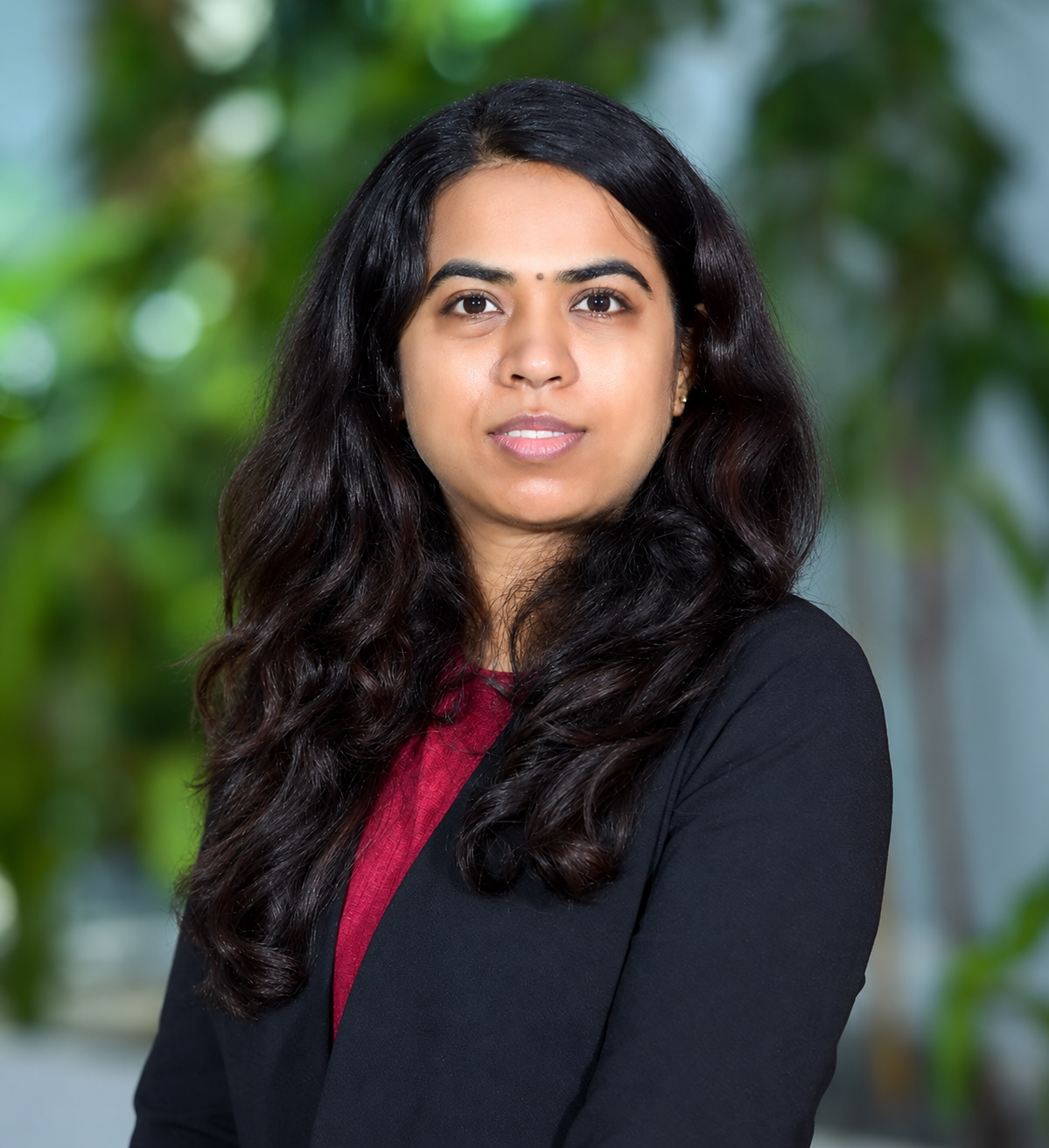}}]{Sobia Jangsher}
is an Assistant Professor in the School of Electronic Engineering at Dublin City University (DCU), Ireland, and a Funded Investigator with the CONNECT Research Centre. She received her Ph.D. degree from The University of Hong Kong in 2015, where her research focused on optimization for wireless communication systems. Prior to joining DCU in 2023, she was an Assistant Professor at the Institute of Space Technology, Islamabad, Pakistan, from 2015 to 2021, and subsequently a Researcher in the Department of Electrical Engineering and Computer Science at Khalifa University, Abu Dhabi, UAE, from 2021 to 2023. Her research interests include the modelling, design, optimization, and performance evaluation of emerging and future wireless networks. Her work spans intelligent reflecting surface (IRS)-assisted communications, multiple access techniques, small-cell and mobile networks, edge-enabled wireless systems, resource allocation, and AI/ML-enabled wireless communications. She is also interested in energy-efficient neural receivers and low-complexity learning-based physical-layer solutions for reducing computational and energy overhead while maintaining reliable communication performance. She has led and contributed to competitively funded national and international research projects, including interdisciplinary work involving wireless, optical, and future network architectures. Her broader research agenda focuses on resource-efficient, intelligent, and adaptive communication systems for beyond-5G and 6G networks.\end{IEEEbiography}
\begin{IEEEbiography}[{\includegraphics[width=1in,height=1.25in,clip,keepaspectratio]{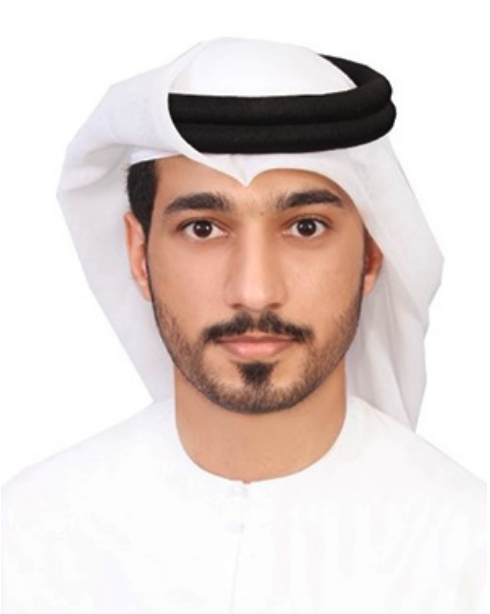}}]{Hamad Yahya}(Member, IEEE) received the M.Sc. degree (with Distinction) in communications and signal processing and the Ph.D. degree in electrical and electronic engineering from The University of Manchester, Manchester, U.K., in 2019 and 2023, respectively. He is currently enrolled in an UAE National Postdoctoral Fellowship Program with Khalifa University. He is also a Visiting Researcher with the Wireless Communications Research Group, Imperial College London, U.K. Prior joining Khalifa University, he worked as a Research Associate with the Communication Systems Research Group, The University of Manchester. His research interests include modeling, performance analysis, and optimization of wireless communications systems. He received the Best Student and the Best Project Prizes from the Department of Electrical and Electronic Engineering, The University of Manchester, where he was the Co-Founder and the Chair of IEEE ComSoc SBC.
\end{IEEEbiography}
\begin{IEEEbiography}[{\includegraphics[width=1in,height=1.25in,clip,keepaspectratio]{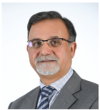}}]{Shihab Jimaa}
is currently an Associate Professor at the department of Computer and Information Engineering and  Associate Dean of UG studies at the College of Computing and Mathematical Sciences. He received his PhD and MSc degrees in Digital Communications from Loughborough University, England, UK, in 1990, and 1986, respectively. Also Dr. Jimaa holds a Post Graduate Certificate in Higher Education (PGCE) from the University of Hertfordshire, UK in 1998. He held several positions as Research Fellow, Senior Lecturer, and Assistant Professor at various institutions include Loughborough University/UK, Warwick University/UK, Near-East University/Cyprus, Amman University/Jordan, and the University of Hertfordshire/UK. His research interests include energy efficient wireless communication systems and networks, channel estimation and equalization, Atrial fibrillation data analysis and validation, and signal processing for digital communications/biomedical engineering. He successfully supervised 30 MSc, 3 MPhil, and 4 PhD students. Dr. Jimaa is a UK Chartered Electrical Engineer, a Senior Member of the IEEE Communications and Education societies, and a Fellow Member of the Higher Education Academy (HEA)/UK. He has authored 3 book chapters, 1 patent, and more than 100 papers in various referred international conferences and journals.
\end{IEEEbiography}
\end{document}